\documentclass[a4paper,11pt,openany]{book}

\usepackage{indentfirst}
\usepackage[dvips]{graphicx}
\usepackage{amsmath,amssymb}
\usepackage[small]{caption}

\usepackage{graphicx, fancybox}
\usepackage{amsmath,amssymb,cite}
\usepackage[dvipdfm, colorlinks=true, pdfstartview=FitV, linkcolor=black, citecolor=blue, urlcolor=blue]{hyperref}
\usepackage{color}

\newcommand{\vect}[1]{\mathbf{#1}}
\newcommand{\vzero}{\vect{0}}
\newcommand{\vp}{\vect{p}}
\newcommand{\vk}{\vect{k}}
\newcommand{\vgamma}{{\boldsymbol \gamma}}
\newcommand{\vx}{\vect{x}}
\newcommand{\vy}{\vect{y}}

\newcommand{\vv}{\vect{v}}
\newcommand{\vE}{\vect{E}}

\newcommand{\im}{\mathrm{Im}\,}
\newcommand{\re}{\mathrm{Re}\,}
\newcommand{\sgn}{\mathrm{sgn}}
\newcommand{\Slash}[1]{\ooalign{\hfil/\hfil\crcr$#1$}}

\newcommand{\Tr}{\mathrm{Tr}}
\newcommand{\SR}{S^R}
\newcommand{\DR}{D^{R}}

\newcommand{\SF}{S^{S}}
\newcommand{\DF}{D^{S}}

\newcommand{\cp}{g}
\newcommand{\kernel}{X} 
\newcommand{\selfEnergyR}{\varSigma^R}
\newcommand{\Nf}{N_f}
\newcommand{\Cf}{C_f}

\newcommand{\LQCD}{\varLambda_{\text{QCD}}}

\newcommand{\deltam}{\delta m}
\newcommand{\damping}{\zeta}
\newcommand{\pzero}{p^0}
\newcommand{\pw}{(\pzero,\vp)}
\newcommand{\GaugeFixing}{\tilde{G}}

\newcommand{\mb}{m_b}
\newcommand{\mf}{m_f}
\newcommand{\mph}{m_\gamma}
\newcommand{\me}{m_e}
\newcommand{\mq}{m_q}
\newcommand{\mg}{m_g}
\newcommand{\zetaf}{\zeta_f}
\newcommand{\zetab}{\zeta_b}
\newcommand{\zetae}{\zeta_e}
\newcommand{\zetaph}{\zeta_\gamma}
\newcommand{\zetaq}{\zeta_q}
\newcommand{\zetag}{\zeta_g}
\newcommand{\omegae}{\omega_{0e}}
\newcommand{\omegaph}{\omega_{0\gamma}}

\newcommand{\comment}[1]{}

\newcommand{\jind}{j_{\mathrm {ind}}}
\newcommand{\etaind}{\eta_{\mathrm {ind}}}

\newcommand{\Tc}{{\mathrm T}_C}
\newcommand{\R}{R}
\newcommand{\A}{A}
\newcommand{\nf}{n_F}
\newcommand{\nb}{n_B}
\newcommand{\VC}{ \varXi }
\newcommand{\V}{\tilde{\varGamma}}
\newcommand{\thetac}{\theta_C}

\newcommand{\ValidPerturbation}{Strickland:2010tm, Andersen:1999sf, Andersen:1999fw, Blaizot:1999ip, Blaizot:1999ap, Blaizot:2000fc, Blaizot:2003iq}
\newcommand{\HardPhotonDamping}{Aurenche:2000gf, Aurenche:1999tq, Baier:1991em, Kapusta:1991qp, Arnold:2001ba, Arnold:2001ms, Arnold:2002ja}
\newcommand{\AnomalousDamping}{Blaizot:1996az, Blaizot:1996hd, Lebedev:1990kt, Lebedev:1990un, Pisarski:1993rf}
\newcommand{\TransportRP}{Jeon:1994if, Wang:1999gv, Carrington:1999bw, ValleBasagoiti:2002ir, Wang:2002nba, Gagnon:2006hi, Gagnon:2007qt, Hidaka:2010gh}
\newcommand{\TransportBoltzmann}{Arnold:2000dr, Arnold:2003zc, Defu:2005hb}
\newcommand{\TransportTwoPI}{Aarts:2003bk, Aarts:2004sd, Aarts:2005vc}

\newcommand{\HTL}{Frenkel:1989br, Braaten:1990az}

\newcommand{\plasmino}{Klimov:1981ka, Weldon:1982bn, Weldon:1989ys}
\newcommand{\UltrasoftGluon}{Blaizot:1999fq, Blaizot:1999xk, Bodeker:2002gy, Bodeker:1999ud, Bodeker:1999ey, Bodeker:2000da, Bodeker:1998hm} 
\newcommand{\HTLResum}{Pisarski:1988vd, Braaten:1989kk, Braaten:1989mz, Braaten:1990it}
\newcommand{\HTLResumQuarkDamping}{Braaten:1992gd, Kobes:1992ys}
\newcommand{\GaugeIndependence}{Kobes:1990xf, Kobes:1990dc, Rebhan:2001wt}
\newcommand{\HTLVlasov}{Blaizot:1993be, Blaizot:1993zk, Blaizot:1992gn}
\newcommand{\ThreepeakSD}{Harada:2008vk, Harada:2007gg, Qin:2010pc, Nakkagawa:2012ip, Nakkagawa:2011ci}
\newcommand{\CSC}{Barrois:1977xd, Bailin:1983bm, Iwasaki:1994ij, Alford:1997zt, Rapp:1997zu, Alford:2007xm}
\newcommand{\NJL}{Nambu:1961tp, Nambu:1961fr, Hatsuda:1994pi}
\newcommand{\PNJL}{Fukushima:2003fw, Fukushima:2008wg, Ratti:2005jh}
\newcommand{\Scoupling}{Munster:1980iv, deForcrand:2009dh, Nakano:2009bf}
\newcommand{\Keldysh}{Schwinger:1960qe, Keldysh:1964ud}
\newcommand{\PinchSingularity}{Niemi:1983ea, Matsumoto:1983by, Blaizot:1999fq, Blaizot:1999xk, Bodeker:2002gy, Bodeker:1999ud, Bodeker:1999ey, Bodeker:2000da, Bodeker:1998hm, Jeon:1994if, Wang:1999gv, Carrington:1999bw, ValleBasagoiti:2002ir, Wang:2002nba, Gagnon:2006hi, Gagnon:2007qt, Hidaka:2010gh, Arnold:2000dr, Arnold:2003zc, Defu:2005hb, Aarts:2003bk, Aarts:2004sd, Aarts:2005vc, Carrington:2009xf, Carrington:2007ea} 
\newcommand{\BackgroundFieldMethod}{DeWitt:1967ub, DeWitt:1975ys, Abbott:1980hw, Meissner:1986tr, Hansson:1987um, Hansson:1987un, Abbott:1981ke}
\newcommand{\DampingProblem}{Kalashnikov:1979cy, Hansson:1987um, Hansson:1987un, Kobes:1987bi, Elze:1987rh}
\newcommand{\ImaginaryChemicalPotential}{Roberge:1986mm, Sakai:2008um}
\newcommand{\Goldstino}{Kratzert:2003cr, Lebedev:1989rz, Boyanovsky:1983tu, Aoyama:1984bk, Gudmundsdottir:1986uq, Gudmundsdottir:1986ur }
\newcommand{\TransitionTemperature}{Bazavov:2009zn, Bernard:1996cs, Cheng:2007jq, Kogut:1982rt, Fodor:2004nz} 
\newcommand{\HDL}{Altherr:1992mf, Vija:1994is, Manuel:1995td}
\newcommand{\Neutrino}{Boyanovsky:2005hk, Giudice:2003jh, Notzold:1987ik}
\newcommand{\SUSYColdAtom}{Yu:2007xb, Shi:2009ak, Snoek:2006nf, Yu:2010zv, Olemskoi:2009, Snoek:2005, Lozano:2007, Imambekov:2006} 
\newcommand{\IsospinChemicalPotential}{Kogut:2004zg, Sinclair:2006zm, deForcrand:2007uz, Detmold:2012wc}

\begin{document}
\frontmatter
\pagestyle{empty}
\begin{center}

{\Large 

Ph.D.~thesis

\vskip 1.5cm

{\bf Diagrammatic and Kinetic Equation Analysis of Ultrasoft Fermionic Sector in Quark-Gluon Plasma
} 

\vskip 1.5cm

{\bf Daisuke Satow}

\vskip 1.5cm

{\it Nuclear Theory Group\\
Department of Physics\\
Graduate School of Science\\
Kyoto University\\ }
\vskip 1cm
Accepted in January 2013
}

\vskip 2cm

\begin{figure}[h]
\centering
\includegraphics[scale=0.8]{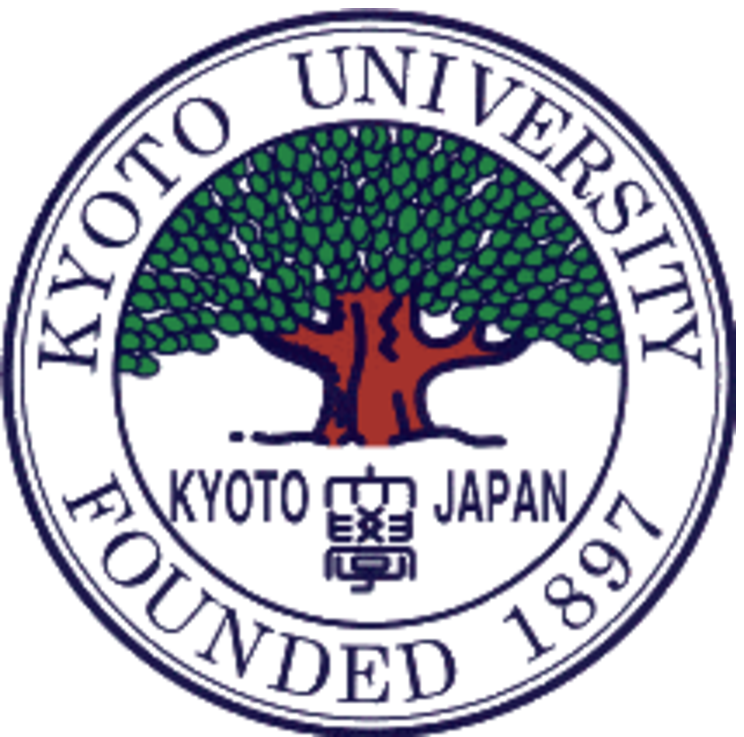}
\end{figure}

\end{center}

\chapter*{Abstract}
\setcounter{page}{1}
At so high temperature ($T$) that the coupling constant ($g$) is small and the masses of the particles are negligible, different scheme has to be applied in each energy scale in the analysis of the quark-gluon plasma (QGP).
In the soft energy region ($p\sim gT$), the simple perturbative expansion called the hard thermal loop (HTL) approximation can be applied, and that approximation expects the existence of the bosonic and fermionic collective excitations called plasmon and plasmino.
On the other hand, in the ultrasoft energy region ($p\sim g^2T$), the HTL approximation is inapplicable due to infrared singularity, so the question whether there are any excitation modes in that energy region has not been studied well.

In this thesis, we analyze the quark spectrum whose energy is ultrasoft in QGP, using the resummed perturbation theory which enables us to successfully regularize the infrared singularity.
Since the Yukawa model and QED are simpler than QCD but have some similarity to QCD, we also work in these models.
As a result, we establish the existence of a novel fermionic mode in the ultrasoft energy region, and obtain the expressions of the pole position and the strength of that mode.
We also show that the Ward-Takahashi identity is satisfied in the resummed perturbation theory in QED/QCD.

Furthermore, we derive the linearized and generalized Boltzmann equation for the ultrasoft fermion excitations in the Kadanoff-Baym formalism, and show that the resultant equation is equivalent to the self-consistent equation in the resummed perturbation theory. 
We also derive the equation which determines the $n$-point functions with external lines for a pair of fermions and $(n-2)$ bosons with ultrasoft momenta, by considering the non-linear response regime using the gauge symmetry. 
We also derive the Ward-Takahashi identity from the conservation law of the electromagnetic current in the Kadanoff-Baym formalism.

\tableofcontents 


\mainmatter
\chapter{Introduction} 
\label{chap:intro}

\thispagestyle{headings}
\pagestyle{headings}

\section{Quark-gluon plasma} 
\label{sec:intro-QGP}

Hadrons are composed of the fermions called quarks and the gauge bosons called gluons, and the dynamics of these particles are described by the quantum chromodynamics (QCD). 
The Lagrangian of QCD is as follows:
\begin{align}
{\cal L}=-\frac{1}{4}F^{\mu \nu}_a F^a_{\mu \nu}+\sum^{\Nf}_{\alpha=1}\overline{\psi}_\alpha(i\Slash{D}-m_\alpha)\psi_\alpha.
 \hspace{30pt} (a=1,2...N^2-1)
\end{align} 
Here $F^{\mu \nu}_a=\partial^\mu A_a^\nu -\partial^\nu A_a^\mu-\cp f_{abc}A^\mu_bA^\nu_c$ is the field strength, $\psi_\alpha$ the quark field, $\cp$ the coupling constant, $D^\mu\equiv\partial^\mu+i\cp A^{\mu a}t_a$ the covariant derivative, $f_{abc}$ the structure constant of $SU(N)$, $m_\alpha$ the current quark mass, respectively.
The subscript $\alpha$ in the quark field stands for the flavor. 
In the real world, $N=3$, $\Nf=6$ (up, down, charm, strange, top, and bottom), and the values for $m_\alpha$ are as follows\footnote{The values written here are those of running masses in $\overline{\text{MS}}$ scheme.
For $u$, $d$, and $s$ quarks, the scale is $\kappa\sim 2$GeV while for $c$, $b$, and $t$ quarks, the scale is set to that of the mass of each particle.
}~\cite{Beringer:1900zz}: 
$m_u\simeq$ 2 MeV, $m_d\simeq$ 6 MeV, $m_c\simeq$ 1.2 GeV, $m_s\simeq$ 95 MeV, $m_b\simeq$ 4.1 GeV, $m_t\simeq$ 175 GeV.

Let us briefly review the present theoretical understandings on the phase structure of QCD matter. 
Since the quark number is conserved in each flavor, there exist the chemical potential ($\mu_i$) which are related to the conserved quark number.
It is possible to analyze the case that the chemical potentials of each flavor are different~\cite{\IsospinChemicalPotential}, but in the following we will review only the case that they have the same value: $\mu_i=\mu$.
We can make rather reliable predictions in the case of extremely high temperature or chemical potential, because of the asymptotic freedom.
At high chemical potential, smallness of $\cp$ justifies us to expect that the system should be a Fermi liquid with a Fermi sphere of the quarks.
However, there is an attractive interaction between the two quarks in the color anti-triplet channel via gluon exchange.
This attractive interaction causes an instability of the Fermi surface as is well known, and it results in the permanent formation of the Cooper pair composed of quarks leading to the superconducting phase called color superconducting (CSC) phase~\cite{\CSC}. 
Also in the case that temperature is extremely high, the quarks and the gluons are expected to be deconfined~\cite{Collins:1974ky}. 
That phase is called quark-gluon plasma (QGP)~\cite{Yagi:2005yb}.

Since $\cp$ is not so small when temperature and chemical potential are not extremely large\footnote{The values for $\cp$ determined by experiments in some energy scales are as follows~\cite{Beringer:1900zz}: $\cp$(1 GeV)$\simeq 2.5$, $\cp$(10 GeV)$\simeq 1.5$, and $\cp$(100 GeV)$\simeq 1.2$.}, it is difficult to what phase is realized in that region. 
Nevertheless, the first principle calculation using the Monte Carlo simulation, which is called lattice QCD, is possible when the chemical potential is zero.
The lattice results tell us that the transition from the confined phase to the deconfined one is crossover, not the phase transition in the usual sense, and that the transition temperature ($T_c$) is approximately 200 MeV~\cite{\TransitionTemperature}.
Analyses in the small-$\mu$ region have been tried by using some methods such as the Taylor expansion method~\cite{Allton:2002zi} and the imaginary chemical potential method~\cite{\ImaginaryChemicalPotential}. 

\begin{figure}[t] 
\begin{center}
\includegraphics[width=0.6\textwidth]{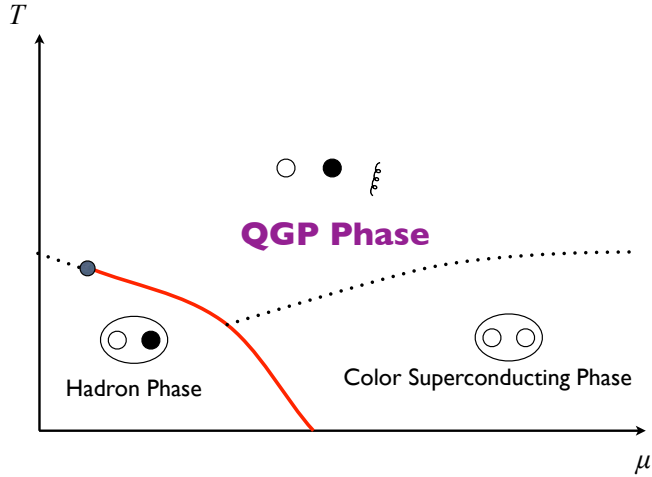}
\caption{schematic figure of a possible phase diagram of QCD.
The vertical axis is temperature axis while horizontal one is the chemical potential axis.}
\label{fig:PhaseDiagram}
\end{center}
\end{figure}

In the case that the chemical potential is finite, the lattice calculation is unreliable at present, so we have to use some effective models such as the Nambu-Jona-Lasino (NJL) model~\cite{\NJL} and its improved version~\cite{\PNJL}, the chiral random matrix model~\cite{Shuryak:1992pi}, and the strong coupling expansion~\cite{\Scoupling}. 
Many of the analyses~\cite{Asakawa:1989bq, Barducci:1989wi, Wilczek:1992sf, Berges:1998rc} suggest the following:
There is a second-order critical point at finite $T$ and $\mu$. 
The first order phase transition line exist, whose one end is connected with the critical point and the other is on the $\mu$ axis.
Summarizing the suggestions above and the expectations based on the lattice QCD and the asymptotic freedom, a schematic figure of a possible phase diagram of QCD is drawn in Fig.~\ref{fig:PhaseDiagram}. 

In this thesis, we focus on the analysis of the QGP phase.
QGP is considered to be realized by heavy ion collision experiment~\cite{Muller:2006ee, Gyulassy:2004zy}, which is performed in the Relativistic Heavy Ion Collider in Brookhaven national laboratory and Large Hadron Collider in CERN. 
QGP was also realized in the early universe since it is considered that temperature was higher than $T_c$ when the age of universe is $\sim 10^{-5}$ seconds or younger.
For this reason, the properties of QGP are relevant to the analysis of the early universe.
QGP is the only experimentally realizable fermion-boson system in which the masses are negligible compared with temperature. 
Because of this property, fermionic collective excitations appear in QGP, which is the main topic of this thesis.
We note that the Yukawa model and QED at high temperature have that property, so we also treat these models, which are simpler than QCD, in this thesis.
Throughout this thesis, the Yukawa model means the model in which a fermion (denoted by $\psi$) is coupled with a scalar field $\phi$.
The Lagrangian is
\begin{align}
\label{eq:Yukawa-Lagrangian}
{\cal L}[\phi, \psi, \overline{\psi}]= \frac{1}{2}(\partial^\mu\phi)^2+\overline{\psi}i(\Slash{\partial}+i\cp\phi)\psi.
\end{align}
We do not include the possible self-coupling of the scalar fields for simplicity.

\section{Hierarchy of energy scales}  
\label{sec:intro-review} 
\thispagestyle{headings} 

In the following, we consider the case that the temperature ($T$) is so high that the current quark mass and the nonperturbative effect are negligible: $T\gg m_i$, $\LQCD$.
Here $\LQCD$ is a scale parameter, which is of order 200 MeV.
The inequality $T\gg\LQCD$ implies that $\cp\ll 1$ so that the perturbative expansion in terms of the coupling constant is applicable.
We consider the case that the chemical potential is zero ($\mu= 0$).
Some of $m_i$ are very large, so it seems that the temperature at which the massless approximation is valid is too high to be realized.
However, the heavy quarks are almost decoupled from the system if their masses are much larger than $T$, so that approximation is expected to be valid for light quarks in that case.
One may also have doubt on the validity of the perturbative expansion:
the coupling constant is not small enough for justification of the perturbative expansion in realistic temperature.
Nevertheless, there are some discussions~\cite{\ValidPerturbation} which suggest that the perturbative expansion is at least partially applicable if the temperature is higher than $\sim 2.5 T_c$.
For example, the entropy calculated from approximately self-consistent scheme which is motivated by the existence and the properties of the collective excitations predicted from the perturbation theory, agrees very well with that calculated from the lattice QCD~\cite{Blaizot:1999ip, Blaizot:1999ap, Blaizot:2000fc, Blaizot:2003iq}. 

\begin{figure}[t] 
\begin{center}
\includegraphics[width=\textwidth]{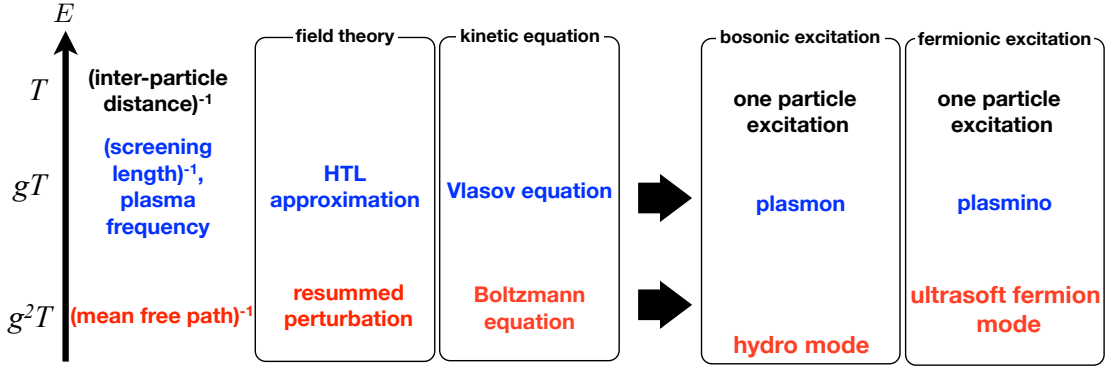}
\caption{The energy hierarchy of high temperature QCD.
The characteristic quantities, the correspondence between the methods in the field theory and the kinetic equations, and the bosonic and the fermionic modes obtained from these theories in each scale are shown.
The vertical axis denote the energy scale.} 
\label{fig:intro}
\end{center}
\end{figure}

In the temperature region described above, the perturbative method suggests that the system has multi-energy scale structure~\cite{Blaizot:2001nr}, which will be reviewed in the following subsections and summarized in Fig.~\ref{fig:intro}.
We note that the structure appears not only in QCD but also in the Yukawa model and QED.

Among the properties of QGP, we focus on the spectrums of the quarks and the gluons.
These quantity are quite important to elucidate the picture how the quarks and the gluons, which are the basic building blocks of QCD, behave in the system.
Naively, that problem seems to be trivial in the weak coupling regime:
A naive expectation might be that the spectrum of the particle seems to be almost the same as that of free particle since $\cp$ is small.
We will see that it is not the case when the energy of the particle is $\sim \cp T$ or $\cp^2T$ in the following subsections.

\subsection{Hard scale} 
\label{ssc:intro-hard}
\subsubsection{Properties of hard particles}

The energy scale $T$ is called hard scale.
One example of appearance of this scale in the theory is the average inter-particle distance, which can be shown as follows~\cite{lebellac}:
For simplicity, we show in the case of the boson in the Yukawa model.
The number density of the boson in the free limit is
\begin{align} 
 \nonumber
 \overline{n}_B&= \int\frac{d^3\vk}{(2\pi)^3} \nb(|\vk|) \\
 \nonumber
&=\frac{1}{2\pi^2}\int^\infty_0 d|\vk| |\vk|^2 \nb(|\vk|)\\
\label{eq:intro-boson-number}
&=\frac{\zeta(3)}{\pi^2}T^3,
\end{align}
where $\nb(k^0)\equiv {1/(\exp(k^0/T)-1)}$ is the Bose-Einstein distribution function, and $\zeta(s)$ is the Riemann zeta function.
$\zeta(3)$ is approximately equal to 1.202.
We note that the volume occupied by one particle is $(\overline{n}_B)^{-1}$.
Since $\overline{n}_B^{1/3}$ is of order $T$, the average inter-particle distance is of order $T^{-1}$. 

The de Broglie wavelength of the hard particle, whose momentum is of order $T$, is also of order $T^{-1}$.
This fact suggests that the properties of the hard particle is not so different from those of the free particle because the overlap among the particles is not large enough to compensate the smallness of the coupling constant.
We show that this suggestion is valid in the following.
One modification of the dispersion relation of the hard particle is the momentum-independent mass called asymptotic thermal mass~\cite{Thoma:1994yw,  Flechsig:1995ju, Blaizot:1999ap, Blaizot:2000fc, Blaizot:2001nr}, which is of order $\cp T$.
The expressions of the asymptotic thermal masses for the fermions and the bosons, in the Yukawa model and QED/QCD, are as follows:
\begin{align} 
\label{eq:mass-fermion}
\mf^2&= \frac{\cp^2 T^2}{8},\\
\label{eq:mass-boson}
\mb^2&= \frac{\cp^2 T^2}{6} , \\
\label{eq:mass-electron}
\me^2&=\frac{\cp^2T^2}{4}, \\
\label{eq:mass-photon}
\mph^2&= \frac{\cp^2 T^2}{6} , \\
\label{eq:mass-quark}
\mq^2&=\frac{\cp^2T^2}{4}\Cf, \\
\label{eq:mass-gluon}
\mg^2&= \frac{\cp^2T^2}{6}\left(N+\frac{\Nf}{2}\right).
\end{align}
Here $\mf^2$ ($\mb^2$), $\me^2$ ($\mph^2$), and $\mq^2$ ($\mg^2$) are the squares of the masses for the fermion (boson) in the Yukawa model~\cite{Thoma:1994yw}, electron (photon) in QED~\cite{Flechsig:1995ju}, quark (gluon) in QCD~\cite{Flechsig:1995ju, Blaizot:1999ap, Blaizot:2000fc, Blaizot:2001nr}, respectively, and we have introduced $\Cf\equiv (N^2-1)/(2N)$.
We use $\cp$ as a coupling constant not only in QCD but also in the Yukawa model and QED instead of the standard notation ($e$ in the case of QED), to make it clear that the same order counting appears as that in QCD in this thesis.

To demonstrate that the momentum-independent masses appear as a result of the perturbative analysis, here we perform an analysis at the one-loop order in the Yukawa model.
In the present and the next chapter, we employ the real-time formalism in Keldysh basis~\cite{\Keldysh}. 
The retarded self-energy of a massless fermion (the boson) $\varSigma^\R(p)$ ($\varPi^\R(p)$) at $\mu=0$ is defined as follows:
\begin{align} 
\label{eq:intro-propagator-selfenrgy-fermion}
\SR(p)&= \frac{-1}{(p^0+i\epsilon)\gamma^0-\vp\cdot\vgamma-\varSigma^\R(p)},\\
\DR(p)&= \frac{-1}{p^2+ip^0\epsilon-\varPi^\R(p)},
\end{align} 
where $\SR(p)$ ($\DR(p)$) is the retarded fermion (boson) propagator.
The pole positions of the fermion and the boson are determined by the conditions that the inverse of the retarded propagators are zero:
\begin{align} 
\label{eq:intro-fermion-polecondition}
(p^0+i\epsilon-\varSigma^0(p))^2-|\vp|^2\left(1-\frac{\varSigma^V(p)}{|\vp|}\right)^2&=0 ,\\
\label{eq:intro-boson-polecondition}
p^2+ip^0\epsilon-\varPi^\R(p)&=0 .
\end{align}
Note that since $\mu=0$ and the fermion is massless, $\varSigma^\R(p)$ can be expressed in terms of the scalar functions $\varSigma^0(p)$ and $\varSigma^V(p)$ as $\varSigma^\R(p)=\gamma_0 \varSigma^0(p)-\hat{\vp}\cdot\vgamma\varSigma^V(p)$ with $\hat{\vp}\equiv\vp/|\vp|$.
Thus we see that the self-energy gives the modification of the pole position due to the interaction.

\begin{figure}[t]
\begin{center}
\includegraphics[width=0.4\textwidth]{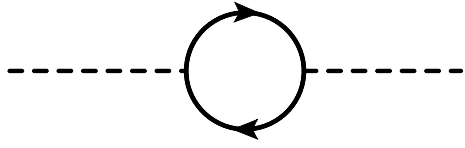}
\caption{The boson self-energy at the one-loop level.
The solid and dashed line are the propagator of the fermion and the boson, respectively.
}
\label{fig:boson-selfenergy}
\end{center}
\end{figure}

First, we derive the expression for the asymptotic thermal mass of the boson.
The boson retarded self-energy at the one-loop order is given by
\begin{align}
\label{eq:intro-boson-selfenergy}
\begin{split}
\varPi^\R_{\text{bare}}(p)&= i\cp^2\int\frac{d^4k}{(2\pi)^4}\Tr\Bigl[\SF_0(-k)\SR_0(p+k)
+\SR_0(-k)\SF_0(p+k)\Bigr] ,
\end{split}
\end{align}
whose diagrammatic expression is shown in Fig.~\ref{fig:boson-selfenergy}. 
$S_0^{R,S}(p+k)$ are the bare propagators of the fermion which are defined as
\begin{align}
\label{eq:bare-propagatorS}
\SR_0(k)&=\frac{-\Slash{k}}{k^2+ik^0\epsilon},\\
\label{eq:bare-propagatorSF}
\SF_0(k)&=\left(\frac{1}{2}-\nf(k^0)\right)i\Slash{k} \rho^0(k).
\end{align} 
Here we have introduced the Fermi-Dirac distribution function $\nf(k^0)\equiv 1/(\exp(k^0/T)+1)$ and the free spectral function $\rho^0(k)$ which is given by 
\begin{align}
\begin{split}
\rho^0(k)&\equiv 2\pi\sgn(k^0)\delta(k^2)\\
&=\frac{2\pi}{2|\vk|}(\delta(k^0-|\vk|)-\delta(k^0+|\vk|)). 
\end{split}
\end{align}
Since the dominant contribution to $\varPi^\R_{\text{bare}}(p)$ comes from the region $k\sim T$, as will be confirmed later, $\varPi^\R_{\text{bare}}(p)\sim \cp^2T^2$, which is smaller than $|\vp|^2\sim T^2$.
For this reason, $p^0$ which satisfies the pole condition (Eq.~(\ref{eq:intro-boson-polecondition})) is expected to be approximately equal to $\pm|\vp|$, so $p^2$ is negligible in the calculation of the asymptotic thermal mass.
By neglecting $p^2$, we get
\begin{align}
\label{eq:intro-thermalmass-boson-final}
\begin{split}
\varPi^\R_{\text{bare}}(p)&\simeq 4\cp^2\int\frac{d^4k}{(2\pi)^4}\left(\nf(k^0)-\frac{1}{2}\right)\rho^0(k)\\
&=\frac{2\cp^2}{\pi^2}\int^\infty_0 d|\vk||\vk|\nf(|\vk|)\\
&= \frac{\cp^2 T^2}{6},
\end{split}
\end{align}
where we have used Eq.~(\ref{eq:integral-formula-f1}).
From this expression and Eq.~(\ref{eq:intro-boson-polecondition}), we confirm Eq.~(\ref{eq:mass-boson}).

\begin{figure}
\begin{center}
\includegraphics[width=0.4\textwidth]{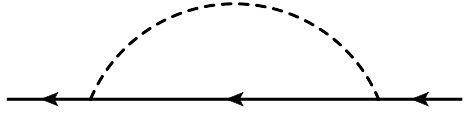}
\caption{Diagrammatic representation of the fermion self-energy at one-loop order.
The notations are the same as Fig.~\ref{fig:boson-selfenergy}.
}
\label{fig:fermion-selfenergy}
\end{center}
\end{figure}

Next we evaluate the asymptotic thermal mass of the fermion.
The diagram for the retarded fermion self-energy at the one-loop order is drawn in Fig.~\ref{fig:fermion-selfenergy}.
Its expression is
\begin{align}
\label{eq:one-loop-bare}
\begin{split}
\selfEnergyR_{\text{bare}}(p)&= i\cp^2\int\frac{d^4k}{(2\pi)^4}\Bigl[\DF_0(-k)\SR_0(p+k)
+\DR_0(-k)\SF_0(p+k)\Bigr],
\end{split}
\end{align}
where $D_0^{R,S}(-k)$ are the bare propagators of the scalar boson defined as
\begin{align}
\label{eq:bare-propagatorER}
\DR_0(k)&=\frac{-1}{k^2+ik^0\epsilon},\\
\DF_0(k)&=\left(\frac{1}{2}+\nb(k^0)\right)i \rho^0(k).
\label{eq:bare-propagatorE}
\end{align}
Inserting Eqs.~(\ref{eq:bare-propagatorS}), (\ref{eq:bare-propagatorSF}), (\ref{eq:bare-propagatorER}), and (\ref{eq:bare-propagatorE}) into (\ref{eq:one-loop-bare}), we obtain
\begin{align}
\label{eq:intro-selfenergy-fermion-oneloop}
\begin{split}
\selfEnergyR_{\text{bare}}(p)&= \cp^2\int\frac{d^4k}{(2\pi)^4} \rho^0(k)
\Bigl[\left(\frac{1}{2}+n_B(k^0)\right)\frac{\Slash{k}+\Slash{p}}{p^2+2p\cdot k+i(k^0+\pzero)\epsilon}\\
&\quad+\left(-\frac{1}{2}+n_F(k^0)\right)
\frac{\Slash{k}}{-p^2+2p\cdot k+i(k^0-p^0)\epsilon}\Bigr],
\end{split}
\end{align}
where we have used the on-shell conditions for the bare particles, $k^2=0$, and $(k+p)^2=0$ in $\DF_0(-k)$ and  $\SF_0(p+k)$.
To obtain the expression of the asymptotic thermal mass, we only have to calculate $\{\Slash{p}, \selfEnergyR_{\text{bare}}(p)\}$, which is easier than calculation of $\selfEnergyR_{\text{bare}}(p)$ when $p$ is hard.
Let us see this.
By using the fact $\selfEnergyR_{\text{bare}}(p)\sim \cp^2T$,  Eq.~(\ref{eq:intro-fermion-polecondition}) becomes
\begin{align}
\label{eq:intro-fermion-polecondition-calculate}
\begin{split}
p^2+i\epsilon p^0-\{\Slash{p}, \selfEnergyR_{\text{bare}}(p)\}&=0,
\end{split}
\end{align}
where the terms which are of order $\cp^4T^2$ have been neglected.
Thus we see that the square of the asymptotic thermal mass is equal to $\{\Slash{p}, \selfEnergyR_{\text{bare}}(p)\}$.
By neglecting $p^2$ as in the boson case, $\{\Slash{p}, \selfEnergyR_{\text{bare}}(p)\}$ reads
\begin{align}
\begin{split}
\{\Slash{p}, \selfEnergyR_{\text{bare}}(p)\}&\simeq
 \cp^2\int\frac{d^4k}{(2\pi)^4}\kernel(k),
\end{split}
\end{align}
where 
\begin{equation}
\label{eq:Kernel-definition}
\kernel(k)=\rho^0(k)(\nf(k^0)+\nb(k^0)).
\end{equation}
Note that $\kernel(k)$ is independent of $p$.
By using the formulae Eqs.~(\ref{eq:integral-formula-f1}) and (\ref{eq:integral-formula-b1}), $\{\Slash{p}, \selfEnergyR_{\text{bare}}(p)\}$ becomes
\begin{align}
\label{eq:intro-fermion-thermalmass-final}
\begin{split}
\{\Slash{p}, \selfEnergyR_{\text{bare}}(p)\} &= \frac{\cp^2}{2\pi^2}\int^\infty_0 d|\vk| |\vk| (\nf(|\vk|)+\nb(|\vk|)) \\
 &=\frac{\cp^2 T^2}{8}.
 \end{split}
\end{align}
From this expression and Eq.~(\ref{eq:intro-fermion-polecondition-calculate}), we confirm Eq.~(\ref{eq:mass-fermion}).


The other modification of the dispersion relation of the hard particles is the damping rate, which comes from the imaginary part of the self-energies.
In the Yukawa model, the imaginary parts of the self-energies at the leading order have the following forms:
\begin{align}
\label{eq:intro-imaginary-selfenergy-boson}
\left.\im\varPi^\R(p)\right|_{p^2=0}&= -2 p^0 \zetab(|\vp|),\\
\label{eq:intro-imaginary-selfenergy-fermion}
\left.\{ \Slash{p}, \im \selfEnergyR(p)\}\right|_{p^2=0}&= -2 p^0 \zetaf(|\vp|),
\end{align}
where $\zetaf$ and $\zetab$, which will be found to be the damping rate of the fermion and the boson at the leading order later, are the real numbers.
The leading contribution to $\zetab$ and $\zetaf$ is found to come from the two-loop diagrams containing the collision effect among the hard particles, not from the one-loop diagrams~\cite{Thoma:1994yw}. 
The damping rates of the hard particles are of order $\cp^4T\ln (1/\cp)$ (the fermion and the boson in the Yukawa model~\cite{Thoma:1994yw}, photon~\cite{\HardPhotonDamping}) or $\cp^2 T \ln(1/\cp)$ (electron, quark, gluon)~\cite{\AnomalousDamping}. 
The damping rates of the electron, the quark, and the gluon at the leading order are known to be independent of momentum~\cite{Blaizot:1996az}.
The expressions of the damping rates are obtained for the electron ($\zetae$), the quark ($\zetaq$), and the gluon ($\zetag$), and only at the leading-log order\footnote{In this thesis, we do not regard $\ln(1/\cp)$ as a large quantity.
Therefore $\ln(1/\cp)$ which appears in the order estimate of the damping rate is sometimes omitted. }~\cite{\AnomalousDamping}:
\begin{align} 
\zetae&=\cp^2T\left[\frac{1}{4\pi}\ln\left(\frac{1}{\cp}\right) +O(1)\right] ,\\
\zetaq&=\cp^2T\Cf\left[\frac{1}{4\pi}\ln\left(\frac{1}{\cp}\right) +O(1)\right] , \\
\zetag&=\cp^2TN\left[\frac{1}{4\pi}\ln\left(\frac{1}{\cp}\right) +O(1)\right] .
\end{align}

The asymptotic thermal mass and the damping rate indeed modify the pole position of the hard particle.
In the case of the Yukawa model, Eqs.~(\ref{eq:intro-fermion-polecondition}), (\ref{eq:intro-boson-polecondition}), (\ref{eq:intro-thermalmass-boson-final}), (\ref{eq:intro-fermion-thermalmass-final}), (\ref{eq:intro-imaginary-selfenergy-boson}), and (\ref{eq:intro-imaginary-selfenergy-fermion}) lead to
\begin{align} 
\label{eq:intro-dispersion-hard}
(p^0)^2&\simeq|\vp|^2+m^2_i -2i\zeta_i(|\vp|) p^0, 
\end{align} 
where $i=f$ or $b$, and $\zeta_i$ is the damping rate of each particle.
The terms of the asymptotic thermal mass and the damping rate, which give the corrections to the dispersion relation and come from the self-energy, are of order $\cp^2T^2$ and much smaller than the momentum term, which is of order $T^2$.
This fact is consistent with the expectation that the medium correction to the properties of the hard particle is small.
In this sense, we can regard the fermion and the boson with the hard energy as independent particle excitations.

\subsubsection{Relevance of hard particles to other quantities}

We make a comment on relevance of the hard particle to the physical quantities which is not sensitive to the infrared energy region. 
Whereas the number density of the hard particle is of order $T^3$ due to the large phase space as was shown before, that of the fermion (boson) whose energy is of order $\cp T$ is much smaller and of order $\cp^3T^3$ ($\cp^2T^3$)~\cite{lebellac, Blaizot:2001nr}, as will be shown in the case of the boson in the Yukawa model in the following:
First we introduce the cutoff parameter $\varLambda$ which satisfies $\varLambda\ll T$.
From Eq.~(\ref{eq:intro-boson-number}), the number density of the free particle whose energy is less than $\varLambda$ is
\begin{align}
\begin{split}
 \overline{n}_B|_{k<\varLambda}
&=\frac{1}{2\pi^2}\int^\varLambda_0 d|\vk| |\vk|^2 \nb(|\vk|) \\
&\sim T\varLambda^2.
\end{split}
\end{align} 
Here we have used $\nb(x)\simeq T/x$ for $x\ll T$.
This order estimate implies that the dominant contribution comes from $|\vk|>\varLambda$ because $\overline{n}_B$ is of order $T^3$ and much larger than $\overline{n}_B|_{k<\varLambda}$.
By setting $\varLambda\sim \cp T$, we see that the contribution from $|\vk|\lesssim \cp T$ is of order $\cp^2 T^3$.
This fact implies that most of the particles have the hard energy.
For this reason, it seems that the hard particles determine physical quantities which are not sensitive to the infrared region at the leading order.
Actually, the leading contribution to some thermodynamic quantities (pressure and energy~\cite{Kapusta:2006pm, lebellac}) and dynamical quantities (transport coefficients~\cite{\TransportRP, Arnold:2000dr, Arnold:2003zc, Defu:2005hb, Aarts:2003bk, Aarts:2004sd, Aarts:2005vc, Carrington:2009xf, Carrington:2007ea}, self-energies of the fermion~\cite{\HTL, Klimov:1981ka, Weldon:1982bn, Weldon:1989ys} and the boson~\cite{\HTL, Weldon:1982aq}), comes from a part in which the energy of the particle in the loop integral is hard.
We demonstrate this in the case of the hard and on-shell boson self-energy.
By using Eq.~(\ref{eq:intro-thermalmass-boson-final}), the contribution to $\varPi^\R_{\text{bare}}(p)$ from the energy region $k<\varLambda$ is of order
\begin{align}
\begin{split}
\varPi^\R_{\text{bare}}(p)|_{k<\varLambda}&\simeq \frac{2\cp^2}{\pi^2}
\int^\varLambda_0 d|\vk||\vk|\nf(|\vk|)\\
&\sim \cp^2\varLambda^2.
\end{split}
\end{align}
Here we have used $\nf(x)\simeq 1/2$ for $x\ll T$.
This order estimate means that the contribution from the hard particle, $\sim \cp^2T^2$, is much larger than that from the particle with smaller energy, $\sim \cp^2\varLambda$.

Nevertheless, we note that there are some quantities which are sensitive to the energy region $\ll T$.
Thus in analyzing such quantities, it is necessary to take into account the contribution from the soft particles even at the leading order. 
Examples of such quantities will be introduced in the next subsection. 

\subsection{Soft scale}  
\subsubsection{Properties of soft particles and hard thermal loop approximation}

The energy scale $\cp T$ is called soft scale.
In this scale, the self-energies and the inverses of the bare propagators have the same order of the magnitude in Eqs.~(\ref{eq:intro-fermion-polecondition}) and (\ref{eq:intro-boson-polecondition}), as will be shown later. 
This fact suggests that the medium effects can not be neglected even at the weak coupling when the energy scale is soft ($p\sim\cp T$).
Thus, bosonic and fermionic collective excitations can appear in this energy region.
There are transverse and longitudinal excitations~\cite{Weldon:1982aq} in the bosonic sector, and the latter is called plasmon and does not exist in the vacuum. 

\begin{figure}[t] 
\begin{center}
\includegraphics[width=0.45\textwidth]{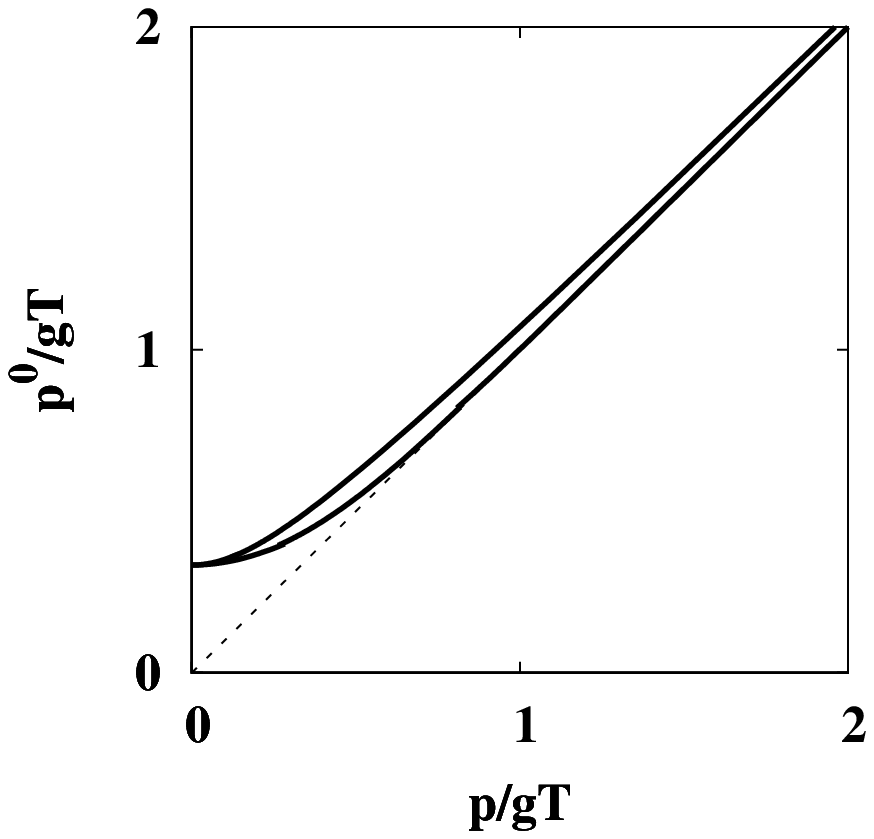}
\includegraphics[width=0.45\textwidth]{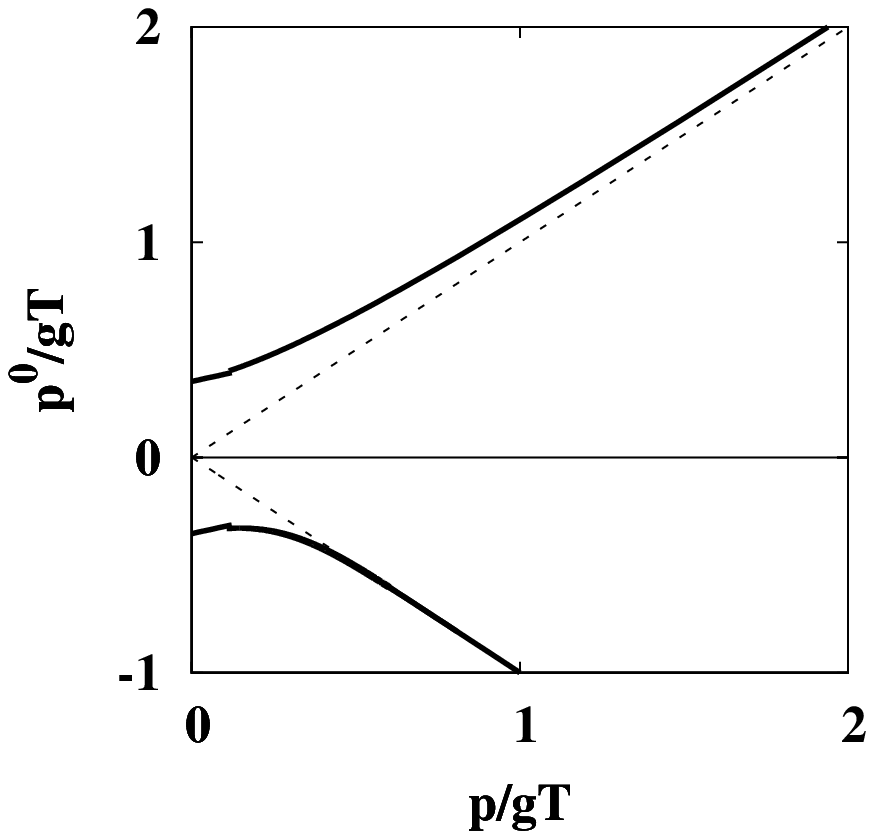} 
\caption{The dispersion relations in the bosonic sector and the fermionic sector.
Since the structure is the same in QED and QCD, we plot the dispersion relations in QED.
The vertical axis is the energy $\pzero$, while the horizontal axis is the momentum $|\vp|$.
The dotted lines denote the light cone.
Left panel: 
The dispersion relation in the bosonic sector.
The solid line with larger energy represents the transverse mode, while the other the longitudinal mode (plasmon).
The residue of the longitudinal mode becomes exponentially small for $|\vp|\gg \cp T$, so the plot of that mode does not represent physical excitation in that momentum region.
Right panel: 
The dispersion relation in the fermionic sector. 
The solid line in the positive energy region corresponds to the normal fermion, while that in the negative energy region corresponds to the antiplasmino.
Note that since we focus on the fermion sector, the antiplasmino appears instead of the plasmino.
The residue of the antiplasmino becomes exponentially small for $|\vp|\gg \cp T$, so the plot of the antiplasmino does not represent physical excitation in that momentum region.
} 
\label{fig:HTL-dispersion}
\end{center}
\end{figure}

To confirm that fact and to see the expression of the dispersion relations and the residues of the collective excitations, we here calculate the photon self-energy at the one-loop order in QED, adopting the Coulomb gauge.
The retarded photon self-energy at the one-loop level is given by
\begin{align}
\begin{split}
\varPi^{\R\mu\nu}_{\text{bare}}(p)&= i\cp^2\int\frac{d^4k}{(2\pi)^4}
\Tr\Bigl[\gamma^\mu\SF_0(-k)\gamma^\nu \SR_0(p+k)+\gamma^\mu\SR_0(-k)\gamma^\nu \SF_0(p+k)\Bigr]\\
&\simeq \cp^2\int\frac{d^4k}{(2\pi)^4} \nf(k^0)\rho^0(k)
\Tr\left[\gamma^\mu\left(\frac{2k^\nu\Slash{k}+\Slash{k}\gamma^\nu\Slash{p}}{2k\cdot p+p^2}
-\frac{2k^\nu\Slash{k}-\Slash{p}\gamma^\nu\Slash{k}}{2k\cdot p-p^2}\right)\right],
\end{split} 
\end{align}
where we have neglected the contribution from $T$-independent part since it can be eliminated by the renormalization of photon wave function in the vacuum. 
We note that not only the quark but also the gluon contributes to the gluon self-energy in QCD in contrast to the case of QED.
Since we are focusing on the soft region, we can utilize the useful condition $p\ll k$.
It is justified since $p\sim \cp T$ while $k\sim T$, which was shown in the previous subsection. 
By neglecting the terms which are of order or much smaller than $\cp^2pk$, we arrive at 
\begin{align} 
\label{eq:HTL-boson-selfenergy-result}
\begin{split} 
\varPi^{\R\mu\nu}_{\text{bare}}(p)&\simeq 4\cp^2\int\frac{d^4k}{(2\pi)^4} \nf(k^0)\rho^0(k)
\left[\frac{1}{k\cdot p}\left(p^\mu k^\nu+k^\mu p^\nu-\frac{p^2}{k\cdot p}k^\mu k^\nu \right)-g^{\mu\nu}\right] \\
&= 4\cp^2\int\frac{d^4k}{(2\pi)^4} \nf(k^0)\rho^0(k)
\frac{1}{k\cdot p}\left(p^\mu k^\nu+k^\mu p^\nu-\frac{p^2}{k\cdot p}k^\mu k^\nu \right)
-\frac{3\omegaph^2}{2}g^{\mu\nu} ,
\end{split}
\end{align}
where $\omegaph^2\equiv \cp^2T^2/9$.
This approximation is called the hard thermal loop (HTL) approximation~\cite{\HTL, Taylor:1990ia, Braaten:1991gm, Frenkel:1991ts}.
Here we decompose the photon propagator into the transverse and the longitudinal component in the Coulomb gauge:
\begin{align}
\label{eq:HTL-photon-propagator}
D^\R_{\mu\nu}(p)&= -\left(\frac{P^T_{\mu\nu}(p)}{p^2-\varPi^{\R T}(p)}+\frac{u_\mu u_\nu}{|\vp|^2+\varPi^{\R L}(p)}\right),
\end{align} 
where $\varPi^{\R T}(p)\equiv P^{T ij}(p)\varPi^{\R}_{ij}(p) /2$ and $\varPi^{\R L}(p)\equiv -\varPi^\R_{00}(p)$ are the transverse and the longitudinal components of the retarded self-energy. 
Here we omitted the subscript ``bare'' for simplicity.
From Eq.~(\ref{eq:HTL-boson-selfenergy-result}), the retarded self-energy in each component is calculated as
\begin{align}
\label{eq:HTL-selfenergy-transverse}
\varPi^{\R T}(p)&\simeq \frac{3\omegaph^2}{2}
\left[\frac{(p^0)^2}{|\vp|^2}+\frac{p^0}{2|\vp|}\left(1-\frac{(p^0)^2}{|\vp|^2}\right)\ln\frac{p^0+|\vp|}{p^0-|\vp|}\right], \\ 
\label{eq:HTL-selfenergy-longitudinal}
\varPi^{\R L}(p)&\simeq 3\omegaph^2\left(1-\frac{p^0}{2|\vp|}\ln\frac{p^0+|\vp|}{p^0-|\vp|}\right),
\end{align}
where we have used Eq.~(\ref{eq:integral-formula-cos2}) in the computation of the transverse component.
Thus, the dispersion relations in the transverse and longitudinal sectors read
\begin{align}
\label{eq:HTLresult-dispersion-transverse}
(p^0)^2-|\vp|^2-\frac{3\omegaph^2}{2}
\left[\frac{(p^0)^2}{|\vp|^2}+\frac{p^0}{2|\vp|}\left(1-\frac{(p^0)^2}{|\vp|^2}\right)\ln\frac{p^0+|\vp|}{p^0-|\vp|}\right]&= 0,\\
\label{eq:HTLresult-dispersion-longitudinal}
|\vp|^2+3\omegaph^2\left(1-\frac{p^0}{2|\vp|}\ln\frac{p^0+|\vp|}{p^0-|\vp|}\right)&= 0,
\end{align}
respectively.
Since Eqs.~(\ref{eq:HTLresult-dispersion-transverse}) and (\ref{eq:HTLresult-dispersion-longitudinal}) are invariant for the transformation $p^0\rightarrow -p^0$, we see that the positive energy solution and the negative energy solution are degenerated.
We analyze only the positive energy solutions from now on.
The dispersion relations of the excitations in the transverse ($\omega_T(|\vp|)$) and the longitudinal sector ($\omega_L(|\vp|)$) are plotted in the left panel of Fig.~\ref{fig:HTL-dispersion}.
The longitudinal excitation is called plasmon.
By using $p\sim\cp T$, we see that the terms coming from the inverse of the free propagator and the terms coming from the self-energy in Eqs.~(\ref{eq:HTLresult-dispersion-transverse}) and (\ref{eq:HTLresult-dispersion-longitudinal}) have the same order of magnitude in contrast to the case of the hard particle.
In this sense, the medium effect can not be regarded as a weak perturbation in the soft region.
The damping rates of the two excitations are zero in the HTL approximation, and their leading contribution comes from the two-loop diagrams.
They are of order $\cp^2 T\ln(1/\cp)$~\cite{Blaizot:1996az}, so the damping rate of the bosonic excitations are much smaller than the excitation energies,  which is of order $\cp T$.
The expressions for the residues of the two excitations are given by
\begin{align}
Z_T(|\vp|)&= \frac{\omega_T(|\vp|)(\omega^2_T(|\vp|)-|\vp|^2)}{3\omegaph^2\omega^2_T(|\vp|)-(\omega^2_T(|\vp|)-|\vp|^2)^2},\\
Z_L(|\vp|)&= \frac{\omega_L(|\vp|)(\omega^2_L(|\vp|)-|\vp|^2)}{|\vp|^2(|\vp|^2+3\omegaph^2-\omega^2_L(|\vp|))}.
\end{align}

The asymptotic forms of the dispersion relations for momenta which satisfies $|\vp|\ll \cp T$ are as follows:
\begin{align}
\omega^2_T(|\vp|)&\simeq \omegaph^2+\frac{6}{5}|\vp|^2,\\
\omega^2_L(|\vp|)&\simeq \omegaph^2+\frac{3}{5}|\vp|^2.
\end{align}
Both expressions coincide at $|\vp|=0$ because it is impossible to distinguish between transverse and longitudinal excitations in that case.
From this expressions, we see that the energies of both branches are $\omegaph$ at $|\vp|=0$.
Thus, we understand the physical meaning of $\omegaph$: 
the plasma frequency.
On the other hand, for $|\vp|$ which satisfies $\cp T\ll |\vp|\ll T$, the dispersion relations are approximated as 
\begin{align} 
\label{eq:HTL-result-dispersion-largep-transverse}
\omega^2_T(|\vp|)&\simeq |\vp|^2+\frac{3}{2}\omegaph^2,\\
\omega_L(|\vp|)&\simeq |\vp|\left(1+2e^{-[2|\vp|^2/(3\omegaph^2)+1]}\right).
\end{align}
We see that both dispersion relation approaches the light cone as the momentum becomes large.
However,we note that the residue of the plasmon is exponentially small for large momenta:
\begin{align}
Z_L(|\vp|)&\simeq \frac{4|\vp|}{3\omegaph^2} e^{-2|\vp|^2/(3\omegaph^2)-1} .
\end{align}
These observations support the expectation that the medium correction is suppressed by powers of $\cp$ when the momentum is much larger than $\cp T$.
Equation~(\ref{eq:HTL-result-dispersion-largep-transverse}) tells us that the mass of the transverse photon with large momenta is $\sqrt{3/2}\omegaph=\cp T/\sqrt{6}$, which coincides the asymptotic mass given in Eq.~(\ref{eq:mass-photon}).
This coincidence is unexpected since the HTL approximation is not valid when $p\sim T$.

On the other hand, in the fermionic sector, there are also two independent excitations~\cite{\plasmino}: 
one is called normal fermion and the other plasmino. 
The plasmino does not exist in the vacuum as the plasmon does not.
Let us perform the calculation of the self-energy using the HTL approximation in QED, to show the existence and the properties of the fermionic excitations introduced above.
The retarded self-energy is expressed as
\begin{align}
\label{eq:intro-oneloop-fermion-QED}
\varSigma_\text {bare}^R(p) &=i\cp^2\int\frac{d^{4}k}{(2\pi)^{4}} \gamma^\mu \bigl[\DF_{0\mu\nu}(-k)\SR_0(p+k)
+\DR_{0\mu\nu}(-k)\SF_0(p+k)\bigr]\gamma^\nu,
\end{align}
where $D^{\R, S}_{0\mu\nu}(-k)$ is the bare propagators of the photon defined as 
\begin{align}
\DR_{0\mu\nu}(k)&=\frac{-P^T_{\mu\nu}(k)}{k^2+ik^0\epsilon}-\frac{u_\mu u_\nu}{|\vk|^2},\\ 
\DF_{0\mu\nu}(k)&=\left(\frac{1}{2}+\nb(k^0)\right)iP^T_{\mu\nu}(k) \rho^0(k), 
\end{align}
where $u^\mu=(1,\vect{0})$ and $P^T_{\mu\nu}(k)$ is the projection operator on the transverse direction,
\begin{equation}
P^T_{\mu\nu}(k) \equiv g_{\mu i}g_{\nu j}(\delta_{ij}-\hat{k}_i\hat{k}_j) ,
\end{equation}
with $\hat{k}^i\equiv k^i/|\vect{k}|$.  
The diagram corresponding to Eq.~(\ref{eq:intro-oneloop-fermion-QED}) is shown in Fig.~\ref{fig:fermion-selfenergy}.
Before proceeding the calculation, let us show that the term which contains $u_\mu u_\nu$ in $\DR_{0\mu\nu}(k)$ does not contribute to the result in the HTL approximation.
The contribution from that term becomes
\begin{align} 
\label{eq:HTL-longitudinal-contribution}
& -\cp^2\int\frac{d^{4}k}{(2\pi)^{4}} \gamma^0 \frac{1}{|\vk|^2}\nf(k^0)\Slash{k} \rho^0(k) \gamma^0.
\end{align}
Here we have used $p\ll k$ and neglected the $T=0$ contribution.
After $k$ integration, this expression vanishes.
Thus by using the condition $k\gg p$ as in the bosonic case, Eq.~(\ref{eq:intro-oneloop-fermion-QED}) becomes
\begin{align}
\label{eq:intro-HTL-selfenergy-electron}
\begin{split}
\varSigma_\text {bare}^R(p) &\simeq 2\cp^2\int\frac{d^4k}{(2\pi)^4}\kernel(k)\frac{\Slash{k}}{2p\cdot k+ik^0\epsilon}.
\end{split}
\end{align} 
From this expression, we see that $\varSigma_\text {bare}^R(p)\sim \cp T$ on account of $p\sim \cp T$.
This order-estimate implies that $p\sim \varSigma_\text {bare}^R(p)$, so the computation of $\{\Slash{p}, \varSigma_\text {bare}^R(p)\}$ does not give us the dispersion relation of the electron in contrast to the case of $p\sim T$.
Therefore we proceed the calculation of $\varSigma_\text {bare}^R(p)$ instead of evaluation of $\{\Slash{p}, \varSigma_\text {bare}^R(p)\}$.
Equation~(\ref{eq:intro-HTL-selfenergy-electron}) becomes
\begin{align}
\label{eq:HTL-selfenergy-fermion-result}
\begin{split}
\varSigma_\text {bare}^R(p)&\simeq 
\frac{\cp^2}{8\pi^2}\int^\infty_0 d|\vk| |\vk|(\nf(|\vk|)+\nb(|\vk|))\sum_{s=\pm}\int^1_{-1} d\cos\theta
\frac{\gamma^0-s\cos\theta \hat{\vp}\cdot\vgamma}{p^0-s|\vp|\cos\theta}\\
&= \omega^2_{0e}\left[ \frac{\gamma^0}{2|\vp|}\ln\left(\frac{p^0+|\vp|}{p^0-|\vp|}\right) 
- \frac{\hat{\vp}\cdot\vgamma}{|\vp|}
\left[-1+\frac{p^0}{2|\vp|}\ln\left(\frac{p^0+|\vp|}{p^0-|\vp|}\right)\right]\right],
\end{split}
\end{align} 
In the last line we have used Eqs.~(\ref{eq:integral-formula-f1}) and (\ref{eq:integral-formula-b1}) and introduced $\omega^2_{0e}\equiv \cp^2T^2/8$.
This expression can be rewritten as
\begin{align} 
\varSigma^0_\text {bare}(p)&= \frac{\omega^2_{0e}}{2|\vp|}
\ln\left(\frac{p^0+|\vp|}{p^0-|\vp|}\right),\\ 
\varSigma^V_\text {bare}(p)&= \frac{\omega^2_{0e}}{|\vp|}
\left[-1+\frac{p^0}{2|\vp|}\ln\left(\frac{p^0+|\vp|}{p^0-|\vp|}\right)\right].
\end{align}
By substituting this expression into Eq.~(\ref{eq:intro-propagator-selfenrgy-fermion}), we obtain
\begin{align} 
\label{eq:intro-HTL-propagator-fermion}
\SR(p)&=  -\frac{1}{2}
\left(\frac{\gamma^0-\vgamma\cdot\hat{\vp}}{p^0+i\epsilon-|\vp|-\varSigma^0_\text {bare}(p)+\varSigma^V_\text {bare}(p)}
+\frac{\gamma^0+\vgamma\cdot\hat{\vp}}{p^0+i\epsilon+|\vp|-\varSigma^0_\text {bare}(p)-\varSigma^V_\text {bare}(p)}\right).
\end{align} 
From the spinor structure in the numerators of the two terms in the right-hand side, we see that these terms are eigenstates of  (chirality)/(helicity).
The eigenvalue of the first term is $+1$ while that of the second term is $-1$.
In the vacuum, the fermion number of the former is $+1$ while that of the latter is $-1$.
From Eq.~(\ref{eq:intro-HTL-propagator-fermion}), the dispersion relations of the collective excitations which has the fermion number $+1$ is
\begin{align}
\label{eq:HTL-result-dispersion-fermion}
p^0-|\vp|-\frac{\omegae^2}{2|\vp|}\left[2
+\left(1-\frac{p^0}{|\vp|}\right)\ln\left(\frac{p^0+|\vp|}{p^0-|\vp|}\right)\right]
=0.
\end{align}
The two dispersion relations are obtained as solutions of this equation, which are denoted by $\omega_+(\vp)$ and  $-\omega_-(\vp)$.
These dispersion relations are plotted in the right panel of Fig.~\ref{fig:HTL-dispersion}.
We note that the excitations whose dispersion relations are $\omega_-(\vp)$ and $-\omega_+(\vp)$ also appear in the fermion number $-1$ sector.
The excitation whose dispersion relation is $\omega_+(\vp)$ is called normal fermion, $\omega_-(\vp)$ the plasmino, and $-\omega_-(\vp)$ the antiplasmino, respectively.
The residue of each branch is
\begin{align}
Z_\pm(|\vp|)&= \frac{\omega^2_{\pm}(|\vp|)-|\vp|^2}{2\omegae^2}.
\end{align}

Eq.~(\ref{eq:HTL-result-dispersion-fermion}) can be solved explicitly for $|\vp|$ which satisfies $|\vp|\ll \cp T$ or $\cp T\ll|\vp|\ll T$.
In the case of $|\vp|\ll \cp T$, the dispersion relations become
\begin{align}
\omega_\pm(|\vp|)\simeq \omegae \pm \frac{1}{3}|\vp|.
\end{align} 
From this expression, we understand the physical meaning of $\omegae$:
the counterpart of plasma frequency for fermion.
The residue becomes
\begin{align}
Z_\pm(|\vp|)&\simeq \frac{1}{2}\pm\frac{|\vp|}{3\omegae} .
\end{align}
This expression tells us that both branches have the same strength when they are at rest.
On the other hand, for $\cp T\ll |\vp|\ll T$, the dispersion relations become
\begin{align}
\label{eq:HTLresult-dispersion-plus-largep}
\omega_+(|\vp|)&\simeq |\vp| +\frac{\omegae}{|\vp|},\\
\omega_-(|\vp|)&\simeq |\vp|(1+2 e^{-2|\vp|^2/\omegae^2-1}), 
\end{align}
and the residue is
\begin{align}
Z_+(|\vp|)&\simeq 1+\frac{\omegae^2}{2|\vp|^2}\left[1-\ln\left(\frac{2|\vp|^2}{\omegae^2}\right)\right],\\
Z_-(|\vp|)&\simeq \frac{2|\vp|^2}{\omegae^2} e^{-2|\vp|^2/\omegae^2-1}.
\end{align}
From these equations, we see that the residue of the plasmino becomes exponentially small as $|\vp|$ becomes large.
We also see that the dispersion relation of the normal fermion approaches the light cone and the residue approaches unity, as $|\vp|$ becomes large.
These two facts can be understood from the expectation that the medium effect is suppressed for $T\lesssim p$, as explained in the previous subsection.
Equation~(\ref{eq:HTLresult-dispersion-plus-largep}) can be expressed as $\omega^2_+(|\vp|)\simeq |\vp|^2+2\omegae^2$ if we neglect higher order term, which implies that the mass of the normal fermion with $\cp T\gg|\vp|$ is equal to the asymptotic thermal mass $\me$ at the leading order.
It is nontrivial because the HTL approximation is not valid when $p\sim T$.

We emphasize that the plasmino is a novel excitation which reflects the fact that QGP is a fermion-boson system at ultrarelativistic temperature: 
If each of the mass of the quark and the gluon is not negligible, the fermion self-energy would be suppressed and the plasmino would not appear.

Since $\varSigma^\R(p)\sim \cp T$ as was shown before, we note that the fermion retarded self-energy and the momentum of the soft particle in Eq.~(\ref{eq:intro-propagator-selfenrgy-fermion}) have the same order of magnitude ($\sim\cp T$), as in the case of the boson. 
The damping rates of these excitations are zero in the HTL approximation.
By evaluating the two-loop diagrams containing the effect of the collision, they are found to be of order $\cp^2T \ln(1/\cp)$~\cite{Blaizot:1996az}.
Thus the damping rates of the fermionic excitations are much smaller than the excitation energy $\omega_\pm(|\vp|)$.

The normal fermion and the plasmino exists not only in $\mu=0$ case but also in the case that $\mu$ is finite~\cite{\HDL}.
It is not straightforward to give an interpretation of the two fermionic excitations at finite $T$ since it does not have their classical counterparts, but at zero temperature and finite chemical potential case, it is possible to clarify the state which forms the plasmino~\cite{Blaizot:1993bb}.
From that analysis, it was suggested that the state of the plasmino is superposition of an antifermion state and the state composed of a boson and a hole. 
The extension of such analysis to the finite temperature case is an interesting task.
We also note that there is another attempt of interpretation in terms of the idea of resonant scattering~\cite{Kitazawa:2006zi}.

In addition to collective excitations, the HTL approximation also leads to the Debye screening~\cite{Weldon:1982aq}.
The screening mass is 
\begin{align}
\sqrt{\lim_{|\vp|\rightarrow 0}\varPi^{\R L}(0,\vp)}=\frac{\cp T}{\sqrt{3}},
\end{align}
which is of order $\cp T$.

\subsubsection{Relevance of soft particles to other quantities}

The knowledge of the spectra of the soft excitations is necessary not only to establish the picture of the particles, but also to calculate the quantities which are sensitive to the soft energy region.
Examples include the gluon damping rate at rest.
That quantity was first calculated by using the bare perturbation expansion and found to be dependent on the gauge-fixing~\cite{\DampingProblem}, though that quantity should be gauge-independent~\cite{\GaugeIndependence}.
The solution was given from the following observation:
Since properties such as the dispersion relation of the soft particles are different largely from that in the free limit because of the medium effect, we have to use the resummed propagators (Eqs.~(\ref{eq:HTL-photon-propagator}) and (\ref{eq:intro-HTL-propagator-fermion})) which include the information of the result of the HTL approximation instead of the bare propagators, when soft particles appear in the loop integral.
Vertex function also needs to be resummed in such a way.
This resummation is called HTL resummation~\cite{\HTLResum}, and it is essentially important to get a sensible result.
By using the HTL resummation, the sensible and gauge-independent result was obtained~\cite{\HTLResum} for the gluon damping rate at rest.
That method was also applied to the analysis of the quark damping rate at zero spatial momentum~\cite{\HTLResumQuarkDamping}.

\subsubsection{Vlasov equation}

The HTL approximation is a diagrammatic technique for quantities at equilibrium.
Nevertheless, due to the linear response theory, the HTL results should be reproduced from the analysis of the nonequilibrium time evolution at the leading order in the case where the system is close to the equilibrium state. 
Actually, the HTL approximation corresponds to the linearized Vlasov equation, which is a collisionless kinetic equation~\cite{\HTLVlasov}, as is shown in the case of the photon self-energy in the following: 
The Vlasov equation which describes the time evolution of the electron distribution function under the background electromagnetic field reads
\begin{align} 
\label{eq:Vlasov} 
v\cdot\partial_X\tilde{n}_F (\vk, X)&= \cp (E^i(X)+\epsilon^{ijk}v^jB^k(X)) \partial^i_k \tilde{n}_F (\vk, X)
\end{align}
where $v^\mu\equiv(1,\hat{\vk})$ is the four-velocity of a massless particle, $\tilde{n}_F (\vk, X)$ the electron distribution function at the nonequilibrium state, $E_i(X)\equiv -\partial_i A_0(X)+\partial_0 A_i(X)$ the background electric field, $B_i(X)\equiv -\epsilon_{ijk}\partial_{Xj}A_{k}(X)$ the background magnetic field with $A_\mu(X)$ being the background gauge field, respectively. 
The left-hand side describes the time evolution in the free limit, and is called the drift term.
The right-hand side contains the effect of the interaction between the electron and the background electromagnetic field, and is called the force term.
By linearizing Eq.~(\ref{eq:Vlasov}), we get
\begin{align} 
v\cdot\partial_X \delta \nf(\vk, X)&= \cp \vv\cdot\vE(X) \frac{d}{d|\vk|} \nf(|\vk|),
\end{align} 
where $\delta \nf(\vk, X)\equiv \tilde{n}_F(\vk, X)-\nf(|\vk|)$ is the deviation of the electron distribution function from the value at equilibrium. 
We note that the background magnetic field does not contribute in this order.
The induced current $\jind^\mu(X)$ can be expressed in terms of $\delta \nf(\vk, X)$ as follows:
\begin{align} 
\begin{split}
\jind^\mu(X)&= -4\cp \int \frac{d^3\vk}{(2\pi)^3}v^\mu\delta \nf(\vk, X) \\
&= 4\cp^2 \int \frac{d^3\vk}{(2\pi)^3} \frac{v^\mu v_i}{v\cdot\partial_X} \frac{d}{d|\vk|} \nf(|\vk|)
(-\partial_{0}A_i(X)+\partial_{i}A_0),
\end{split}
\end{align} 
where the degeneracy of the spin of electron and the contribution from the positron have been taken into account.
By using this expression, we get the following expression of the polarization tensor by performing the Fourier transformation ($X\rightarrow p$):
\begin{align} 
\varPi^{\R\mu\nu}(p)&= -\frac{2\cp^2}{\pi^2}\int^\infty_0 d|\vk| |\vk|^2\frac{d\nf(|\vk|)}{d|\vk|}
\left(-\delta^{\mu 0}\delta^{\nu 0}
+\frac{p^0}{2}\int^1_{-1}d\cos\theta\frac{v^\mu v^\nu}{p\cdot v+i\epsilon} \right).
\end{align} 
Here we have used the relation in the linear response theory, $\jind^\mu(p)=\varPi^{\R\mu\nu}(p)A_\nu(p)$, and introduced $\cos\theta=\hat{\vp}\cdot\vv$. 
The infinitesimal number $+i\epsilon$ comes from the condition that the background field is introduced adiabatically: $\delta\nf(\vk,X)$ and $\vE(X)$ vanish when $X^0\rightarrow-\infty$.
This expression yields Eqs.~(\ref{eq:HTL-selfenergy-transverse}) and (\ref{eq:HTL-selfenergy-longitudinal}) after the $k$ integration, which means that the linearized Vlasov equation reproduces the photon self-energy in the HTL approximation.

We can also generalize the Vlasov equation to the case where the background field is fermionic~\cite{\HTLVlasov}. 
In this situation, we are to analyze the time evolution of the amplitude of the process in which a hard fermion becomes a hard boson and its inverse process, instead of the distribution function whose time evolution was analyzed above (see Chapter~\ref{chap:kinetic}).
By performing such analyses, the fermion retarded self-energy Eq.~(\ref{eq:HTL-selfenergy-fermion-result}) in the HTL approximation, is reproduced.

\subsection{Ultrasoft scale} 
\label{ssc:intro-ultrasoft}

\subsubsection{Resummed perturbation and Boltzmann equation}
The energy scale $\cp^2T$ is called ultrasoft scale.
This scale appears in the damping rate, or equivalently, the inverse of the relaxation time of the hard particle~\cite{\AnomalousDamping}, as was shown in Sec.~\ref{ssc:intro-hard}.
For this reason we expect that the collision effect becomes important in the ultrasoft region ($p\lesssim \cp^2T$). 
To demonstrate that the collision effect is not negligible in the ultrasoft region, we consider the linearized Boltzmann equation, which is the kinetic equation containing the collision effect, in the relaxation time approximation.
The situation is the same as in Eq.~(\ref{eq:Vlasov}): the particle is electron, and the background field is the electromagnetic one.
That equation reads
\begin{align} 
\left(v\cdot\partial_X +\frac{1}{\tau}\right)\delta \nf(\vk, X)&= \cp \vv\cdot\vE(X) \frac{d}{d|\vk|} \nf(|\vk|). 
\end{align}
The second term in the left-hand side is the collision term in the relaxation time approximation.
Here $\tau$ is a typical relaxation time for the hard electron, whose inverse has the same order of magnitude as the damping rate of the hard electron ($\zetae$), which is of order $\cp^2T\ln(1/\cp)$\footnote{For the photon in QED and the particles in the Yukawa model, the mean free path is of order of $(\cp^4T)^{-1}$, so it seems that we do not need to take into account the interaction among the hard particles in the ultrasoft region.
However, the difference of the thermal masses of the fermion and the boson, which is of order $\cp^2T^2$, plays the similar effect to that of the mean free path, as will be shown in Chapter~\ref{chap:ultrasoft}, so we can not neglect the interaction effect when we analyze the fermion self-energy with ultrasoft momentum. }~\cite{\AnomalousDamping}.
From this equation, we see that the collision term is not negligible in the case of ultrasoft region ($\partial_X\lesssim \cp^2T $) since that term and the drift term have the same order of magnitude.

In fact, the Boltzmann equation should be used instead of the Vlasov equation in the calculation of the gluon $n$-point function in this energy region~\cite{\UltrasoftGluon}.
We will show in Chapter~\ref{chap:kinetic} that the other interaction effects such as the asymptotic thermal mass should be taken into account as well as the collision, in the analysis of the ultrasoft fermion propagator.
We note that the asymptotic thermal mass and a part of the collision effect correspond to the real and the imaginary part of Eq.~(\ref{eq:intro-dispersion-hard}), respectively.
The Boltzmann equation in the analysis of the gluon $n$-point function can be translated into a diagrammatic language, and the resultant diagrammatic method is not a simple one-loop approximation (HTL approximation), but the resummed perturbation which resums the damping rate of the hard particles and sums up the ladder diagrams~\cite{\UltrasoftGluon}.

In addition to the gluon $n$-point function, there are other quantities whose calculation need considering the interaction effect among the hard particles.
For example, the transport coefficient is such quantity.
That quantity can be obtained from the Kubo formula, by taking the zero energy limit of the Green function as follows~\cite{Kapusta:2006pm}:
The shear viscosity $\eta$, the bulk viscosity $\zeta$, the electrical conductivity $\sigma$ in QED, and the flavor diffusion constant $D_{\alpha\beta}$ in QCD are given by
\begin{align} 
\eta&= \frac{1}{20}\lim_{\omega\rightarrow 0} \frac{1}{\omega}
\int d^4x e^{i\omega x^0}\langle[\pi_{lm}(x), \pi_{lm}(0)] \rangle, \\
\zeta&= \frac{1}{2}\lim_{\omega\rightarrow 0} \frac{1}{\omega}
\int d^4x e^{i\omega x^0}\langle[P(x), P(0)] \rangle, \\
\sigma&= \frac{1}{6}\lim_{\omega\rightarrow 0} \frac{1}{\omega}
\int d^4x e^{i\omega x^0}\langle[j_{i}(x), j_{i}(0)] \rangle, \\
D_{\alpha\beta}&= \frac{1}{6}\lim_{\omega\rightarrow 0} \frac{1}{\omega}
\int d^4x e^{i\omega x^0}\langle[j^\alpha_{i}(x), j^\gamma_{i}(0)] \rangle \varXi^{-1}_{\gamma\beta},
\end{align}
respectively, where $P$ is the local pressure, $\pi_{lm}(x)\equiv T_{lm}-\delta_{lm}P$ the traceless part of the energy-momentum tensor with $T_{lm}$ being the energy-momentum tensor, $j_i(x)$ the electromagnetic current, $j^\alpha_{i}(x)$ the flavor current, respectively, and we have introduced $\varXi_{\alpha\beta}\equiv \frac{\partial }{\partial \mu_{\beta}} \langle j^0_{\alpha}\rangle$.
Because the zero energy limit is taken, the energy scale of that quantity is much less than $\cp^2T$.
For this reason, the analysis of that quantity also needs including the interaction effect among the hard particles.
Accordingly, the transport coefficient can be calculated either by the Boltzmann equation~\cite{\TransportBoltzmann} or by the resummed perturbation\footnote{That quantity was also computed using the $n$-particle irreducible formalism~\cite{\TransportTwoPI, Carrington:2009xf, Carrington:2007ea}.
All of these methods produce the same result in the leading order of the coupling constant.}~\cite{\TransportRP}.
Using the correspondence between that perturbation theory and the Boltzmann equation, the resummation scheme in the resummed perturbation theory is interpreted with the language of the kinetic theory.

\subsubsection{Suggestion on existence of ultrasoft fermion mode} 
When the energy scale is much below $\cp^2T$, the hydrodynamics works well, and the bosonic hydrodynamic modes such as the phonon and the diffusion mode appear. 
On the other hand, the fermionic sector whose energy is of order $\cp^2T$ has not been well investigated, due to the difficulty of taking into account of the interaction among the hard particles.
For this reason, it has not been studied well whether there are any modes in this region.

Nevertheless, there have been some suggestive works for supporting the existence of such an ultrasoft fermion mode at finite temperature.
Some of them are based on supersymmetry, and others are based on analyses using effective model, as will be shown later.
First we introduce the works related to supersymmetry.
Historically, the ultrasoft fermion mode at finite $T$ was found in supersymmetric models as Nambu-Goldstone fermions called goldstino associated with spontaneous breaking of supersymmetry at $T\not=0$~\cite{Buchholz:1997mf}, by using Ward-Takahashi identity and a diagrammatic technique~\cite{\Goldstino}. 
The complete calculation at the leading order was first performed by Lebedev and Smilga~\cite{Lebedev:1989rz}.
Because the supercurrent and the energy-momentum tensor are in the same supermultiplet, the goldstino was also regarded as the supersymmetric analogue of the phonon, phonino.
Here we note that the analysis in~\cite{Lebedev:1989rz} was performed in the temperature region $T\ll  m/\cp$, where $m$ is the bare mass. 
It implies that their analysis is only valid for $\cp T\ll m$, but not for $m\ll \cp T$ for which our analysis in the present paper is concerned.
The analysis was extended to QCD at so high temperature that the coupling constant is weak~\cite{Lebedev:1989ev}, in which a supersymmetry is still assigned at  the vanishing coupling, and hence, the supersymmetry is, needless to say, explicitly broken by the interaction.
Thus, there exists no exact fermionic zero mode but only a pseudo-zero mode does.
Although these analyses~\cite{\Goldstino, Lebedev:1989ev} are suggestive, it is still obscure whether a genuine ultrasoft fermion mode exists when supersymmetry is absent, in particular, at extremely high temperature.

Here we note that there have been suggestions of the existence of ultrasoft fermion mode at finite $T$ even without supersymmetry.
It was shown in one-loop calculations~\cite{Kitazawa:2006zi} that when a fermion is coupled with a massive boson with  mass $m$, the spectral function of the fermion gets to have  a novel peak in the far-low-energy region in addition to the normal fermion and the antiplasmino, when $T\sim m$, irrespective of the type of boson; it means that the spectral function of the fermion has a three-peak structure in this temperature region. 
It was suggested that such a three-peak structure may persist even at the high temperature limit in the sense $m/T\rightarrow 0$, for the massive vector boson on the basis of a gauge-invariant formalism, again, at the one-loop order~\cite{Satow:2010ia}. 
Thus, one may expect that the novel excitation may exist in the far-infrared region also for a fermion coupled with a massless boson, although the one-loop analysis admittedly may not be applicable at the ultrasoft momentum region.
There are also the works suggesting the existence of the ultrasoft fermion mode using the Schwinger-Dyson 
equation~\cite{\ThreepeakSD}; we note, however,  that it is difficult to keep gauge symmetry in the Schwinger-Dyson approach at finite $T$.
The analysis of the quark spectrum around $T_c$ using the NJL model~\cite{\NJL} was also performed~\cite{Kitazawa:2005mp}, and as a result, the existence of the ultrasoft fermion mode was suggested due to the coupling between the quark and the mesons.

Finally let us give a generic argument supporting the existence of an ultrasoft fermion mode at finite $T$ on the basis of the symmetry of the self-energy for a massless fermion.
It was shown that the chiral, parity, and the charge symmetry make the fermion retarded propagator have the following structure~\cite{Weldon:1999th}:
\begin{align}
\begin{split} 
S^\R(p^0,\vp)&= -\frac{1}{2}\left(\frac{\gamma^0-\vgamma\cdot\hat{\vp}}{S_+(p^0,\vp)}
+\frac{\gamma^0+\vgamma\cdot\hat{\vp}}{S_-(p^0,\vp)}\right).
\end{split}
\end{align}
Here we have introduced $S_\pm(p^0,\vp)\equiv p^0+i\epsilon \mp |\vp|-\varSigma^0(p)\pm\varSigma^V(p)$, and these functions satisfy
\begin{align}
\label{eq:symmetry-S1}
S_-(p^0,\vp)&=-(S_+(-p^0{}^*,\vp))^*,\\
\label{eq:symmetry-S2}
S_+(p^0,\vzero)&=S_-(p^0,\vzero). 
\end{align}
By setting $|\vp|=0$, we get
\begin{align}
\begin{split}
S^\R(p^0,\vzero)&= -\frac{\gamma^0}{S_+(p^0,\vzero)}.
\end{split}
\end{align}
Eqs.~(\ref{eq:symmetry-S1}) and (\ref{eq:symmetry-S2}) imply
\begin{align}
S_+(p^0,\vzero)&=-(S_+(-p^0{}^*,\vzero))^*.
\end{align}
By using $S_+(p^0,\vzero)=p^0-\varSigma^0(p^0 ,\vzero)$, we get
\begin{align}
\label{eq:symmetry-sigma}
\begin{split}
\re\varSigma^0(-p^0 ,\vzero)=-\re\varSigma^0(p^0 ,\vzero),
\end{split}
\end{align}
where $p^0$ is real.
This property implies that $p^0-\re\varSigma^0(p^0 ,\vzero)$, which is the real part of the inverse of the retarded fermion propagator, is zero at $p^0=0$ if there is no singularity\footnote{In the HTL approximation~\cite{\HTL}, the singularity appears at the origin.
That is the reason why that approximation can not suggest the fermion mode around the origin.}.
For this reason, it is suggested that the fermion retarded propagator always have the pole around the origin, provided that the imaginary part of the fermion retarded self-energy is small enough.
This argument suggests that the existence of the ultrasoft pole may be a universal phenomenon at high temperature in the theory composed of massless fermion coupled with a boson, as long as chiral, parity, and charge symmetry exist.
We note that the attenuation of the pole at the origin in the case that the fermion has finite bare mass~\cite{Kitazawa:2007ep}, is consistent with that argument because the finite fermion mass breaks the chiral symmetry.

\subsubsection{What we do in the thesis}
In this thesis, we analyze the properties in the ultrasoft region, focusing on the fermion spectrum:
we calculate the ultrasoft fermion spectrum by using the resummed perturbation which enables the calculation of the fermion propagator with ultrasoft momentum, and show that a novel fermionic mode exists in that energy region.
We also obtain the expressions for the dispersion relation, the damping rate, and the strength of that mode.

As explained before, there is an equivalence between the diagrammatic method and the kinetic equation such as the equivalence between HTL approximation and Vlasov equation, or the resummed perturbation treating the gluon $n$-point function and Boltzmann equation.
Because of this equivalence, we expect that the resummed perturbation used in the analysis of the ultrasoft fermion is expected to be equivalent to some kinetic equation which contains the interaction effect among the hard particles.
We derive the generalized Boltzmann equation that is equivalent to the basic equation of the resummed perturbation theory treating the ultrasoft fermion, from the Kadanoff-Baym equation, which describes the time evolution of the nonequilibrium system based on the field theory.

\section{Outline of the thesis}
\label{sec:intro-outline}
This thesis is organized as follows: 
Chapter~\ref{chap:ultrasoft} is devoted to the analysis on the spectrum of the ultrasoft fermion using the resummed perturbation theory in the Yukawa model and QED/QCD.
The analysis in the Yukawa model and QED is based on Ref.~\cite{Hidaka:2011rz}.
As a result of the analysis, we show that a novel fermionic mode which we call ultrasoft fermion mode exists in that energy region, and obtain the expressions of the pole position and the strength of that mode.
We also show that the resultant fermion propagator and the vertex function satisfy the Ward-Takahashi identity in QED/QCD. 
In Chapter~\ref{chap:kinetic}, we derive the linearized and generalized Boltzmann equation for ultrasoft fermion excitations, from the Kadanoff-Baym equation in a Yukawa model and QED.
We show that this equation is equivalent to the self-consistent equation in the resummed perturbation theory used in the analysis of the ultrasoft fermion spectrum at the leading order. 
Furthermore, we derive the equation that determines the $n$-point function with external lines for a pair of fermions and $(n-2)$ bosons with ultrasoft momenta in QED.
We also showed that the Ward-Takahashi identity is satisfied, and that identity can be derived the conservation law of the current.
The analysis in that chapter is based on Ref.~\cite{Satow:2012ar}.
Finally we summarize this thesis and give the concluding remarks in Chapter~\ref{chap:summary}.
In Appendix~\ref{app:temporal}, we show that the results obtained in the Coulomb gauge in this thesis, also can be obtained in the temporal gauge.
We write some formulae used in the text in Appendix~\ref{app:formula}.

\chapter{Ultrasoft Fermion Mode} 
\label{chap:ultrasoft}
\thispagestyle{headings}

Though there are some suggestion on existence of the ultrasoft mode as is written in the previous chapter, It is not a simple task to establish that fermionic mode exist in the ultrasoft region on the complete leading order calculation because of the infrared divergence called pinch singularity~\cite{\PinchSingularity} that breaks a naive perturbation theory, as will be briefly reviewed in the next section.
We remark that the same difficulty arises in the calculation of transport coefficients~\cite{\TransportRP, Arnold:2000dr, Arnold:2003zc, Defu:2005hb, Aarts:2003bk, Aarts:2004sd, Aarts:2005vc, Carrington:2009xf, Carrington:2007ea} and the gluon self-energy~\cite{\UltrasoftGluon} in the ultrasoft energy region.
Therefore, in this chapter, we analyze the fermion propagator in the ultrasoft energy region in Yukawa model and QED/QCD using a similar diagrammatic technique in Refs.~\cite{Lebedev:1989ev, Lebedev:1989rz} to regularize the pinch singularity.
We shall show that the retarded fermion propagator has a pole at $\pzero = \pm |\vp|/3-i\zeta$  ($\zeta\sim \cp^4T\ln \cp^{-1}$ for Yukawa model and $\sim \cp^2T\ln \cp^{-1} $ for QED/QCD) with the residue $Z\sim \cp^2$ for ultrasoft momentum $\vp$ taking into account the ladder summation. 

This chapter is organized as follows:
In Sec.~\ref{sec:ultrasoft-yukawa}, we discuss the ultrasoft fermion mode in Yukawa model as  a simple example without supersymmetry.
In Sec.~\ref{sec:ultrasoft-QED}, we examine  the ultrasoft fermion mode in QED.
We  analytically sum up the  ladder diagrams giving the vertex correction in the leading order, and find the existence of the ultrasoft fermion mode as in the Yukawa model. 
 We shall also show that the resultant ultrasoft fermion propagator and the vertex satisfy the Ward-Takahashi identity. 
We perform the analysis in QCD using the method which is similar to that in QED in Sec.~\ref{sec:ultrasoft-QCD}.
Section~\ref{sec:ultrasoft-summary} is devoted to a brief summary of this chapter.

The analysis in Secs.~\ref{sec:ultrasoft-yukawa} and \ref{sec:ultrasoft-QED} is based on Ref.~\cite{Hidaka:2011rz}.

\section{Yukawa model}
\label{sec:ultrasoft-yukawa}

Let us start with the Yukawa model, which is the simplest model to study the ultrasoft fermion mode.
Generalization to gauge theory will be discussed in Sec.~\ref{sec:ultrasoft-QED} and \ref{sec:ultrasoft-QCD}.
We calculate the fermion retarded self-energy and obtain the fermion retarded Green function with an ultrasoft momentum $p\lesssim \cp^2T$.
We first see that the naive perturbation theory breaks down in this case. 
Then, we shall show that a use of a dressed propagator gives a sensible result in the perturbation theory and that  the resulting fermion propagator has a new pole in the ultrasoft region.

\subsection{Pinch singularity}

The retarded self-energy in the one-loop level is given by Eq.~(\ref{eq:one-loop-bare}), and its diagrammatic representation is shown in Fig.~\ref{fig:fermion-selfenergy}. 
For small $p$, the self-energy, which is expressed as Eq.~(\ref{eq:intro-selfenergy-fermion-oneloop}) after some manipulations, is reduced to
\begin{equation}
\begin{split}
\selfEnergyR_{\text{bare}}(p)&= \cp^2\int\frac{d^4k}{(2\pi)^4}\kernel(k)\frac{\Slash{k}}{2p\cdot k+ik^0\epsilon},
\label{eq:bareSelfEnergy}
\end{split}
\end{equation}
where $\kernel(k)$ is defined in Eq.~(\ref{eq:Kernel-definition}).
This approximation is equivalent to the HTL approximation~\cite{\HTL}.
The HTL approximation is, however, only valid for $p\sim \cp T$, and not applicable in the ultrasoft momentum region.
In fact, the retarded self-energy in the one-loop level obtained with use of the bare
propagators is found to diverge when $p\rightarrow 0$,
since the integrand contains $1/ p\cdot k$.
This singularity is  called ``pinch singularity''~\cite{\PinchSingularity}.

The origin of this singularity is traced back to the use of the bare propagators because the singularity is caused by the fact that the dispersion relations of the fermion and the boson are the same and the damping rates are zero in these propagators.
For this reason,  one may suspect that this singularity can be removed by adopting the dressed propagators taking into account the asymptotic masses and damping rates of the quasiparticles, as will be shown to be the case shortly.

\subsection{Resummed perturbation}

Since the leading contribution comes from the hard ($k \sim T$) internal and almost on-shell ($k^2\approx 0$)
momentum\footnote{Here we note that the case where the internal momenta are soft ($k\sim gT$) or smaller is not relevant:
In fact, the HTL-resummed propagators~\cite{\HTLResum} should be used for soft momenta.
However, the dispersion relations of the fermion and the boson obtained from these propagators 
are different from each other, so the pinch singularity will not appear in this case.},
we are led to employ the following dressed propagators for the fermion and boson:
\begin{align}
\label{eq:resum-propagator}
\SR(k)\simeq&-\frac{\Slash{k}}{k^2-m^2_f+2i\zeta_f k^0},\\
\SF(k)\simeq&\left(\frac{1}{2}-n_F(k^0)\right)\Slash{k}\frac{4i\zeta_fk^0}{(k^2-m^2_f)^2+4\zeta^2_f(k^0)^2},\\
\DR(k)\simeq&-\frac{1}{k^2-m^2_b+2i\zeta_b k^0},\\
\DF(k)\simeq&\left(\frac{1}{2}+n_B(k^0)\right)\frac{4i\zeta_bk^0}{(k^2-m^2_b)^2+4\zeta^2_b(k^0)^2} ,
\end{align}
where the expressions of $\mf$ and $\mb$ are given in Eqs.~(\ref{eq:mass-fermion}) and (\ref{eq:mass-boson}).
The damping rates of the hard particles, $\zeta_f$ and $\zeta_b$, are of order $\cp^4T\ln\cp^{-1}$. The logarithmic enhancement for the damping rate  is caused by the soft-fermion exchange, which is analogous to that of the hard photon~\cite{\HardPhotonDamping}.
Note that these resummed propagators are the same as those used in~\cite{Lebedev:1989ev}, except for the smallness of the damping rates:
We remark that such a smallness is not the case in QED/QCD, where the damping rate is anomalously large and of order $g^2T\ln g^{-1}$ (``anomalous damping'') \cite{\AnomalousDamping}. 

Using these dressed propagators, we obtain
\begin{equation}
\label{eq:one-loop-pc}
\varSigma^R(p)\simeq \int\frac{d^4k}{(2\pi)^4}\tilde{\kernel}(k)
\frac{\Slash{k}}{1+2\tilde{p}\cdot k/\delta m^2 }
\end{equation}
for small $p$,
where $\delta m^2\equiv \mb^2-\mf^2=g^2T^2/24$, $\zeta\equiv \zeta_f+\zeta_b$, 
$\tilde{\kernel}(k)\equiv (g^2/\delta m^2) \kernel(k)$, and $\tilde{p}^\mu=(\pzero+i\zeta, \vect{p})$.
We have used the modified on-shell condition of  the quasi-particles,
$k^2-m_f^2+2i\zeta_f k^0=0$ and $k^2-m_b^2+2i\zeta_f b^0=0$, 
to obtain the denominator of the integrand in Eq.~(\ref{eq:one-loop-pc}).
We have also neglected $m_b$, $m_f$, $\zeta_b $, and $\zeta_f $ in $\kernel(k)$,
since the leading contribution comes from hard momenta $k\sim T$.
It is worth emphasizing that thanks to $\delta m^2$ and $\zeta$, 
$\varSigma^R(p)$ given in Eq.~(\ref{eq:one-loop-pc}) does not diverge in the infrared limit, $p\rightarrow 0$.

Before evaluating Eq.~(\ref{eq:one-loop-pc}),  we introduce the the following dimensionless value:
\begin{equation}
\label{eq:ultrasoft-lambda}
\lambda\equiv \int\frac{d^4k}{(2\pi)^4}\tilde{\kernel}(k). 
\end{equation}
It can be computed as follows:
\begin{align}
\label{eq:ultrasoft-lambda-computation}
\begin{split}
\int\frac{d^4k}{(2\pi)^4}\tilde{\kernel}(k)
&=\frac{\cp^2}{\deltam^2} \int\frac{d^3\vk}{(2\pi)^3}\sum_{s=\pm}\frac{s}{2|\vk|}(\nf(s|\vk|)+\nb(s|\vk|))\\
&=\frac{\cp^2}{2\pi^2\deltam^2} \int^\infty_0 d|\vk| |\vk| (\nf(|\vk|)+\nb(|\vk|)) \\
&= \frac{\cp^2T^2}{8\deltam^2}.
\end{split}
\end{align}
In the last line, we have used Eqs.~(\ref{eq:integral-formula-f1}) and (\ref{eq:integral-formula-b1}).
From Eq.~(\ref{eq:ultrasoft-lambda-computation}), we see that $\lambda$ is of order unity.
This value will characterize the strength of residue of the pole for both Yukawa model and QED/QCD.

We expand the self-energy in terms of $\tilde{p}^{\mu}$ instead of $p^{\mu}$ itself.
This is the key point of our expansion, which enables us to analytically find the pole of the ultrasoft fermion mode.
Then,  the leading contribution is
\begin{align} 
\label{eq:sigma-p1-computation}
\begin{split}
\varSigma^R(p)&\simeq -\int \frac{d^4k}{(2\pi)^4}\tilde{\kernel}(k)\Slash{k}\frac{2\tilde{p}\cdot k}{\delta m^2}\\
&=-\frac{2\cp^2}{(\deltam^2)^2}\int\frac{d^3\vk}{(2\pi)^3}\sum_{s=\pm}\frac{s}{2|\vk|}(\nf(s|\vk|)+\nb(s|\vk|))\\
&~~~\times(s|\vk|\gamma^0-\vk\cdot\vgamma)((p^0+i\zeta)s|\vk|-\vk\cdot\vp) \\
&=-\frac{\cp^2}{2\pi^2(\deltam^2)^2}\int^\infty_0 d|\vk| |\vk|^3 (\nf(|\vk|)+\nb(|\vk|)) \\
&~~~\times\int^1_{-1} d \cos\theta((p^0+i\zeta) \gamma^0+\cos^2\theta \vp\cdot\vgamma).
\end{split}
\end{align}
By using Eqs.~(\ref{eq:integral-formula-f3}) and (\ref{eq:integral-formula-b3}), we get
\begin{align}
\label{eq:sigma-p1}
\varSigma^R(p)&\simeq-\frac{1}{Z}\left((\pzero+i\zeta)\gamma^0+v \vp\cdot \vgamma\right),
\end{align}
with $Z\equiv \cp^2/(8\lambda^2\pi^2)$ and $v=1/3$. 
Note that the zeroth-order term is absent,  which implies that there is no mass term. 
Actually, it is guaranteed by the symmetries, as discussed in Sec.~\ref{ssc:intro-ultrasoft}.
Thus, we obtain the fermion propagator in the ultrasoft region as
\begin{equation}
\label{eq:result-propagator}
\begin{split}
\SR\pw&\simeq\frac{1}{\selfEnergyR\pw}\\
&=-\frac{Z}{2}\left(\frac{\gamma^0-\hat{\vp}\cdot\vgamma }{\pzero+v|\vp|+i\zeta}+\frac{\gamma^0+\hat{\vp}\cdot\vgamma}{\pzero-v|\vp|+i\zeta}\right).
\end{split}
\end{equation}
Here we have used $\selfEnergyR\pw\gg \Slash{p}$ and decomposed the fermion propagator into the fermion number $+1$ and $-1$ sectors
in the second line.
These two sectors are symmetric under the transformation $\vp\leftrightarrow -\vp$ and $v\leftrightarrow -v$,
so we analyze only the fermion number $+1$ sector in the following.

\begin{figure}[t]
\begin{center}
\includegraphics[width=0.4\textwidth]{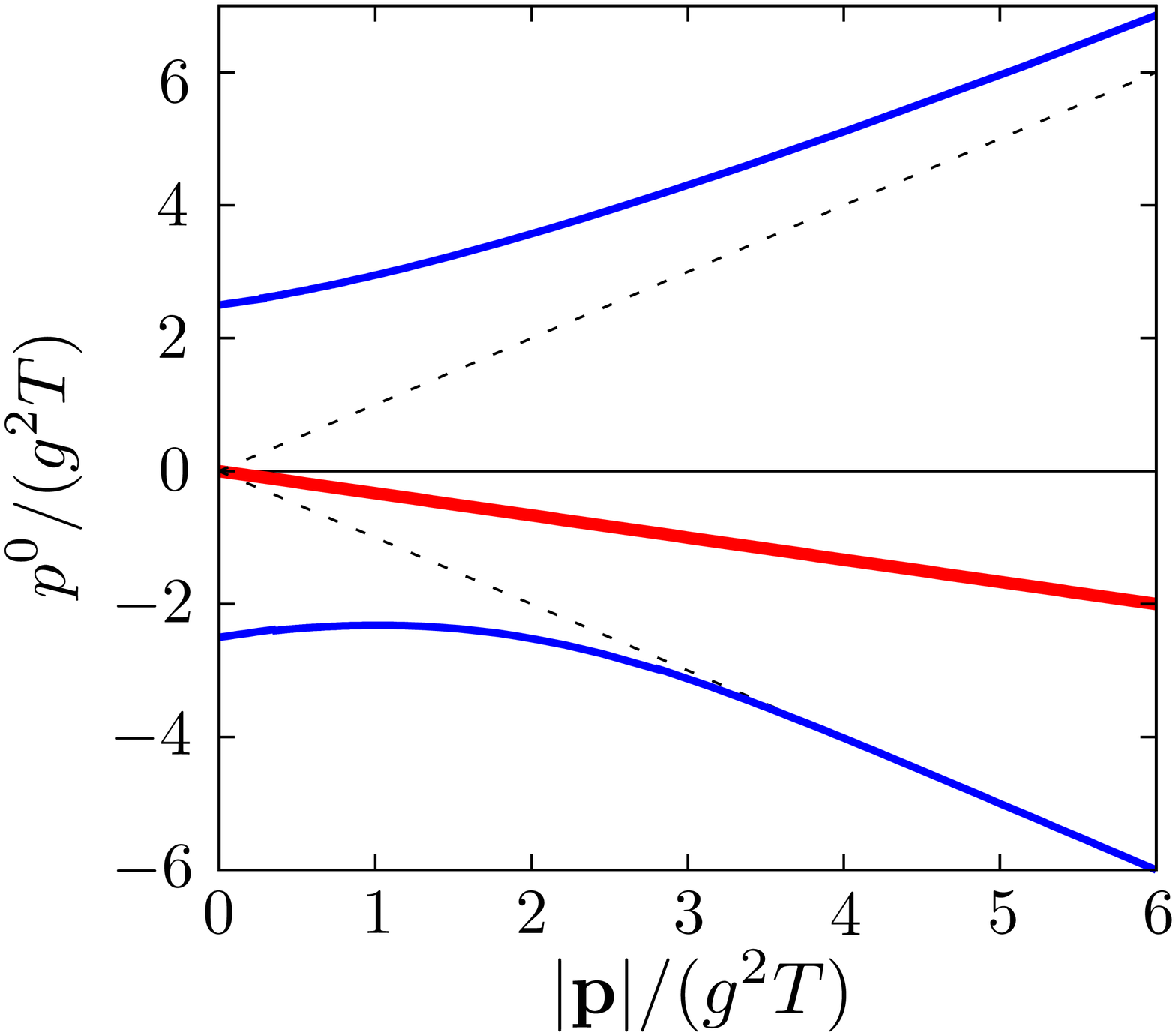} 
\includegraphics[width=0.45\textwidth]{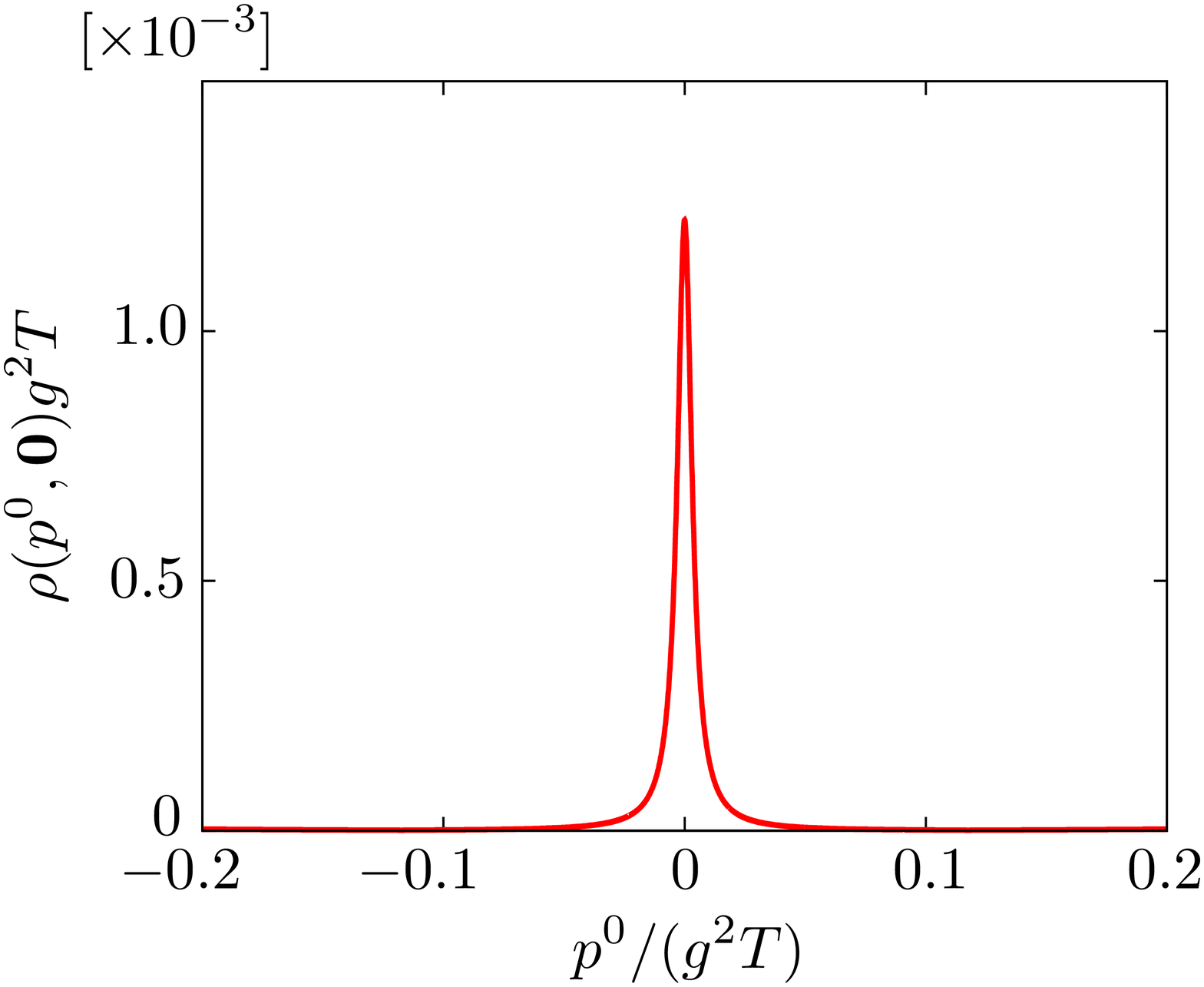}
\caption{Left panel: The dispersion relation in the fermionic sector.
In both of the two figures, the coupling constant is set to $\cp=0.1$.
The vertical axis is the energy $\pzero$, while the horizontal axis is the momentum $|\vp|$.
The solid (blue) lines correspond to the normal fermion and the antiplasmino, while the bold solid (red) one to the ultrasoft mode.
Note that since we focus on the fermion sector, the antiplasmino appears instead of the plasmino.
The dotted lines denote the light cone.
Since our analysis on the ultrasoft mode is valid only for $|\vp|\ll \cp^2T$, 
the plot for $|\vp|\gtrsim \cp^2T$ may not have a physical meaning.
The residue of the antiplasmino becomes exponentially small 
for $|\vp|\gg \cp T$, 
so the plot of the antiplasmino does not represent physical excitation for
 $|\vp|\gg \cp T$, either.
Right panel: The spectral function in the fermion sector, Eq.~(\ref{eq:spectral}), as a function of energy $\pzero$ at zero momentum.
}
\label{fig:dispersion}
\end{center}
\end{figure}

From Eq.~(\ref{eq:result-propagator}), we find a pole at 
\begin{align}
\label{eq:result-dispersion}
\pzero=- v|\vp|-i\zeta.
\end{align}
The dispersion relation of the real part, $\re\pzero=-v|\vp|$, is shown 
in the left panel of Fig.~\ref{fig:dispersion} together with the HTL results \cite{\plasmino} for comparison,
where the coupling constant is chosen as $\cp=0.1$.
The imaginary part of the pole reads
\begin{equation}
\label{eq:result-width}
\zeta \sim \cp^4T\ln g^{-1},
\end{equation}
which is much smaller than those of the normal fermion and the antiplasmino~\cite{Thoma:1994yw}. 
Since the real part and the imaginary part of the pole are finite for $|\vp|\neq 0$, this mode is a damped oscillation mode. 
The residue of the pole is evaluated to be
\begin{align}
\label{eq:result-residue}
Z =\frac{\cp^2}{8\lambda^2\pi^2}=\frac{\cp^2}{72\pi^2}\sim \cp^2,
\end{align}
which means that the mode has only a weak strength in comparison with those of the normal fermion 
and the antiplasmino, whose residues are order of unity.
It is worth mentioning that 
such smallness of the residue is actually compatible with the results in the HTL approximation:
The sum of the residues of the normal fermion and the antiplasmino modes obtained
in the HTL approximation is unity and thus the sum rule of the spectral function of the fermion
is satisfied in the leading order.
Therefore, one could have anticipated that the residue of the ultrasoft mode 
can not be the order of unity but should be of higher order.

The pole given by Eq.~(\ref{eq:result-dispersion}) gives rise to a new peak in the spectral 
function of the fermion as
\begin{align}\label{eq:spectral}
\rho_+\pw= \frac{Z}{\pi}\im \frac{-1}{\pzero+v|\vp|+i\zeta},
\end{align}
which is depicted in the right panel of Fig.~\ref{fig:dispersion}, where $|\vp|$ is set to zero.
Since the expression of $\zeta$ for the Yukawa model is not available in the literature,
we simply adopt $\zeta=\cp^4T\ln \cp^{-1}/(2\pi)$ in the figure.

\subsection{Suppression of ladder diagrams}
\label{ssc:ultrasoft-yukawa-ladder}

\begin{figure}
\begin{center}
\includegraphics[width=0.9\textwidth]{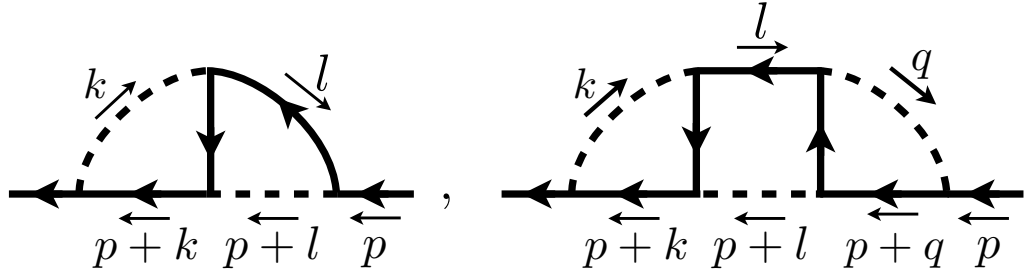}
\caption{Some of the ladder class diagrams.
The solid and dashed line are the dressed propagators of the fermion and the scalar boson, respectively.
}
\label{fig:ladder}
\end{center}
\end{figure}

So far, we have considered the one-loop diagram. 
We need to check that the higher-order loops are suppressed by the coupling constant.
This task would not be straightforward because,  $\delta m^2\sim\cp^2 T^2$
appears in the denominator, as seen in  Eq.~(\ref{eq:one-loop-pc}),
which could make invalid the naive loop expansion. 
The possible diagrams contributing in the leading order are ladder diagrams 
shown in Fig.~\ref{fig:ladder} because the pair of the fermion and the boson propagators gives a contribution of order $1/g^2$, and the vertex gives $g^2$.
However, there is a special suppression mechanism in the present case with the scalar coupling.

For example, let us evaluate the first diagram in Fig.~\ref{fig:ladder},
at  small $p$.
The self-energy is evaluated to be
\begin{align}
\begin{split}
\simeq&\int\frac{d^4k}{(2\pi)^4}\tilde{\kernel}(k)\int\frac{d^4l}{(2\pi)^4}\tilde{\kernel}(l)
\frac{\Slash{k}(\Slash{k}-\Slash{l})\Slash{l}}{(2k\cdot l )}
\frac{2\tilde{p}\cdot (k-l)}{\delta m^2} .
\end{split}
\end{align}
Since there are four vertices and two pairs of the propagators whose momenta are almost the same, 
the formula would apparently yield the factor, $\tilde{\kernel}(k)\tilde{\kernel}(l)\sim\cp^4\times(\delta m^2)^{-2}\sim g^0$.
One can easily verify that this order estimate would remain the same in any higher-loop diagram, 
so any ladder diagram seems to contribute in the leading order as explained.
However, this is not the case for Yukawa model with the scalar coupling. 
An explicit evaluation of the numerators of the fermion propagators gives
$\Slash{k}(\Slash{k}-\Slash{l})\Slash{l}=\Slash{l}k^2-\Slash{k}l^2$, which
turns out to be of  order $\cp^2$.
This is because the internal line is almost on-shell, i.e., 
$k^2$, $l^2\sim \cp^2T^2$,
which comes from the asymptotic masses squared. 
An analysis shows that 
the same suppression occurs in the higher-order diagrams 
such as the second diagram in Fig.~\ref{fig:ladder}.
 Thus, the ladder diagrams giving a vertex correction do not contribute in the leading order in the scalar coupling, and hence, the one-loop diagram in Fig.~\ref{fig:fermion-selfenergy} with the dressed propagators solely suffices to give the self-energy in the leading-order.

We remark that a similar suppression occurs in the effective three-point-vertex at $p\sim \cp T$~\cite{Thoma:1994yw}.
We also note that this suppression mechanism is quite similar to that found in a supersymmetric model for an intermediate temperature region in the sense that $\cp T\ll m$~ \cite{Lebedev:1989rz}, whereas we are dealing with extremely high-$T$ case.
It should be emphasized that this suppression of the vertex correction is not the case in QED/QCD, where all the ladder diagrams contribute in the leading order and must be summed over~\cite{Lebedev:1989ev},  as will be shown in the following sections.

\section{QED} 
\label{sec:ultrasoft-QED}

Next we explore whether the ultrasoft fermion mode also exists in QED at high $T$. 
One might expect that the analysis would  be done in much the same way as in the Yukawa model.
It turns out, however, that the analysis is more complicated and involved. 
It is necessary to sum up the contributions from all the ladder diagrams even apart from the complicated helicity structure of the photon.
In this section, we successfully perform the summation of the ladder diagrams in an analytic way, and obtain the fermion propagator that is valid in the ultrasoft region.
Then we evaluate the pole in the ultrasoft region explicitly and examine the properties of the ultrasoft fermion mode in QED.
We also discuss whether the resummed vertex satisfies the Ward-Takahashi identity.

\subsection{Resummed perturbation}

\subsubsection{One-loop calculation}
First, we evaluate the contribution from the one-loop diagram.
The dressed propagators with hard momenta read
\begin{align}
\SR(k) &\simeq \frac{-\Slash{k}}{k^2-\me^2+2i\zetae k^0} , \\
\label{eq:qed-photon-retarded-propagator}
\DR_{\mu\nu}(k)& \simeq \frac{-P^T_{\mu\nu}(k)}{k^2-\mph^2+2i\zetaph k^0},\\
\SF(k)&\simeq\left(\frac{1}{2}-n_F(k^0)\right)\Slash{k}\frac{4i\zetae k^0}{(k^2-\me^2)^2+4\zetae^2(k^0)^2},\\
\label{eq:qed-photon-s-propagator}
\DF_{\mu\nu}(k)&\simeq\left(\frac{1}{2}+n_B(k^0)\right)\frac{4i\zetaph k^0 P^T_{\mu\nu}(k)}{(k^2-\mph^2)^2+4\zetaph^2(k^0)^2} .
\end{align}
The expressions of $m_e$ and $m_\gamma$ are given in Eqs.~(\ref{eq:mass-electron}) and (\ref{eq:mass-photon}).
The damping rates of electron and photon are estimated as $\zeta_e\sim\cp^2T\ln \cp^{-1}$~\cite{\AnomalousDamping} and $\zeta_\gamma\sim\cp^4T\ln g^{-1}$ \cite{\HardPhotonDamping}. 
Note that $\zetae$ is much larger than that in the Yukawa model, 
which is called ``anomalous damping'' \cite{\AnomalousDamping}.
This large electron damping makes the damping rate of the ultrasoft mode much larger than in the Yukawa model.
Here we have adopted the Coulomb gauge, in which the analysis becomes simple thanks to the transversality of the photon propagator.
We note that the $u^\mu u^\nu/|\vk|^2$ term has been omitted in Eqs.~(\ref{eq:qed-photon-retarded-propagator}) and (\ref{eq:qed-photon-s-propagator}) because that term vanishes after the $k^0$ integral, as in Eq.~(\ref{eq:HTL-longitudinal-contribution}).

By using these resummed propagators, 
the one-loop contribution in the ultrasoft region is evaluated as 
\begin{equation}
\label{eq:pairPropagator}
\begin{split}
\varSigma_\text {one-loop}^R(p) &=i\cp^2\int\frac{d^{4}k}{(2\pi)^{4}} \gamma^\mu \bigl[\DF_{\mu\nu}(-k)\SR(p+k)
+\DR_{\mu\nu}(-k)\SF(p+k)\bigr]\gamma^\nu\\
&\simeq2\int\frac{d^{4}k}{(2\pi)^{4}}\tilde{\kernel}(k)\frac{\Slash{k}}{1+2\tilde{p}\cdot k/\deltam^2} ,
\end{split}
\end{equation}
where $\deltam^2\equiv\mph^2-\me^2=-\cp^2T^2/12$, $\damping\equiv\zetae+\zetaph\simeq\zetae$.  
Here we have used the same notation for $\delta m^2$ and $\zeta$ as those in the Yukawa model, 
although their parametrical expressions are different from each other.
The factor two in the last line of Eq.~(\ref{eq:pairPropagator}) comes from two degrees of freedom of photon polarization.
At the one-loop order, we obtain
\begin{equation}
\begin{split}
\varSigma^R_\text{one-loop}(p) 
&= -\frac{16\pi^2}{\cp^2}\lambda^2 \big( (p^0+i\zeta)\gamma^0+v\vp\cdot \vgamma \big).
\end{split}
\end{equation}
We note that this expression has the same structure as that for the Yukawa model;
see Eq.~(\ref{eq:sigma-p1}).

\begin{figure}
\begin{center}
\includegraphics[width=0.5\textwidth]{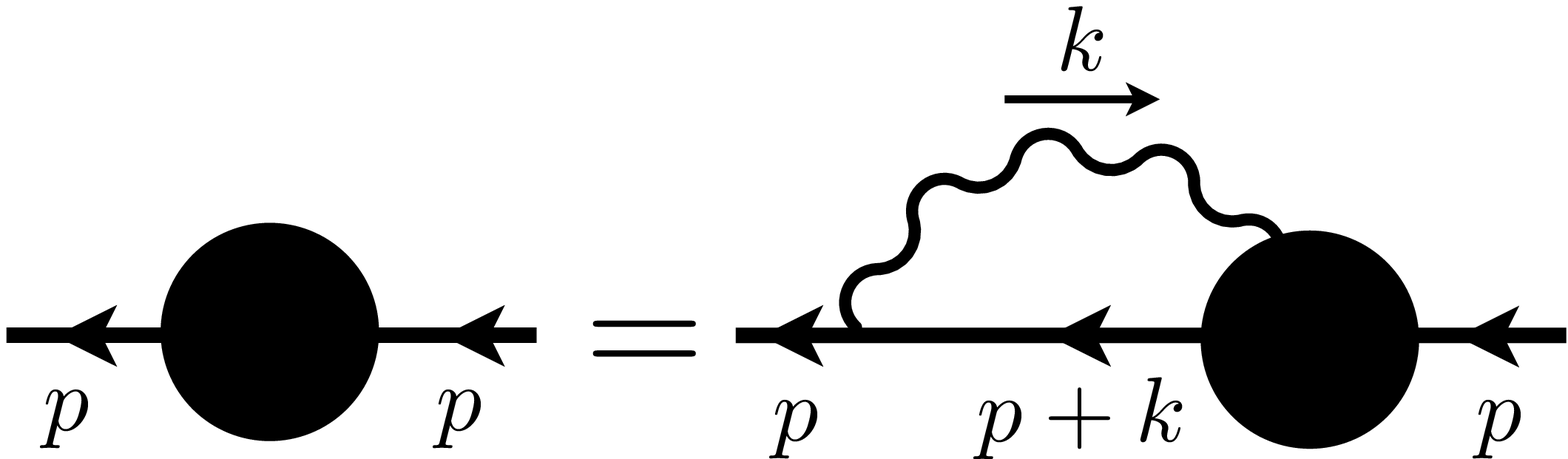}\\
\includegraphics[width=0.85\textwidth]{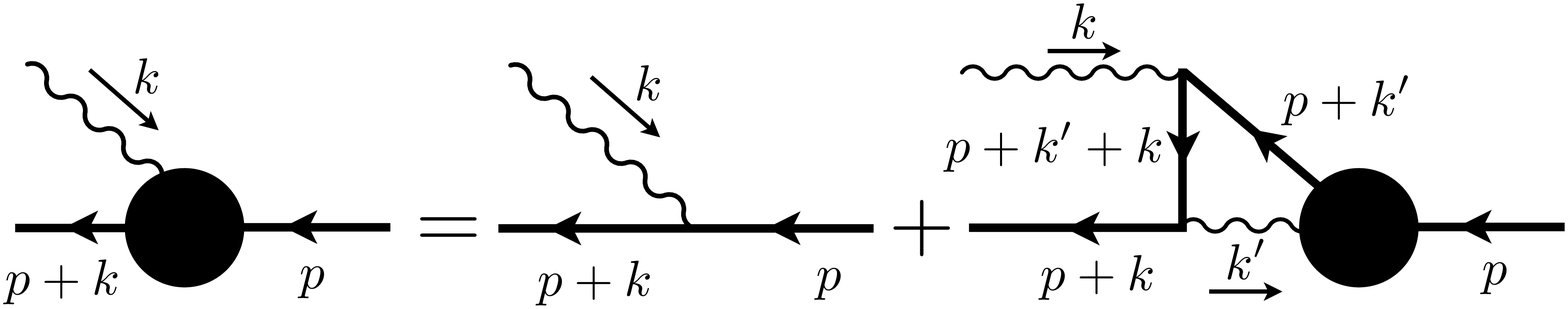}
\caption{Resummed self-energy and the self-consistent equation for the vertex function.
}
\label{fig:selfconsistentEquation}
\end{center}
\end{figure}

\subsubsection{Ladder summation}
As already mentioned, the ladder diagrams contribute to the leading-order in contrast to the case of the Yukawa model.
In this subsection, we sum up all the ladder diagrams, and obtain the analytical expressions of the pole position and the residue of the ultrasoft mode. 

For this purpose,
we introduce the vertex $\cp\varGamma^\mu(p,k)$
 defined through the following self-energy:
\begin{equation}
\varSigma^R(p) =
\int\frac{d^{4}k}{(2\pi)^{4}}\tilde{\kernel}(k)\frac{\gamma^{\mu}\Slash{k}P^T_{\mu\nu}(k)
\varGamma^{\nu}(p,k)}{1+2\tilde{p}\cdot k/\deltam^2}.
\label{eq:selfEnergy}
\end{equation}
Here the vertex contains the contributions from all the ladder diagrams by imposing the following self-consistent equation for the vertex function: 
\begin{align}
\label{eq:vertex}
\varGamma^\mu(p,k)&=\gamma^{\mu}  +\int\frac{d^{4}k'}{(2\pi)^{4}}
\tilde{\kernel}(k')\gamma^\nu\frac{\Slash{p}+\Slash{k}+\Slash{k}'}{(p+k+k')^2}
\frac{\gamma^\mu(\Slash{p}+\Slash{k}')}{1+2\tilde{p}\cdot k'/\deltam^2}
 P^T_{\nu\rho}(k') \varGamma^\rho(p,k').
\end{align}
Equations (\ref{eq:selfEnergy}) and (\ref{eq:vertex}) are represented diagrammatically in Fig.~\ref{fig:selfconsistentEquation}.
We have used the same approximation as that used in the derivation of Eq.~(\ref{eq:bareSelfEnergy}) for the propagator of fermion and photon.
We should remark here that this summation scheme using the self-consistent equation was first constructed in Ref.~\cite{Lebedev:1989ev}. 
However, we also note that the equation has never been solved  either analytically nor numerically.
In the following, we solve this self-consistent equation analytically for small $\tilde{p}$, and show that the dispersion relation does not change from that in the one-loop order even after incorporating all the ladder diagrams.

At small $\tilde{p}$, Eq.~(\ref{eq:vertex}) reduces to 
\begin{align}
\varGamma^\mu(p,k)
&=\gamma^\mu +\int\frac{d^4k'}{(2\pi)^4}\tilde\kernel(k')
\frac{P^T_{\nu\rho}(k')}{1+2(\tilde{p}\cdot k')/\deltam^2}
\frac{(k^\nu \gamma^\mu+k'^\mu\gamma^\nu)\Slash{k}'\varGamma^\rho(p,k')}{k\cdot k'},
\label{eq:vertex2}
\end{align}
where we have dropped the term which is proportional to $\Slash{k}$ because it only gives higher order contribution after being multiplied by $\Slash{k}$ in the numerator of Eq.~(\ref{eq:selfEnergy}), because $\Slash{k}^2=k^2\sim g^2T^2$.

Let us solve the self-consistent equation (\ref{eq:vertex2}).
We expand the vertex function as
\begin{equation}
\label{eq:ultrasoft-QED-vertex-expand}
\varGamma^\mu(p,k) = \varGamma^\mu_0(k) +\delta \varGamma^\mu(p,k) ,
\end{equation}
where $\varGamma^\mu_0(k)$ is of order unity and $\delta\varGamma^\mu(p,k)$ is of order $\tilde{p}/(\cp^2T)$.

We first evaluate $\varGamma^\mu_0(k)$, which can be decomposed as follows:
\begin{equation}
\varGamma^\mu_0(k)=\gamma^\mu A(k) + k^\mu B(k)+u^\mu C(k) ,
\label{eq:varGamma}
\end{equation}
where $A$, $B$, and $C$ are $4\times 4$ matrices.
Then the self-consistent equation for $\varGamma_0^\mu(k)$ becomes
\begin{equation}
\label{eq:ultrasoft-QED-BS-0}
\begin{split}
\gamma^\mu A(k)+ k^\mu B(k)+u^\mu C(k) 
=\gamma^\mu
+\gamma^{\mu}\int\frac{d^4k'}{(2\pi)^4}\tilde\kernel(k')
\frac{\Slash{k}' k^\nu P^T_{\nu\rho}(k') \gamma^\rho}{k\cdot k'}A(k') ,
\end{split}
\end{equation}
where $B(k)$ and $C(k)$ in the right hand side vanish due to transversality of the photon propagator:
$P^T_{\mu\nu}(k) k^\nu= P^T_{\mu\nu}(k) u^\nu=0$.
By assuming that $A(k)$ is a constant, the integral becomes
\begin{align} 
\begin{split}
&\int\frac{d^4k'}{(2\pi)^4}\tilde\kernel(k')
\gamma^\mu\frac{\Slash{k}' k^\nu P^T_{\nu\rho}(k') \gamma^\rho}{k\cdot k'}A \\
&= \frac{\cp^2}{\deltam^2}\int\frac{d^3\vk'}{(2\pi)^3}\sum_{s=\pm}\frac{s}{2|\vk'|}(\nf(s|\vk'|)+\nb(s|\vk'|))
\gamma^\mu\frac{(s|\vk'|\gamma^0-\vk'\cdot\vgamma) }{k^0s|\vk'|-\vk\cdot\vk'}|\vk| \\
&~~~\times(\hat{\vk}\cdot\vgamma-(\hat{\vk}\cdot\hat{\vk}')(\hat{\vk}'\cdot\vgamma)) A  \\
&= \frac{\cp^2}{\deltam^2}\int\frac{d^3\vk'}{(2\pi)^3}\frac{1}{|\vk'|}(\nf(|\vk'|)+\nb(|\vk'|))
\gamma^\mu\frac{k^0\gamma^0-(\hat{\vk}'\cdot\vgamma)(\vk\cdot\hat{\vk}') }{(k^0)^2-(\vk\cdot\hat{\vk}')^2}|\vk|\\
&~~~\times(\hat{\vk}\cdot\vgamma-(\hat{\vk}\cdot\hat{\vk}')(\hat{\vk}'\cdot\vgamma)) A.
\end{split}
\end{align}
By imposing $k^2=0$ as before, this expression becomes
\begin{align} 
\begin{split} 
\label{eq:integral}
&\frac{\cp^2}{\deltam^2}\int\frac{d^3\vk'}{(2\pi)^3}\frac{1}{|\vk'|}(\nf(|\vk'|)+\nb(|\vk'|))\gamma^\mu
\frac{k^0\gamma^0 \vk\cdot\vgamma}{|\vk|^2} A \\
&=\frac{\cp^2}{2\pi^2\deltam^2}\int^\infty_0 d|\vk'| |\vk'|(\nf(|\vk'|)+\nb(|\vk'|))\gamma^\mu\frac{k^0\gamma^0 \vk\cdot\vgamma}{|\vk|^2} A  \\ 
&= \left(-\gamma^\mu\lambda 
+\frac{2k^\mu}{k^0}\gamma^0\lambda\right)A,
\end{split}
\end{align}
where we have used Eqs.~(\ref{eq:integral-formula-f1}) and (\ref{eq:integral-formula-b1}), and dropped  $\Slash{k}$ term in the last line. 
Then, from Eq.~(\ref{eq:ultrasoft-QED-BS-0}), we find
\begin{align}
A = \frac{1}{1+\lambda}\vect{1},
\quad B(k) = \frac{1}{k^0}\frac{2\lambda}{1+\lambda}\gamma^0,
\quad C(k)=0,
\label{eq:varGammaABC}
\end{align}
where $\vect{1}$ is the unit $4\times 4$ matrix. 

Next we evaluate $\delta\varGamma(p,k)$.
From Eq.~(\ref{eq:vertex2}), expanding $\varGamma^\mu(p,k)$ in terms of $2\tilde{p}\cdot k'/\delta m^2$, we find the self-consistent equation for $\delta\varGamma(p,k)$ as
\begin{equation}
\begin{split}
\delta\varGamma^\mu(p,k)=& \int\frac{d^4k'}{(2\pi)^4}
\tilde\kernel(k')P^T_{\nu\rho}(k')
\frac{(k^\nu \gamma^\mu+k'^\mu\gamma^\nu)\Slash{k}'}{k\cdot k'} 
 \left[-A\gamma^\rho\frac{2\tilde{p}\cdot k'}{\deltam^2}+\delta\varGamma^\rho(p,k')\right].
\end{split}
\label{eq:deltaVarGamma}
\end{equation}
It is not easy to analytically solve Eq.~(\ref{eq:deltaVarGamma}) directly.
So we instead calculate the following function:
\begin{equation}
\label{eq:deltaPi}
\delta\varPi(p)=
\int\frac{d^4k}{(2\pi)^4}\tilde\kernel(k)P^T_{\mu\nu}(k)  \gamma^\mu \Slash{k}\delta \varGamma^\nu(p,k) .
\end{equation}
Then Eq.~(\ref{eq:deltaVarGamma}) leads to the following closed equation,
\begin{equation}
\begin{split}
\delta\varPi(p)=&
-4 A\int\frac{d^4k}{(2\pi)^4}\tilde\kernel(k)\int\frac{d^4k'}{(2\pi)^4}\tilde\kernel(k')P^T_{\mu\nu}(k)  \gamma^\mu \Slash{k}
\frac{k'^\nu\Slash{k}'}{k\cdot k'} 
\frac{\tilde{p}\cdot k'}{\deltam^2}\\
&\quad +\int\frac{d^4k'}{(2\pi)^4}
\left[ 
\int\frac{d^4k}{(2\pi)^4}\tilde\kernel(k)P^T_{\mu\nu}(k)  \gamma^\mu \Slash{k}\frac{k'^\nu}{k\cdot k'}\right]
\tilde\kernel(k') P^T_{\rho\lambda}(k')\gamma^\rho\Slash{k}'\delta\varGamma^\lambda(p,k') \\
=&-\lambda A\varSigma_\text{one-loop}(p) -\lambda\delta\varPi(p).
\end{split} 
\end{equation}
Here we have used Eqs.~(\ref{eq:integral}) and (\ref{eq:deltaPi}) in the second line.
The solution to this equation is readily found to be
$\delta\varPi(p)=-\lambda A^2\varSigma_\text{one-loop}(p)$, where Eq.~(\ref{eq:varGammaABC}) is used for $A$.
Then, the self-energy is evaluated as 
\begin{equation}
\label{eq:selfenergy-1st}
\begin{split}
\varSigma^R(p)&= \int\frac{d^4k}{(2\pi)^4}\tilde\kernel(k)P^T_{\nu\rho}(k)  \gamma^\nu \Slash{k}
\left(-\frac{2\tilde{p}\cdot k}{\deltam^2}\varGamma^\rho_0(k)+\delta \varGamma^\rho(p,k)\right)\\
&=A \varSigma_\text{one-loop}(p)
+\delta\varPi(p) \\
&=-\frac{1}{Z}(\gamma^0(p^0+i\zeta)+v\vp\cdot \vgamma),
\end{split}
\end{equation}
where the residue is
\begin{align}
Z&=\frac{\cp^2}{16\pi^2 \lambda^2}(1+\lambda)^2
=\frac{\cp^2}{144\pi^2}.
\end{align}
The pole of the ultrasoft fermion mode has  the velocity $v=1/3$, 
damping rate $\damping$, and the residue $Z$.
The dispersion of the mode is the same as that in the one-loop level 
whereas the residue is changed.
This is our main result for QED. 

\subsection{Ward-Takahashi identity}
In this subsection, we examine whether the summation scheme, Eqs.~(\ref{eq:selfEnergy}) and (\ref{eq:vertex}), is consistent with the $U(1)$ gauge symmetry. 
Concretely, we check that our resummed vertex function and self-energy satisfy the Ward-Takahashi (WT) identity in the leading order of the coupling constant.

The WT identity reads
\begin{equation}
\begin{split}
k^\mu \varGamma_{ \mu}(p,k)
&=\Slash{p}+\Slash{k}-\varSigma^R(p+k) -\Slash{p}+\varSigma^R(p).
\label{eq:WTidentity}
\end{split}
\end{equation}
Since $\varGamma_{ \mu}(p,k)$ contains two separated scales $k\sim T$ and $p\lesssim \cp^2T$, we need to treat them carefully.
For the hard part, $\varSigma^R(p+k)\simeq \varSigma^R(k)$ is of order $\cp^2T$, which is negligible compared to $\Slash{k}$.
In addition, the momentum dependent part $\varSigma^R(p+k)-\varSigma^R(k)\sim g^2 p$ is also negligible compared to $\varSigma^R(p)$.
Therefore the WT identity reduces to  
\begin{equation}
k_\mu \varGamma^{ \mu}(p,k)=\Slash{k}+\varSigma^R(p).
\label{eq:mod-WT}
\end{equation}

On the other hand, multiplying Eq.~(\ref{eq:vertex}) by $k_\mu$, we have
\begin{equation}
\begin{split}
k_\mu\varGamma^\mu(p,k)
&=(\Slash{p}+\Slash{k})-\Slash{p}  + \int\frac{d^{4}k'}{(2\pi)^{4}}
\tilde \kernel(k')\gamma^\nu\frac{ \Slash{p}+\Slash{k}+\Slash{k}'}{(p+k+k')^2} \\
&\quad\qquad\qquad\times\frac{((\Slash{p}+\Slash{k}+\Slash{k}')-\Slash{p}-\Slash{k}')(\Slash{p}+\Slash{k}')}{1+2\tilde{p}\cdot k'/\deltam^2}
P^T_{\nu\rho}(k') \varGamma^\rho(p,k')\\
&=\Slash{k} + \int\frac{d^{4}k'}{(2\pi)^{4}}
\tilde \kernel(k')\gamma^\nu\left(\Slash{p}+\Slash{k}'-\frac{ \Slash{p}+\Slash{k}+\Slash{k}'}{2k\cdot k'}(p+k')^2\right) \\
&\quad\qquad\qquad\times\frac{1}{1+2\tilde{p}\cdot k'/\deltam^2}
P^T_{\nu\rho}(k') \varGamma^\rho(p,k')\\
&\simeq\Slash{k}+\varSigma^R(p),
\end{split}
\end{equation}
where we have dropped the terms of order $g^2T$, and used Eq.~(\ref{eq:selfEnergy}) in the last line.
This expression coincides with Eq.~(\ref{eq:mod-WT}), so our self-consistent equation satisfies the WT identity.
We note that this proof was made without using the expansion in terms of $\tilde{p}/ \cp^2T$.
For this reason, Eq.~(\ref{eq:mod-WT}) is generally valid for $p\lesssim \cp^2T$, as well as for $\tilde{p}\ll \cp^2T$.

Next, we check whether the explicit solution Eq.~(\ref{eq:varGammaABC}) of the self-consistent equation (\ref{eq:vertex}) at zeroth order in $\tilde{p}/(\cp^2T)$ satisfies the WT identity to see the consistency with the gauge symmetry of the following two conditions adopted to obtain Eq.~(\ref{eq:varGammaABC}):
One is that terms proportional to $\Slash{k}$ in $\varGamma(p,k)$ is dropped, since they are negligible in the self-energy due to $\Slash{k}\Slash{k}=k^2\sim g^2T^2$.
The other is that we imposed the on-shell condition $k^2=0$, because the internal photon in the self-energy is almost on-shell.
Using the same conditions, we expect that the vertex function satisfies the WT identity in the leading order of the coupling constant.
In fact, we have for the zeroth order in $\tilde{p}$ 
\begin{equation}
\label{eq:ultrasoft-WT-zeroth}
\begin{split}
k_\mu \varGamma_0^{\mu}(k)& \simeq \Slash{k}A+k^2B(k)\simeq0 .
\end{split}
\end{equation}
Here  we have dropped $\Slash{k}$ and $k^2$ in the last equality.

Finally, we check that the equation determining the vertex function Eq.~(\ref{eq:deltaVarGamma}),
which is first order in $\tilde{p}$, satisfies the WT identity.
By multiplying this equation by $k_\mu$, we obtain
\begin{equation}
\label{eq:ultrasoft-WT-first}
\begin{split}
k_\mu\delta\varGamma^\mu(p,k)&= \int\frac{d^4k'}{(2\pi)^4}
\tilde\kernel(k')P^T_{\nu\rho}(k')\gamma^\nu\Slash{k}'
\left[-A\gamma^\rho\frac{2\tilde{p}\cdot k'}{\deltam^2}+\delta\varGamma^\rho(p,k')\right]\\
&=\varSigma^R(p),
\end{split}
\end{equation}
where  we have dropped $\Slash{k}$ as in the previous equation and used Eq.~(\ref{eq:selfenergy-1st}).
Therefore, our analytic solution of the self-consistent equation satisfies the WT identity in the leading order of the coupling constant.

We note that without the summation of the ladder diagrams, the WT identity is not satisfied when the external momentum of fermion is ultrasoft.
By contrast, the ladder summation was unnecessary in the Yukawa model, in which the gauge symmetry is absent.

\section{QCD} 
\label{sec:ultrasoft-QCD}

\begin{figure}
\begin{center}
\includegraphics[width=0.9\textwidth]{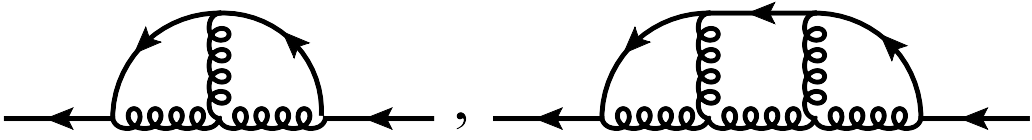} 
\caption{Some of the ladder class diagrams which do not appear in the case of the Yukawa model and QED.
The solid and curly lines are the dressed propagators of the quark and the gluon, respectively.
}
\label{fig:ladder-QCD}
\end{center}
\end{figure}

In this section, we show the existence of the ultrasoft fermion mode, and obtain the expressions of the dispersion relation, the damping rate, and the strength in QCD, in a similar way to that in QED.
The differences between QED and QCD are as follows:
First, the damping rate of the hard gluon is much larger than that of hard photon.
This is because the gluon can collides via the soft gluons~\cite{Lebedev:1990un, Pisarski:1993rf} owing to the self-coupling while the photon can not~\cite{\HardPhotonDamping}.
Second, new kind of the ladder diagrams shown in Fig.~\ref{fig:ladder-QCD} contribute in QCD in addition to the ladder diagrams drawn in Fig.~\ref{fig:ladder}, again due to the self-coupling of the gluons.
Despite these differences, we will see that the properties of the ultrasoft fermion mode are the same qualitatively in both gauge theories.


\begin{figure}
\begin{center}
\includegraphics[width=0.9\textwidth]{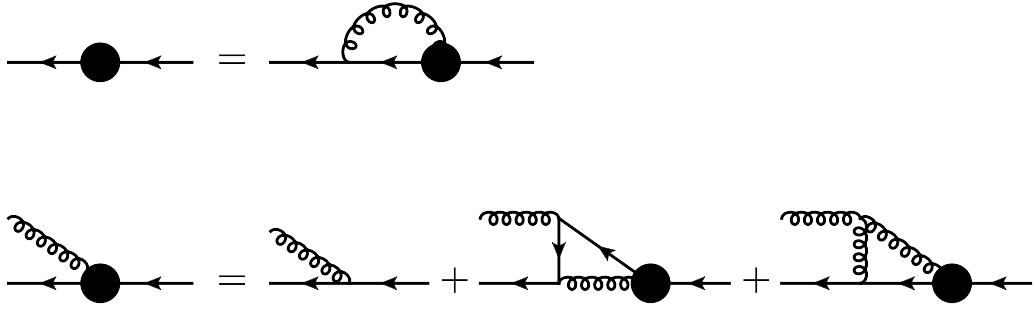} 
\caption{Diagrammatic expressions for Eqs.~(\ref{eq:ultrasoft-QCD-vertex}) and (\ref{eq:selfEnergy-QCD}).
The notation is the same as in Fig.~\ref{fig:ladder-QCD}.
}
\label{fig:BS-QCD}
\end{center}
\end{figure}

To sum up the ladder diagrams, we use the following self-consistent equation for the quark-gluon vertex function $\varGamma^\mu_a(p,k)$, which can be derived as in the QED case:
\begin{align}
\begin{split}
\label{eq:ultrasoft-QCD-vertex} 
\varGamma^\mu_a(p,k)&=\gamma^{\mu} t_a  +\int\frac{d^{4}k'}{(2\pi)^{4}}
\tilde{\kernel}(k')\gamma^\nu t_b \frac{\Slash{p}+\Slash{k}+\Slash{k}'}{(p+k+k')^2}
\frac{\gamma^\mu t_a(\Slash{p}+\Slash{k}')}{1+2\tilde{p}\cdot k'/\deltam^2}
 P^T_{\nu\rho}(k') \varGamma^\rho_b(p,k') \\
&~~~- i\int\frac{d^{4}k'}{(2\pi)^{4}}\tilde{\kernel}(k')
\frac{\gamma^\sigma t_c(\Slash{p}+\Slash{k}')}{1+2\tilde{p}\cdot k'/\deltam^2}P^T_{\nu\rho}(k')\varGamma^{\rho}_{b}(p, k')
\left(\frac{P^T_{\alpha\sigma}(k'-k)}{(k'-k)^2}+\frac{u_\alpha u_\sigma}{|\vk'-\vk|^2} \right) \\
&~~~\times f_{abc}(g^{\nu\alpha}(k-2k')^\mu+g^{\mu\nu}(k+k')^\alpha+g^{\mu\alpha}(k'-2k)^\nu)
.
\end{split} 
\end{align}
Here we have used the Coulomb gauge as in the analysis in QED, and introduced $\deltam^2\equiv\mg^2-\mq^2= \cp^2T^2 [N/24+1/(8N)+\Nf/12]$ and $\zeta\equiv\zetaq+\zetag$.
The expression for $\mq$ and $\mg$ are given by Eqs.~(\ref{eq:mass-quark}) and (\ref{eq:mass-gluon}).
$\zetaq$ and $\zetag$ are of order $\cp^2T\ln(1/\cp)$, which is much larger than $\zetaph$ due to the self-interaction of the gluon.
We note that the presence of the third term in the right-hand side, which is absent in the case of QED, is caused by the self-coupling of the gluons.
The quark self-energy is written in terms of $\varGamma^\mu_a(p,k)$ as
\begin{align}
\begin{split}
\varSigma^R(p) =
\int\frac{d^{4}k}{(2\pi)^{4}}\tilde{\kernel}(k)\frac{\gamma^{\mu}t_a\Slash{k}P^T_{\mu\nu}(k)
\varGamma^{\nu}_a(p,k)}{1+2\tilde{p}\cdot k/\deltam^2}.
\label{eq:selfEnergy-QCD}
\end{split}
\end{align}
The diagrams for these two equations are drawn in Fig.~\ref{fig:BS-QCD}.

At small $\tilde{p}$, Eq.~(\ref{eq:ultrasoft-QCD-vertex}) becomes
\begin{align}
\begin{split}
\label{eq:ultrasoft-QCD-vertex-2}
\varGamma^\mu(p,k) t_a&=\gamma^{\mu} t_a  +\int\frac{d^{4}k'}{(2\pi)^{4}}
\tilde{\kernel}(k')  \frac{k^\nu\gamma^\mu+\gamma^\nu k'{}^\mu}{k\cdot k'}
\frac{  \Slash{k}'}{1+2\tilde{p}\cdot k'/\deltam^2}
 P^T_{\nu\rho}(k')t_b t_a t_b\varGamma^\rho(p,k') \\
&~~~- i\int\frac{d^{4}k'}{(2\pi)^{4}}\tilde{\kernel}(k')
\frac{\gamma^\sigma \Slash{k}'}{1+2\tilde{p}\cdot k'/\deltam^2}P^T_{\nu\rho}(k')
\left(\frac{P^T_{\alpha\sigma}(k'-k)}{-2k\cdot k'}+\frac{u_\alpha u_\sigma}{|\vk'-\vk|^2} \right) \\
&~~~\times (-2k'{}^\mu g^{\nu\alpha} -2k^\nu g^{\mu\alpha} +g^{\mu\nu}(k+k')^\alpha)
f_{abc} t_c t_b \varGamma^{\rho}(p, k'),
\end{split}
\end{align}
where we have used $k^2$, $k'^2\sim \cp^2T^2$, and the fact that $\Slash{k}$ and $P^T_{\mu \alpha}(k)$ appear in the left of the vertex function in Eq.~(\ref{eq:selfEnergy-QCD}), and assumed that the vertex function has the same color structure as the bare one~\cite{Lebedev:1989ev}: 
$\varGamma^\mu_a(p,k)=\varGamma^\mu(p,k) t_a$, where $\varGamma^\mu(p,k)$ does not have color structure. 
The third term in the right-hand side in Eq.~(\ref{eq:ultrasoft-QCD-vertex-2}) is reduced to
\begin{align}
\begin{split} 
\label{eq:ultrasoft-QCD-BS-gluon}
& -i\int\frac{d^{4}k'}{(2\pi)^{4}}\tilde{\kernel}(k')\frac{P^T_{\nu\rho}(k')}{1+2\tilde{p}\cdot k'/\deltam^2}
\Biggl[\gamma^j \Slash{k}' \frac{P^T_{m j}(k'-k)}{-2k\cdot k'}
 (-2k'{}^\mu g^{\nu m} -2k^\nu g^{\mu m} +g^{\mu\nu}(k+k')^m)\\
&+\gamma^0 \Slash{k}' \frac{ 1}{|\vk'-\vk|^2} 
 (-2k'{}^\mu g^{\nu 0} -2k^\nu g^{\mu 0} +g^{\mu\nu}(k+k')^0)
\Biggr]f_{abc} t_c t_b \varGamma^{\rho}(p, k')\\
&= -i\int\frac{d^{4}k'}{(2\pi)^{4}}\tilde{\kernel}(k')\frac{P^T_{\nu\rho}(k')}{1+2\tilde{p}\cdot k'/\deltam^2}
\Biggl[\gamma^j \Slash{k}' \frac{1}{k\cdot k'}
 (k'{}^\mu g^{\nu m} +k^\nu g^{\mu m} -k^mg^{\mu\nu})\\
 &\times \left(\delta_{mj}-\frac{(k'-k)_m(k'-k)_j}{|\vk'-\vk|^2}\right) 
+\gamma^0 \Slash{k}' \frac{ 1}{|\vk'-\vk|^2} 
 g^{\mu\nu}(k+k')^0
\Biggr]f_{abc} t_c t_b \varGamma^{\rho}(p, k') \\
&= -i\int\frac{d^{4}k'}{(2\pi)^{4}}\tilde{\kernel}(k')\frac{P^T_{\nu\rho}(k')}{1+2\tilde{p}\cdot k'/\deltam^2}
\Biggl[  \frac{k'{}^\mu g^{\nu j} +k^\nu g^{\mu j} -k^j g^{\mu\nu}}{k\cdot k'} \gamma^j \Slash{k}' \\
&~~~ +\frac{(\vk'\cdot\vgamma-\vk\cdot\vgamma)}{|\vk'-\vk|^2 k\cdot k'} \Slash{k}' 
 g^{\mu\nu}(\vk'\cdot\vk-|\vk|^2) +\frac{(k+k')^0\gamma^0 \Slash{k}' }{|\vk'-\vk|^2} 
 g^{\mu\nu}
\Biggr]f_{abc} t_c t_b \varGamma^{\rho}(p, k') . 
\end{split}
\end{align}
Here we have used the fact that $P^T_{\mu \alpha}(k)$ appear in the left of the vertex function in Eq.~(\ref{eq:selfEnergy-QCD}) again.
By using the fact that $\Slash{k}$ appear in the left of the vertex function in Eq.~(\ref{eq:selfEnergy-QCD}) again, the terms which are proportional to $g^{\mu\nu}$ are summed up to yield
\begin{align}
\begin{split} 
& -i\int\frac{d^{4}k'}{(2\pi)^{4}}\tilde{\kernel}(k')\frac{P^T_{\nu\rho}(k')g^{\mu\nu}}{1+2\tilde{p}\cdot k'/\deltam^2}
\Biggl[ \vk\cdot\vgamma(k_0 k'_0-|\vk'|^2)  +\vk'\cdot\vgamma (k_0 k'_0-|\vk|^2)  \Biggr] \\
&~~~\times \frac{\Slash{k}'f_{abc} t_c t_b \varGamma^{\rho}(p, k')}{k\cdot k' |\vk'-\vk|^2} \\
&= -i\int\frac{d^{4}k'}{(2\pi)^{4}}\tilde{\kernel}(k')\frac{P^T_{\nu\rho}(k')g^{\mu\nu}}{1+2\tilde{p}\cdot k'/\deltam^2}
\Biggl[ k_0\gamma_0(k_0 k'_0-|\vk'|^2)+k'_0\gamma_0 (k_0 k'_0-|\vk|^2) \Biggr] \\
&~~~\times \frac{\Slash{k}'f_{abc} t_c t_b \varGamma^{\rho}(p, k')}{k\cdot k' |\vk'-\vk|^2}.
\end{split}
\end{align}
This contribution is found to be negligible if we use $k^2$, $k'{}^2\sim \cp^2T^2$.
Thus, Eq.~(\ref{eq:ultrasoft-QCD-BS-gluon}) becomes
\begin{align}
\begin{split}
& i\int\frac{d^{4}k'}{(2\pi)^{4}}\tilde{\kernel}(k')\frac{P^T_{\nu\rho}(k')}{1+2\tilde{p}\cdot k'/\deltam^2}
\frac{k'{}^\mu \gamma^\nu +k^\nu \gamma^\mu }{k\cdot k'}  \Slash{k}' 
f_{abc} t_c t_b \varGamma^{\rho}(p, k') .
\end{split}
\end{align} 
By using the commutation relation among $t_a$, this expression partially cancels the second term in the right-hand side of Eq.~(\ref{eq:ultrasoft-QCD-vertex-2}).
Then, Eq.~(\ref{eq:ultrasoft-QCD-vertex-2}) takes the following form:
\begin{align}
\begin{split}
\varGamma^\mu(p,k) &=\gamma^{\mu}   +\Cf \int\frac{d^{4}k'}{(2\pi)^{4}}
\tilde{\kernel}(k')  \frac{k^\nu\gamma^\mu+\gamma^\nu k'{}^\mu}{k\cdot k'}
\frac{  \Slash{k}'}{1+2\tilde{p}\cdot k'/\deltam^2}
 P^T_{\nu\rho}(k') \varGamma^\rho(p,k') ,
\end{split}
\end{align}
where we have used $t_bt_b=\Cf$. 

We note that this self-consistent equation is the same as that in QED except for $\Cf$ in the right-hand side:
There are no quantitatively novel effects which result from the fact that the gauge group is non-abelian.
We also note that the same equation was obtained in the temporal gauge~\cite{Lebedev:1989ev}, which is nontrivial if we consider the following facts:
In QED, the photon propagator in the resummed perturbation scheme is always on-shell, so only the transverse component is considered.
In that case, apparently there is no difference between the Coulomb gauge and the temporal gauge in this scheme.
However, in QCD, the off-shell gluon appears due to the self-interaction of the gluon.  
Then, the components other than the transverse one appear in the calculation, and the propagator of those components are different in both gauges.

We can solve the self-consistent equation in the same way as in QED.
The solution is as follows:
We expand the vertex function in terms of $\tilde{p}/\cp^2T$ as in Eq.~(\ref{eq:ultrasoft-QED-vertex-expand}).
The zeroth order solution is
\begin{align} 
A= \frac{1}{1+\Cf\lambda}\vect{1},~~~
B(k)= \frac{1}{k^0}\frac{2\Cf\lambda}{1+\Cf\lambda}\gamma^0,~~~
C(k)=0 .
\end{align}
Here $A$, $B(k)$, and $C(k)$ are defined by Eq.~(\ref{eq:varGamma}), and $\lambda$ is defined by Eq.~(\ref{eq:ultrasoft-lambda}).
The self-consistent equation which determines the vertex function at the first order $\delta\varGamma^\mu(p,k)$ is written as
\begin{align}
\label{eq:ultrasoft-QCD-vertex-first}
\begin{split}
\delta\varGamma^\mu(p,k) &=\Cf \int\frac{d^{4}k'}{(2\pi)^{4}}
\tilde{\kernel}(k')  \frac{k^\nu\gamma^\mu+\gamma^\nu k'{}^\mu}{k\cdot k'}
 \Slash{k}' P^T_{\nu\rho}(k') 
 \left(-2\frac{\tilde{p}\cdot k'}{\deltam^2} A\gamma^\rho+\delta\varGamma^\rho(p,k')\right).
\end{split}
\end{align}
Then $\delta\varPi(p)$ defined by Eq.~(\ref{eq:deltaPi}) becomes
\begin{align}
\begin{split}
\delta\varPi(p)&= \frac{16\pi^2A^2\lambda^3\Cf}{\cp^2} (\gamma^0(p^0+i\zeta)+v\vp\cdot \vgamma).
\end{split}
\end{align}

Owing to Eq.~(\ref{eq:selfEnergy-QCD}), the retarded quark self-energy is found to be
\begin{align}
\begin{split}
\varSigma^\R(p)&= \Cf\int\frac{d^{4}k}{(2\pi)^{4}}\tilde{\kernel}(k)\frac{\gamma^{\mu}\Slash{k}P^T_{\mu\nu}(k)
\varGamma^{\nu}(p,k)}{1+2\tilde{p}\cdot k/\deltam^2}\\
&=-\frac{1}{Z}(\gamma^0(p^0+i\zeta)+v\vp\cdot \vgamma),
\end{split}
\end{align}
where the expression of the residue $Z$ will be given shortly.
We note that this expression is the same as that in QED except for the numerical factor.
Thus, the expression for the pole position of the ultrasoft mode in QCD is the same as in QED, while the residue of that mode is not:
The residue is
\begin{align}
Z= \frac{\cp^2}{16\pi^2\lambda^2\Cf}(1+\Cf\lambda)^2
=\frac{\cp^2 N}{8\pi^2(N^2-1)}\left(\frac{5}{6}N+\frac{1}{2N}+\frac{2}{3}\Nf\right)^2. 
\end{align}

We can also show that the analytic solution of the self-consistent equation satisfies the WT identity in the leading order as in QED, by checking that the counterparts of Eqs.~(\ref{eq:ultrasoft-WT-zeroth}) and (\ref{eq:ultrasoft-WT-first}) are satisfied.

\section{Brief summary}
\label{sec:ultrasoft-summary}

In this chapter, we developed the resummed perturbation theory which was originally constructed in Ref.~\cite{Lebedev:1989ev} and enables us to successfully regularize the infrared singularity.
By using this method, we analyzed the quark spectrum whose energy is ultrasoft in QGP.
Since the Yukawa model and QED are simpler than QCD but have some similarity to QCD, we also worked in these models.
In QED/QCD, the summation of the ladder diagrams had to be done whereas that procedure was unnecessary in the Yukawa model.
That summation was necessary also from the point of view of the gauge symmetry.
As a result, we established the existence of a novel fermionic mode in the ultrasoft energy region, and obtained the expressions of the pole position and the strength of that mode.
The expressions for the dispersion relation, the damping rate, and the residue of the ultrasoft mode in the Yukawa model, QED, and QCD are summarized in Table~\ref{tab:ultrasoft-expression}.
We also showed that the Ward-Takahashi identity is satisfied in the resummed perturbation theory in QED/QCD.

\begin{table}[t] 
\caption{The expressions of the dispersion relation, the damping rate,  and the residue of the ultrasoft mode in the Yukawa model, QED, and QCD.
$\Nf$ and $N$ are set to three.
}
\begin{center}
\begin{tabular}{l c c c}
\hline
 & Yukawa model & QED & QCD \\ \hline \hline
dispersion relation & \multicolumn{3}{c}{$-|\vp|/3$}  \\ 
damping rate & $\zetaf+\zetab$ & $\zetae$ & $\zetaq+\zetag$\\
residue & $\cp^2/(72\pi^2) $ & $\cp^2/(144\pi^2)$ & 
$\cp^249/(48\pi^2)$\\ 
\hline
\end{tabular} 
\end{center}
\label{tab:ultrasoft-expression}
\end{table}

\chapter{Resummation as Generalized Boltzmann Equation}
\label{chap:kinetic}
\thispagestyle{headings}

As is described in Chapter~\ref{chap:intro}, there is a correspondence between some perturbation schemes and the kinetic equations:
the HTL approximation is equivalent to the Vlasov equation~\cite{\HTLVlasov}, and the resummed perturbation theory which enables the analysis of the ultrasoft gluon is equivalent to the Boltzmann equation~\cite{\UltrasoftGluon}.
Therefore it is natural to expect that the resummed perturbation theory which is used in the analysis of the ultrasoft fermion in Chapter~\ref{chap:ultrasoft}, is equivalent to some kinetic equation.
However, we note that the equation we will obtain is not the kinetic equation in the usual sense.
As will be shown later, due to the fact that the excitation we are considering is fermionic, not bosonic, that equation describes the time-evolution of the amplitude of the process in which the hard fermion becomes the hard boson and its inverse process, not that of the distribution function of any particles~\cite{\HTLVlasov}.
We call such equation ``off-diagonal'' kinetic equation.

In this chapter, we derive a off-diagonal and linearized kinetic equation for fermionic excitations with an ultrasoft momentum in the Yukawa model and QED, while the Boltzmann equation is derived in the case of bosonic excitations.
Our equation is systematically derived from the Kadanoff-Baym equation~\cite{kadanoff-baym}, and is equivalent to the self-consistent equation in the resummed perturbation theory~\cite{Hidaka:2011rz, Lebedev:1989ev} used in the analysis of the fermion propagator at the leading order in Chapter~\ref{chap:ultrasoft}.
The derivation helps us to establish the foundation of the resummed perturbation scheme.
The kinetic equation will also give us the kinetic interpretation of the resummation scheme.
Furthermore, we discuss the procedure of analyzing the $n$-point functions ($n\geq 3$) not only two-point functions of the fermion in QED.

This chapter is organized as follows:
Section~\ref{sec:kinetic-yukawa} is devoted to the derivation of the generalized and linearized kinetic equation and the discussion on the kinetic interpretation of the self-consistent equation in the resummed perturbation theory in the Yukawa model, which is the simplest fermion-boson system.
In Sec.~\ref{sec:kinetic-QED}, a similar analyses in QED is done in the Coulomb gauge.
We also show that the Ward-Takahashi identity is valid in this scheme, and that the $n$-point function whose external momenta are ultrasoft can be determined by using the gauge symmetry.
We briefly summarize this chapter in Sec.~\ref{sec:kinetic-summary}.

The analysis in this chapter is based on Ref.~\cite{Satow:2012ar}.
We note that a similar analysis to that in this chapter can be performed also in QCD~\cite{QCD-kinetic}.

\section{Yukawa model}
\label{sec:kinetic-yukawa}

In this section,  we derive a novel linearized kinetic equation from the Kadanoff-Baym equation in the Yukawa model.
We will find the vertex correction is negligible, which makes the analysis simpler than that in gauge theories.
Next, we show that the kinetic equation is equivalent to the resummation scheme in the resummed perturbation theory~\cite{Hidaka:2011rz, Lebedev:1989ev}, and discuss the interpretation of the resummation scheme using the correspondence between the field theoretical method and the kinetic theory.

\subsection{Derivation of the kinetic equation}
\label{ssc:kinetic-yukawa-derivation}
Throughout this chapter, we work in the closed-time-path formalism~\cite{lebellac, Blaizot:2001nr}.
We perform the derivation of the kinetic equation in a similar way used in~\cite{\HTLVlasov, Blaizot:1999fq, Blaizot:1999xk, Blaizot:2001nr} by applying the gradient expansion to the Kadanoff-Baym equation~\cite{kadanoff-baym} and taking into account the interaction effect among the hard particles in the leading order.

Let us consider the following situation to analyze the fermionic ultrasoft excitation:
Before the initial time $t_0$, the system is at equilibrium with a temperature $T$.
Then, a (anti-) fermionic external source $\eta(x)$ ($\overline{\eta}(x)$) and a scalar external source $j(x)$ are switched on. 
As a result, the system becomes nonequilibrium. 
We will consider the case that $j(x)$ and $\overline{\eta}(x)$ vanish and $\eta(x)$ is so weak that the system is very close to the equilibrium, i.e.,  the linear response regime. 
Concretely, we will retain only the terms in the linear order of the fermionic average field $\varPsi$ in the fermionic induced source, which will be introduced later.

\begin{figure}[t]
\begin{center}
\includegraphics[width=0.5\textwidth]{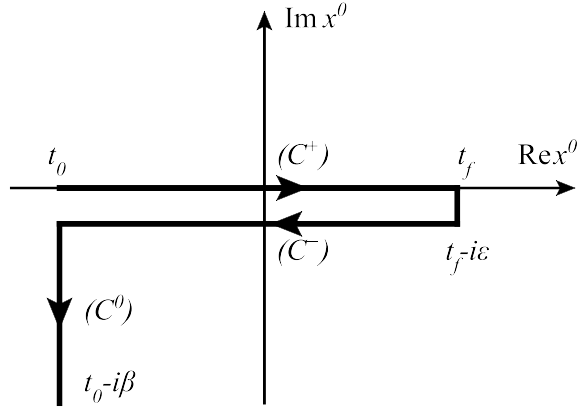}
\caption{The contour path $C$ in the complex $x^0$ plane.}
\label{fig:contour}
\end{center}
\end{figure}

Let us consider the generating functional in the closed time formalism~\cite{Blaizot:2001nr},
\begin{equation} 
\label{eq:yukawa-Z} 
Z[j,\eta,\overline{\eta}]=\int {\cal D}{\phi}\,{\cal D}\overline{\psi}\,{\cal D}\psi \,e^{iS},
\end{equation}
with
\begin{equation}
S=\int_C d^4x
\bigl[  {\cal L}[\phi, \psi, \overline{\psi}]-(j\phi+\overline{\psi}\eta+\overline{\eta}\psi)\bigr],
\end{equation}
where $\phi$ and $\psi$ are the scalar and the fermion fields.
The space-time integral is defined as 
\begin{align}
\int_C d^4x\equiv\int_C dx^0 \int d^3 \vx,
\end{align}
 where $\int_C dx^0$ is the complex-time integral along the contour $C=C^+\cup C^-\cup C^0$ in Fig.~\ref{fig:contour}.
 We will take $t_0\to -\infty$ and $t_f\to \infty$ to factorize out the contribution from the path $C^0$.
 The Lagrangian is given in Eq.~(\ref{eq:Yukawa-Lagrangian}).
By performing an infinitesimal variation with respect to $\phi$ or $\psi$ in Eq.~(\ref{eq:yukawa-Z}), we obtain the following equations of motion:
\begin{align}
\label{eq:yukawa-eom-meanfield-fermion}
i\Slash{D}_x[\varPhi]\varPsi(x)&= \eta(x)+\etaind(x), \\ 
\label{eq:yukawa-eom-meanfield-boson}
-\partial^2\varPhi(x)-\cp{\overline{\varPsi}} \varPsi (x)&=j(x)+\jind(x), 
\end{align}
where $\varPhi\equiv \langle\phi\rangle$ ($\varPsi\equiv\langle \psi\rangle$) is the expectation value of the scalar (fermion) field, and
$\Slash{D}_x[\varPhi]\equiv\Slash{\partial}_x+i\cp\varPhi(x)$.
Here the expectation value for an operator $\mathcal{O}$ is defined as
\begin{equation}
\begin{split}
\langle\mathcal{O}\rangle \equiv \frac{1}{Z}\int {\cal D}{\phi}\,{\cal D}\overline{\psi}\,{\cal D}\psi\, e^{iS}  \mathcal{O}.
\end{split}
\end{equation}
$\etaind(x)\equiv  \cp\langle\phi(x) \psi(x)\rangle_c$ ($\jind(x)\equiv \cp\langle\overline{\psi}(x)\psi(x)\rangle_c$) is the fermionic (scalar) induced source, 
and the subscript $c$ denotes  ``connected,'' i.e.,
\begin{align}
\langle\phi(x) \psi(x)\rangle_c\equiv \langle\phi(x) \psi(x)\rangle-\varPhi(x)\varPsi(x).
\end{align}

By differentiating Eq.~(\ref{eq:yukawa-eom-meanfield-fermion}) with respect to $j(y)$ and Eq.~(\ref{eq:yukawa-eom-meanfield-boson}) with respect to $\overline{\eta}(y)$, we obtain
\begin{align}
\label{eq:yukawa-fluctuation-fermion}
 &i\Slash{D}_x[\varPhi] K(x,y)-gD(x,y)\varPsi(x)=i\frac{\delta \etaind(x)}{\delta j(y)},\\
\label{eq:yukawa-fluctuation-boson}
&\partial^2_xK(y,x)-\cp(\overline{\varPsi}(x)\langle\psi(y)\psi(x)\rangle_c+S(y,x)\varPsi(x))
=i\frac{\delta \jind(x)}{\delta \overline{\eta}(y)}.
\end{align}
Here we have introduced the following propagators:
\begin{align}
D(x,y)&\equiv  \langle\Tc\phi(x)\phi(y)\rangle_c = i\frac{\delta\varPhi(x)}{\delta j(y)},\\
S(x,y)&\equiv \langle\Tc\psi(x)\overline{\psi}(y)\rangle_c=i\frac{\delta \overline{\varPsi}(y)}{\delta\overline{\eta}(x)},\\
K(x,y) &\equiv \langle\Tc\psi(x)\phi(y)\rangle_c = i\frac{\delta \varPsi(x)}{\delta j(y)}=i\frac{\delta \varPhi(y)}{\delta\overline{\eta}(x)},
\end{align}
where $\Tc$ means the path ordering  on the complex-time path $C$:
\begin{align}
\begin{split}
D(x,y)&= \thetac(x^0, y^0)D^>(x,y)+\thetac(y^0, x^0)D^<(x,y),\\
S(x,y)&=  \thetac(x^0, y^0)S^>(x,y)-\thetac(y^0, x^0)S^<(x,y),\\
K(x,y)&= \thetac(x^0, y^0)K^>(x,y)+\thetac(y^0, x^0)K^<(x,y),
\end{split}
\end{align}
with 
\begin{align}
D^>(x,y)&\equiv \langle\phi(x) \phi(y)\rangle_c,\\
D^<(x,y)&\equiv \langle\phi(y)\phi(x)\rangle_c,\\
S^>(x,y)&\equiv \langle\psi(x) \overline{\psi}(y)\rangle_c,\\
S^<(x,y)&\equiv \langle\overline{\psi}(y)\psi(x)\rangle_c,\\
K^>(x,y)&\equiv  \langle\psi(x)\phi(y)\rangle_c,\\
K^<(x,y)&\equiv \langle\phi(y)\psi(x)\rangle_c,
\end{align}
and $\thetac(x,y)$ being the step-function along the path $C$.
In the approximations introduced later, we can see that $K^>(x,y)$ and $K^<(x,y)$ coincide,  which can be checked by $K^{\R}(x,y)\equiv i\theta(x^0, y^0)[K^>(x,y)-K^< (x,y)] \simeq0$. 
For this reason, we simply write these two functions as $K(x,y)$ from now on.
We call $K(x,y)$ ``{\it{off-diagonal propagator}},'' which mixes the fermion and boson, 
while we call $D(x,y)$ and $S(x,y)$ ``{\it{diagonal propagators}}.''
As will be seen in Sec.~\ref{ssc:kinetic-yukawa-correspond},
in the calculation of the ultrasoft fermion self-energy,
the off-diagonal propagator is more relevant than the diagonal ones.

By setting $x^0\in C^+$ and $y^0\in C^-$ in  Eqs.~(\ref{eq:yukawa-fluctuation-fermion}) and (\ref{eq:yukawa-fluctuation-boson}), we obtain 
\begin{align}
\label{eq:yukawa-fermion1}
 &i\Slash{D}_x[\varPhi] K(x,y)-\cp D^<(x,y)\varPsi(x)
 =i\frac{\delta \etaind(x)}{\delta j(y)},\\
 \label{eq:yukawa-boson1}
&-\partial^2_yK(x,y)+\cp(\overline{\varPsi}(y)\langle \psi(y)\psi(x)\rangle_c+S^<(x,y)\varPsi(y))
=i\frac{\delta \jind(y)}{\delta \overline{\eta}(x)}.
\end{align}
Here we have interchanged $x$ and $y$ in the second equation.
Let us evaluate the right-hand side of Eqs.~(\ref{eq:yukawa-fermion1}) and (\ref{eq:yukawa-boson1}) using the chain rule:
\begin{align}
\frac{\delta \etaind(x)}{\delta j(y)} 
&=\int_C d^4 z \left(\frac{\delta \etaind(x)}{\delta \varPsi(z)}\frac{\delta \varPsi(z)}{\delta j(y)}
+\frac{\delta \etaind(x)}{\delta \varPhi(z)}\frac{\delta \varPhi(z)}{\delta j(y)}\right) \notag\\
\label{eq:yukawa-fermion2}
&=\int_C d^4z (\varSigma(x,z) K(z,y)+ \VC(x,z)D(z,y)),\\
\notag
\frac{\partial \jind(y)}{\partial \overline{\eta}(x)} 
& =\int_C d^4 z \left(\frac{\delta \varPhi(z)}{\delta \overline{\eta}(x)} \frac{\delta \jind(y)}{\delta \varPhi(z)}
+\frac{\delta \overline{\varPsi} (z)}{\delta \overline{\eta}(x)}\frac{\delta \jind(y)}{\delta \overline{\varPsi} (z)}\right)\\
\label{eq:yukawa-boson2}
& =\int_C d^4z (\varPi(y,z) K(x,z)+ S(x,z)\VC(z,y)).
\end{align} 
Here we have dropped $(\delta\etaind /\delta \overline{\varPsi} )(\delta \overline{\varPsi} /\delta j)$  and $(\delta \jind/\delta \varPsi) (\delta \varPsi /\delta \overline{\eta})$ since they contain more than one fermionic average field.
We have also used 
\begin{align}
\varSigma(x,y)&\equiv-i\frac{\delta \etaind(x)}{\delta\varPsi(y)},\\
\varPi(x,y)&\equiv -i \frac{\delta \jind(x)}{\delta \varPhi(y)}=\varPi(y,x),
\end{align}
where $\varSigma$ ($\varPi$) is the fermion (scalar) self-energy~\cite{Blaizot:1999fq, Blaizot:1999xk, Blaizot:2001nr}.
We also introduced the {\it off-diagonal} self-energy, 
\begin{align}
\VC(x,y)\equiv-i\frac{\delta \etaind(x)}{\delta \varPhi(y)}=-i\frac{\delta \jind(y)}{\delta \overline{\varPsi} (x)}.
\end{align}

The self-energies are decomposed for arbitrary $x^0$ and $y^0$ on the time path $C$: 
\begin{align}
\varPi(x,y)&=\thetac(x^0, y^0)\varPi^>(x,y)+\thetac(y^0, x^0)\varPi^<(x,y),\\
\varSigma(x,y)&=\thetac(x^0, y^0)\varSigma^>(x,y)-\thetac(y^0, x^0)\varSigma^<(x,y),\\
\VC(x,y)&=\thetac(x^0, y^0)\VC^{>}(x,y) +\thetac(y^0, x^0)\VC^{<}(x,y).
\end{align}
We have not taken into account contact terms, which is negligible in the leading order.  

Here let us rewrite Eqs.~(\ref{eq:yukawa-fermion2}) and (\ref{eq:yukawa-boson2}) in terms of real time integral instead of that on the complex-time-path.
First we evaluate the diagonal self-energy term:
\begin{align}
\label{eq:app-ana-1}
\begin{split}
&\int_C d^4z \varSigma(x,z) K(z,y)\\
&\quad=\int^{x^0}_{t^0} d^4z \varSigma^>(x,z) K(z,y) 
- \int^{y^0}_{x^0} d^4z \varSigma^<(x,z) K(z,y)\\
&\qquad-\int^{t^0-i\beta}_{y^0} d^4z \varSigma^<(x,z) K(z,y)\\
&\quad=\int^{x^0}_{t^0} d^4z (\varSigma^>(x,z)+\varSigma^<(x,z)) K(z,y) 
-\int^{t^0-i\beta}_{t^0} d^4z \varSigma^<(x,z) K(z,y)\\
&\quad\simeq-i\int^{\infty}_{-\infty} d^4z \varSigma^{R}(x,z) K(z,y).
\end{split}
\end{align}
In the last line we have taken $t^0\rightarrow-\infty$ and introduced the retarded fermion self-energy
\begin{align}
\varSigma^\R(x,y)\equiv i \theta(x^0, y^0)[\varSigma^>(x,y)+\varSigma^<(x,y)].
\end{align}
We used the fact that the term integrated on $C^0$ becomes negligible in this limit~\cite{Blaizot:2001nr}.
In the same way, we get
\begin{align}
\label{eq:app-ana-2}
\int_C d^4z \varPi(y,z) K(x,z)\simeq&-i\int^{\infty}_{-\infty} d^4z  \varPi^{A}(z,y)K(x,z),
\end{align}
where we have introduced the advanced scalar self-energy 
\begin{align}
\varPi^A(x,y)\equiv -i\theta(y^0, x^0)[\varPi^>(x,y)-\varPi^<(x,y)] .
\end{align}

Next, we evaluate the off-diagonal self-energy term.
The off-diagonal self-energy term of Eq.~(\ref{eq:yukawa-fermion2}) becomes
\begin{align}
\label{eq:app-ana-3}
\begin{split}
&\int_C d^4z  \VC(x,z)D(z,y)\\
&\quad=\int^{x^0}_{t^0} d^4z \VC^>(x,z) D^<(z,y) 
+ \int^{y^0}_{x^0} d^4z \VC^<(x,z) D^<(z,y)
+\int^{t^0-i\beta}_{y^0} d^4z\VC^<(x,z) D^>(z,y)\\
&\quad= \int^{x^0}_{t^0} d^4z (\VC^>(x,z)-\VC^<(x,z)) D^<(z,y) \\
&\qquad+ \int^{y^0}_{t^0} d^4z \VC^<(x,z)  (D^<(z,y)-D^>(z,y))
-\int^{t^0-i\beta}_{t^0} d^4z \VC^<(x,z) D^>(z,y)\\
&\quad\simeq-i\int^{\infty}_{-\infty} d^4z (\VC^{R}(x,z) D^<(z,y)
+\VC^<(x,z)D^\A(z,y)),
\end{split}
\end{align}
where the advanced boson propagator $D^\A(z,y)\equiv -i\theta(y^0-z^0)[D^>(z,y)-D^<(z,y)]$ and the retarded off-diagonal self-energy $\VC^{R}(x,z)\equiv i\theta(x^0, z^0)[\VC^>(x,z)-\VC^< (x,z)] $ have been introduced.
We stop here and discuss the structure of the off-diagonal self-energy in the leading order.

\begin{figure}[t]
\begin{center}
\includegraphics[width=0.4\textwidth]{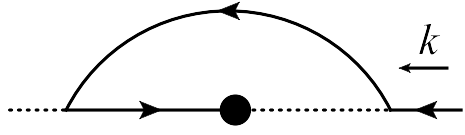} 
\caption{The off-diagonal self-energy $\VC(k, X)$ in the leading order.
The propagator that is composed of the solid line and the dashed line with the black blob is the off-diagonal propagator. 
The other notations are the same as Fig.~\ref{fig:fermion-selfenergy}.}
\label{fig:vertex-correction}
\end{center}
\end{figure}

As we sill see later, we utilize the off-diagonal self-energy in the leading order in the linear response regime, which is given by 
\begin{align}
\VC(x,y)=\cp^2 S^{0}(x,y)K(y,x),
\end{align}
where $S^{0}(x,y)$ is the free fermion propagator at equilibrium.
The diagrammatic expression of this equation is shown in Fig.~\ref{fig:vertex-correction}.
Thus, the components of $\VC$ are given by
\begin{align}
\VC^{\gtrless}(x,y)=
\pm \cp^2 S^{0 \gtrless}(x,y)K(y,x).
\end{align}
Here we perform the Wigner transformation, 
\begin{align}
f(k,X)\equiv\int d^4s e^{ik\cdot s}f\left(X+\frac{s}{2}, X-\frac{s}{2}\right),
\end{align}
where $s\equiv x-y$, $X\equiv(x+y)/2$, and $f(x,y)$ is an arbitrary function.
Then, we get
\begin{align}
\label{eq:app-analyticity-vertex-correction-k}
\VC^\gtrless(k,X)=\pm \cp^2\int\frac{d^4k'}{(2\pi)^4}S^{0 \gtrless}(k+k')K(k', X),
\end{align} 
with 
\begin{align}
S^{0 >}(k)&\equiv \Slash{k}\rho^0(k)(1-\nf(k^0)), \\
\label{eq:kinetic-yukawa-thermalspectrum-fermion}
S^{0<}(k)&= \Slash{k}\rho^0(k)\nf(k^0).
\end{align}
As will be seen later, $K(k', X)$ contains $\delta(k'^2)$.
Thus, since we focus on the on-shell case $k^2\simeq 0$, which will be confirmed later, $(k+k')^2\simeq 2k\cdot k'\neq 0$. 
For this reason, $S^{0 \gtrless}(k+k')\simeq 0$, which implies $\VC^{\gtrless}(k,X)\simeq 0$, so  the only nonzero function of the off-diagonal self-energy appearing at Wigner-transformed Eq.~(\ref{eq:app-ana-3}) is $\VC^\R(k,X)\simeq \VC^\A(k, X)$.
Therefore, we drop the second term in Eq.~(\ref{eq:app-ana-3}) because that term becomes negligible after the Wigner transformation, and hence the equation becomes
\begin{align}
\label{eq:app-ana-4}
\begin{split}
&\int_C d^4z  \VC(x,z)D(z,y)
\simeq-i\int^{\infty}_{-\infty} d^4z \VC^\R(x,z) D^<(z,y).
\end{split}
\end{align}
In the same way, we get
\begin{align}
\label{eq:app-ana-5}
\int_C d^4z  S(x,z)\VC(z,y)\simeq i\int d^4z  S^<(x,z)\VC^\R(z,y).
\end{align}

Combining Eqs.~(\ref{eq:app-ana-1}), (\ref{eq:app-ana-2}), (\ref{eq:app-ana-4}), and (\ref{eq:app-ana-5}), we get
\begin{align}
\label{eq:yukawa-fermion3}
i\Slash{D}_x[\varPhi] K(x,y)-\cp D^<(x,y)\varPsi(x)
&=\int ^\infty_{-\infty}d^4z (\varSigma^\R(x,z) K(z,y)+ \VC^\R(x,z)D^<(z,y)),\\
\label{eq:yukawa-boson3}
-\partial^2_yK(x,y)+\cp S^<(x,y)\varPsi(y)
&=\int ^\infty_{-\infty} d^4z (\varPi^\A(z,y) K(x,z)- S^<(x,z)\VC^\R(z,y)).
\end{align}
Here we have dropped $\overline{\varPsi}(y)\langle\psi(y)\psi(x) \rangle_c$ because $\langle\psi(y)\psi(x) \rangle_c$ contains more than one $\varPsi$.
Equations~(\ref{eq:yukawa-fermion3}) and (\ref{eq:yukawa-boson3}) are the Kadanoff-Baym equations from which the kinetic equation is derived.

After performing the Wigner transformation, Eqs.~(\ref{eq:yukawa-fermion3}) and (\ref{eq:yukawa-boson3}) become
\begin{align}
\notag
&\left(-i\Slash{k}+\frac{\Slash{\partial}_X}{2}+i\cp\varPhi(X)\right) K(k,X)+i\cp D^<(k,X)\varPsi(X)\\
\label{eq:yukawa-fermion4}
&\quad=i (-\varSigma^\R(k,X) K(k,X)-\VC^\R(k,X)D^<(k,X)),\\
\notag
&(k^2-ik\cdot\partial_X)K(k,X)+\cp S^<(k,X)\varPsi(X)\\
\label{eq:yukawa-boson4}
&\quad=\varPi^\A(k,X) K(k,X)- S^<(k,X)\VC^\R(k,X).
\end{align}
Here we have used the following transformation law under the Wigner transformation,
\begin{align}
f(x)g(x,y)&\rightarrow f(X)g(k,X)
-\frac{i}{2}(\partial_X f)\cdot (\partial_k g)+..., \\
f(y)g(x,y)&\rightarrow f(X)g(k,X)
+\frac{i}{2}(\partial_X f)\cdot (\partial_k g)+..., \\
\int^\infty_{-\infty} d^4z g(x,z)h(z,y)&\rightarrow g(k,X)h(k,X)
+\frac{i}{2}\{g ,h \}_{\text {P. B.}}+... ,
\end{align}
where 
\begin{align}
\{g, h\}_{\text {P. B.}}\equiv\partial_k g\cdot \partial_X h-\partial_X g\cdot \partial_k h
\end{align}
 is the Poisson bracket, and neglected higher-order terms that contain $\partial_X$ since we focus on the case that the inhomogeneity of the average field is $\partial_X\sim \cp^2 T$, while a typical magnitude of $k$ is of order $T$.
This expansion is called gradient expansion~\cite{\HTLVlasov, Blaizot:1999fq, Blaizot:1999xk, Blaizot:2001nr}.
We retained the second terms in the left-hand sides of Eqs.~(\ref{eq:yukawa-fermion4}) and (\ref{eq:yukawa-boson4}) because the first terms, which seem to be the leading terms in the gradient expansion, will cancel out in the next manipulation.

By multiplying Eq.~(\ref{eq:yukawa-fermion4}) by $(-i\Slash{k}+\Slash{\partial}_X/2+i\cp\varPhi(X)+i\varSigma^\R(k,X) )$ and adding Eq.~(\ref{eq:yukawa-boson4}), we get
\begin{align}
\label{eq:yukawa-A}
\begin{split}
&(2ik\cdot\partial_X -2\cp\Slash{k}\varPhi(X) -\{\Slash{k}, \varSigma^\R(k,X)\}+\varPi^\A(k,X)) K(k,X)\\
&= \cp(\Slash{k}D^<(k,X)+S^<(k,X))\V(k,X).
\end{split}
\end{align}
Here we have introduced $\cp\V(k,X)\equiv \cp\varPsi(X)+\VC^\R(k,X)$ and neglected higher order terms which are of order $\cp^4T^2K$ and $\cp^3 T^{-1} \V$.
In the leading order, the coupling dependence in $D^{<}(k)$ and $S^{<}(k)$ is negligible, so that $D^{<}(k)$ and $S^{<}(k)$ are replaced by the propagators at equilibrium and free limit ($\cp=0$):
\begin{align}
\label{eq:kinetic-yukawa-thermalspectrum-boson}
D^{0<}(k)&=\rho^0(k)\nb(k^0).
\end{align}
$S^{0<}(k)$ is given by Eq.~(\ref{eq:kinetic-yukawa-thermalspectrum-fermion}).
We note that though the massless condition $k^2=0$ appears in Eqs.~(\ref{eq:kinetic-yukawa-thermalspectrum-boson}) and (\ref{eq:kinetic-yukawa-thermalspectrum-fermion}) in the present approximation, $k^2$ is expected to be of order $\cp^2T^2$ if one takes into account the interaction at equilibrium.
For this reason, we will use the order estimate $k^2\sim \cp^2T^2$.
We also note that $K$ can not be replaced by that at equilibrium since $K$ vanishes at equilibrium.

We see that $k^2$ terms in the left-hand side of Eq.~(\ref{eq:yukawa-A}) were canceled out and $k\cdot\partial_X\sim \cp^2T^2$ term remains.
Thus, we can neglect the terms which are much smaller than $\cp^2T^2 K$ in the calculation of the leading order.
Following this line, the diagonal self-energies are replaced by those at equilibrium in the leading order, whose diagrams are shown in Figs.~\ref{fig:boson-selfenergy} and \ref{fig:fermion-selfenergy}:
\begin{align} 
\{\Slash{k}, \varSigma^{\R {\mathrm {(eq)}}}(k)\}&= \mf^2,\\ 
\varPi^{\A {\mathrm {(eq)}}}(k)&= \mb^2,
\end{align}
as were calculated in Sec.~\ref{ssc:intro-hard}.
Note that the imaginary parts of the self-energies $\sim g^4T^2 \ln( 1/g)$ and momentum dependence are negligible since they are higher order in the coupling constant.
We have used the on-shell condition, $k^2\simeq 0$, which will be verified later.

The same logic as in the diagonal self-energies case justifies substitution of the off-diagonal self-energy $\VC^\R$ in the leading order:
\begin{align}
\label{eq:yukawa-vertex-correction}
\VC^\R(k,X)=& -\cp^2\int\frac{d^4k'}{(2\pi)^4}S^{0\R}(k+k')K(k',X).
\end{align}
Here $S^{0\R}(k)$ is the free fermion retarded propagator at equilibrium, whose expression is given in Eq.~(\ref{eq:bare-propagatorS}).
We note that the self-energies can not be neglected in contrast to $\partial_X\sim\cp T$ case\footnote{This is because $k\cdot\partial_X K\sim \cp T^2K\gg \{\Slash{k}, \varSigma\}K$, $\varPi K$, $ T^{-1}\VC\sim \cp^2T^2K$.}~\cite{\HTLVlasov}, because $\{\Slash{k}, \varSigma\} K$, $\varPi K$, $ T^{-1}\VC$ have the same order of magnitude as $ \cp^2 T^2K$.

\begin{figure}[t]
\begin{center}
\includegraphics[width=0.5\textwidth]{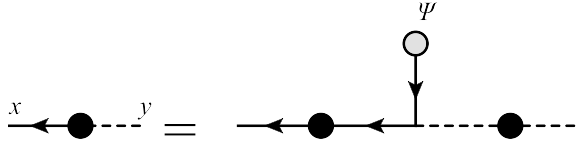}
\caption{The schematic figure of the off-diagonal propagator $K(x,y)$ in the Yukawa model in the leading order in the linear response regime.
The solid (dashed) line with black blob is the resummed fermion (boson) propagator that contains the information on the fermion (boson) self-energy $\varSigma$ ($\varPi$).
The gray blob represents the fermionic average field $\varPsi$.
}
\label{fig:yukawa-K}
\end{center}
\end{figure}

Using these expressions, Eq.~(\ref{eq:yukawa-A}) becomes
\begin{align} 
\label{eq:yukawa-result-1}
\begin{split}
&(2ik\cdot\partial_X-2\cp\Slash{k}\varPhi(X)+\delta m^2)K(k,X)
=\cp\Slash{k}\rho^0(k)(\nb(k^0)+\nf(k^0))\V(k,X),
\end{split}
\end{align}
where $\delta m^2\equiv \mb^2-\mf^2$.
We note that $K(k,X)$ becomes finite only when $k^2=0$ because of $\delta(k^2)$ in the right-hand side.
 We also note that $\Slash{k}K(k,X)\sim \cp^2T K(k,X)$, which is confirmed by multiplying Eq.~(\ref{eq:yukawa-result-1}) by $\Slash{k}$ from the left. 
This property makes the vertex correction term, $\Slash{k}\VC^\R(k, X)$, negligible,
which corresponds to the fact that there is no vertex correction in the analysis using the resummed perturbation theory~\cite{Hidaka:2011rz};
 see Sec.~\ref{ssc:ultrasoft-yukawa-ladder}.
We also find that the $\Slash{k}\varPhi(X)K(k,X)$ term is negligible and thus the effect of the average field of the scalar vanishes in the present approximation.
Thus we get
\begin{align}
\begin{split}
\label{eq:yukawa-result} 
&(2ik\cdot\partial_X+\delta m^2)K(k,X)
=\cp\Slash{k}\rho^0(k)(\nb(k^0)+\nf(k^0))\varPsi(X). 
\end{split}
\end{align}
The schematic figure of $K(k,X)$ is depicted in Fig.~\ref{fig:yukawa-K}.
The solid (dashed) line with the blob stands for the resummed fermion (boson) propagator.

\subsection{Kinetic interpretation}
\label{ssc:kinetic-yukawa-kinetic}
By introducing the ``{\it{off-diagonal density matrix}}'' $\varLambda_\pm(k,X)$ defined as 
\begin{align}
K(k,X)\equiv2\pi\delta(k^2)(\theta(k^0)\varLambda_+(\vk,X)+\theta(-k^0)\varLambda_-(-\vk,X)),
\end{align}
 we arrive at the following generalized and linearized kinetic equation from Eq.~(\ref{eq:yukawa-result}):
\begin{align}
\label{eq:yukawa-kineticeq}
\begin{split}
&\left(2iv\cdot\partial_X\pm\frac{\delta m^2}{|\vk|}\right)\varLambda_\pm(\vk,X)
=\cp\Slash{v}(\nb(|\vk|)+\nf(|\vk|))\varPsi(X).
\end{split}
\end{align}
It should be noted that this equation is not a usual kinetic equation  because $\varLambda(\vk,X)$ can not be interpreted as a distribution function since it is the propagator between different particles in the fermionic background.
Nevertheless, we call this equation ``{\it{generalized kinetic equation}}''  because we can obtain the Boltzmann equation if we analyze the time-evolution of the diagonal propagator instead of the off-diagonal one~\cite{\UltrasoftGluon, Blaizot:2001nr}.
In fact, Eq.~(\ref{eq:yukawa-kineticeq}) has the following points which are similar to the Boltzmann equation: 
\begin{itemize} 
\item The particle is on-shell ($k^2\simeq 0$).
\item The equation has the common structure as the Boltzmann equation: both have the non-interacting part, the interaction part between the hard particle and the average ultrasoft field, and the interaction part among the hard particles,
which correspond to the drift term, the force term, and the collision term in the Boltzmann equation, respectively.
\end{itemize}
When $\partial_X\sim \cp T$, $\delta m^2$ is negligible,  and Eq. (\ref{eq:yukawa-kineticeq}) becomes the counterpart of the Vlasov equation~\cite{\HTLVlasov, Blaizot:2001nr}.
Let us recapitulate the interpretations of each term in Eq. (\ref{eq:yukawa-kineticeq}) except for the $\delta m^2$ term.
The first term in the left-hand side describes the time-evolution of $\varLambda_\pm(\vk,X)$ in the free limit ($\cp=0$), so this term corresponds to the drift term in the Boltzmann equation.
On the other hand, the term in the right-hand side expresses the effect from the average fermionic field.
Hence this term corresponds to the force term in the Boltzmann equation.

Now let us discuss the origin of the $\delta m^2$ term.
The origin of the term is $-\{\Slash{k}, \varSigma^\R(k,X)\}+\varPi^\A(k,X)$ in Eq.~(\ref{eq:yukawa-A}).
The real parts of those terms are of order $\cp^2T^2$ while the imaginary parts are $\cp^4T\ln\cp^{-1}$, so the contribution in the leading order comes from the real parts.
The difference of the real parts of the diagonal self-energies expresses the difference of the dispersion relations of the scalar and the fermion, so we call the $\delta m^2$ term ``{\it mass difference term}.'' 
We note that the ``mass'' here is not the bare one but dynamically generated one thorough the interaction among the hard particles.
$\delta m^2$ term has no counterpart in the usual Boltzmann equation, which describes the time-evolution of the diagonal propagators, $S(k,X)$ and $D(k,X)$.
Therefore we cannot obtain kinetic interpretation of that term in the usual sense.

Here let us see the reason why the mass difference term does not have its counterpart in the diagonal case.
To this end, we derive the equation that corresponds to Eq.~(\ref{eq:yukawa-result}) in the diagonal case.
The equation governing the propagator of the fermion is as follows:
\begin{align}
\label{eq:yukawa-diagonal-0}
\begin{split}
& \Slash{D}_x[\varPhi] S(x,y)+i\cp(K(y,x))^\dagger\gamma^0\varPsi(x)
=\delta^{C(4)}(x-y)+i\frac{\delta \etaind(x)}{\delta \eta(y)}.
\end{split}
\end{align}
Since the second term in the left-hand side contains two $\varPsi$, we neglect that term. 
Now let us calculate the right-hand side.
We set $x^0\in C^+$ and $y^0\in C^-$.
Since the vertex correction term, which contains more than one $\varPsi$,  is negligible, we obtain
\begin{align}
\begin{split}
\frac{\delta \etaind(x)}{\delta \eta(y)}&= \int _C d^4z \varSigma(x,z) S(z,y)\\
&=-\int^{x^0}_{t^0} d^4z \varSigma^>(x,z) S^<(z,y) 
+ \int^{y^0}_{x^0} d^4z \varSigma^<(x,z) S^<(z,y)\\
&\quad-\int^{t^0-i\beta}_{y^0} d^4z \varSigma^<(x,z) S^>(z,y)\\
&=-\int^{x^0}_{t^0} d^4z (\varSigma^>(x,z)+\varSigma^<(x,z)) S^<(z,y)\\
&\quad+ \int^{y^0}_{t^0} d^4z \varSigma^<(x,z) (S^<(z,y)+S^>(z,y)).
\end{split}
\end{align}
By taking the limit $t^0\rightarrow-\infty$, we get
\begin{align}
\begin{split}
&\frac{\delta \etaind(x)}{\delta \eta(y)}
\simeq i\int^{\infty}_{-\infty} d^4z (\varSigma^\R(x,z) S^<(z,y)+ \varSigma^<(x,z) S^\A(z,y)).
\end{split}
\end{align}
Here we have introduced the advanced fermion propagator, 
\begin{align}
S^\A(x,y)\equiv -i\theta(y^0,x^0)(S^>(x,y)+S^<(x,y)).
\end{align}
By performing the Wigner transformation, we get 
\begin{align}
\label{eq:yukawa-diagonal-1}
\begin{split}
&\left[-i\Slash{k}+\frac{\Slash{\partial}_X}{2}+i\cp\left(\varPhi(X)-i\frac{\partial_k}{2}\cdot(\partial_X \varPhi(X))\right)\right] S^<(k,X) \\
 &\quad=-i(\varSigma^\R(k,X)S^<(k,X)+\varSigma^<(k,X)S^\A(k,X)).
\end{split}
\end{align}
The following equation is derived from the conjugate of Eq.~(\ref{eq:yukawa-diagonal-1}) by using $\gamma^0S^>(x,y)\gamma^0=S^<(y,x)$:
\begin{align}
\label{eq:yukawa-diagonal-2}
\begin{split}
& S^<(k,X)\left[i\Slash{k}+\frac{\overleftarrow{\Slash{\partial}_X}}{2}-i\cp\left(\varPhi(X)+i\frac{\overleftarrow{\partial_k}}{2}\cdot(\partial_X \varPhi(X))\right)\right]\\
 &\quad=i(S^\R(k,X)\varSigma^<(k,X)+S^<(k,X)\varSigma^\A(k,X)).
\end{split}
\end{align}
Here we have introduced the retarded fermion propagator, $S^\R$, and the advanced fermion self-energy, $\varSigma^\A$, which are defined as follows:
\begin{align}
S^\R(x,y)&\equiv i \theta(x^0, y^0)(S^>(x,y)+S^<(x,y)), \\
\varSigma^\A(x,y)&\equiv -i\theta(y^0,x^0)(\varSigma^>(x,y)+\varSigma^<(x,y)).
\end{align}
By multiplying Eq.~(\ref{eq:yukawa-diagonal-1}) (Eq.~(\ref{eq:yukawa-diagonal-2})) by 
\begin{align}
&-i\Slash{k}+\frac{\Slash{\partial}_X}{2}+i\cp\left(\varPhi(X)-i\partial_k\cdot\partial_X \frac{\varPhi(X)}{2}\right)-i\varSigma^\R(k,X)\\
&\left(i\Slash{k}+\frac{\Slash{\partial}_X}{2}-i\cp\left(\varPhi(X)+i\partial_k\cdot\partial_X \frac{\varPhi(X)}{2}\right)-i\varSigma^\A(k,X)\right)
\end{align}
 from the left (right), we get
\begin{align}
\notag
&(-k^2-ik\cdot\partial_X+\cp(2\Slash{k}\varPhi(X)-i\Slash{k}\partial_k\cdot(\partial_X \varPhi(X))\\
\notag
&\qquad+\varPhi(X)\Slash{\partial}_X)+\{\Slash{k}, \varSigma^\R(k,X)\}) S^<(k,X)\label{eq:yukawa-diagonal-3}\\
 &\quad=-\Slash{k}\varSigma^<(k,X)S^\A(k,X), \\
 \notag
& S^<(k,X)(-k^2+ik\cdot\overleftarrow{\partial_X}+\cp(2\Slash{k}\varPhi(X)+i\overleftarrow{\partial_k}\cdot(\partial_X \varPhi(X))\Slash{k}\\
\notag
&\qquad-i\overleftarrow{\Slash{\partial}_X}\varPhi(X))+\{\Slash{k}, \varSigma^\A(k,X)\})  \label{eq:yukawa-diagonal-4}\\
 &\quad=-S^\R(k,X)\varSigma^<(k,X) \Slash{k} .
\end{align}
By subtracting Eq.~(\ref{eq:yukawa-diagonal-3}) from Eq.~(\ref{eq:yukawa-diagonal-4}), we get
\begin{align}
\label{eq:yukawa-diagonal-5}
\begin{split}
&\left(2ik\cdot\partial_X-\{\Slash{k},\varSigma^R(k,X)\}\right)S^<(k,X)
+S^<(k,X)\{\Slash{k},\varSigma^\A(k,X)\}\\
&\qquad+2\cp\varPhi(X)[S^<(k,X), \Slash{k}] 
 +i\cp(\partial^\nu_X\varPhi(X))\{\Slash{k},\partial_{k\nu}S^<(k,X)\}\\
&\qquad-i\cp \varPhi(X)\{\gamma_\mu, \partial^\mu_X S^<(k,X)\}\\
 &\quad=\Slash{k}\varSigma^<(k,X)S^\A(k,X)-S^\R(k,X)\varSigma^<(k,X)\Slash{k} . 
\end{split}
\end{align} 
Here we linearize this equation.
By introducing $\delta S(k,X)=S(k,X)-S^{0}(k)$ and $\delta\varSigma(k,X)\equiv\varSigma(k,X)-\varSigma^{\text {(eq)}}(k)$, we arrive at the following equation:
\begin{align}
\label{eq:app-diagonal-kinetic}
\begin{split}
&\left(2ik\cdot\partial_X-\{\Slash{k},\varSigma^{{\text {(eq)}}\R}(k)-\varSigma^{{\text {(eq)}}\A}(k)\}\right)\delta S^<(k,X)\\
&\qquad-\{\Slash{k},\delta\varSigma^R(k,X)\} S^{0<}(k)+S^{0<}(k)\{\Slash{k},\delta\varSigma^{\A}(k,X)\}
+i\cp(\partial^\nu_X\varPhi(X))\{\Slash{k},\partial_{k\nu}S^{0<}(k)\}\\
 &\quad=\Slash{k}\delta \varSigma^<(k,X)S^{0 \A}(k)-S^{0 \R}(k)\delta\varSigma^<(k,X)\Slash{k} \\
 &\qquad+\Slash{k}\varSigma^{{\text {(eq)}} <}(k)\delta S^\A(k,X)-\delta S^\R(k,X)\varSigma^{{\text {(eq)}} <}(k)\Slash{k} . 
\end{split}
\end{align} 
From this equation, we see that the corresponding term to the mass difference in the diagonal kinetic equation becomes 
\begin{align}
\label{eq:kinetic-diagonal-collision-simple}
-\{\Slash{k}, \varSigma^{{\text {(eq)}}\R}(k,X)\}+\{\Slash{k}, \varSigma^{{\text {(eq)}}\A}(k,X)\}=-2i\im\{\Slash{k}, \varSigma^{{\text {(eq)}}\A}(k,X)\}
\end{align}
 instead of $-\{\Slash{k}, \varSigma^{{\text {(eq)}}\R}(k,X)\}+\varPi^{{\text {(eq)}}\A}(k,X)$, so that the real part is canceled out. 
For this reason, the mass difference term is absent in the diagonal case.

We note that the term (Eq.~(\ref{eq:kinetic-diagonal-collision-simple})), which is pure imaginary as a result of the cancellation of the real part, is of order $\cp^4T\ln(1/\cp)$ and thus negligible since $\partial_X\sim\cp^2T$.
We also see that the terms coming from the self-energy have complicated form in the diagonal case, as shown in Eq.~(\ref{eq:app-diagonal-kinetic}), while they are reduced to the simple form, $-\{\Slash{k}, \varSigma^{{\text {(eq)}}\R}(k)\}+\varPi^{{\text {(eq)}}\A}(k)$, in the off-diagonal case in the linear response region.
This difference comes from the following two facts:
\begin{enumerate}
\item Since we linearize the equation in terms of the deviation from the equilibrium state, the terms containing $\delta\varSigma$ and $S^0$ in Eq.~(\ref{eq:app-diagonal-kinetic}) do not have its counter parts in the off-diagonal case [$K(k,X)$ vanishes at equilibrium.]

\item The structure of the right-hand sides of Eqs.~(\ref{eq:yukawa-fermion4}) and (\ref{eq:yukawa-boson4}) after neglecting the vertex correction terms are simpler than those of Eqs.~(\ref{eq:yukawa-diagonal-1}) and (\ref{eq:yukawa-diagonal-2}).
It is because $K^<(x,y)=K^>(x,y)$. 
\end{enumerate}

\subsection{Correspondence between kinetic theory and resummed perturbation theory}
\label{ssc:kinetic-yukawa-correspond}
Here, let us show the relation between our kinetic equation, Eq.~(\ref{eq:yukawa-result}), and the resummation scheme in the resummed perturbation theory~\cite{Hidaka:2011rz} in Chapter.~\ref{chap:ultrasoft}.
For this purpose, we write the fermionic induced source $\etaind$ in terms of the fermion retarded self-energy $\varSigma^R(p)$, by using the relation in the linear response theory~\cite{\HTLVlasov, Blaizot:2001nr} in momentum space as
\begin{align}
\label{eq:yukawa-fermion-induced}
\etaind(p)=\varSigma^\R(p)\varPsi(p).
\end{align}
The expression of $\etaind$ is obtained by using the relation $\etaind(X)=\cp K(X,X)=\cp\int d^4k/(2\pi)^4 K(k,X)$ in the present formalism.
Thus, from Eq.~(\ref{eq:yukawa-result}), $\etaind$ is
\begin{align}
\etaind(X)=\cp^2\int \frac{d^4k}{(2\pi)^4} \frac{\Slash{k}\rho^0(k)(\nb(k^0)+\nf(k^0))}{(2ik\cdot\partial_X+\delta m^2)}\varPsi(X).
\end{align}
By performing the Fourier transformation defined by
\begin{align}
\label{eq:fourier-p}
f(k,p)\equiv\int d^4X e^{ip\cdot X}f(k, X),
\end{align}
 we obtain
\begin{align}
\etaind(p)=\cp^2\int \frac{d^4k}{(2\pi)^4} \frac{\Slash{k}\rho^0(k)(\nb(k^0)+\nf(k^0))}{(2k\cdot p+\delta m^2)}\varPsi(p).
\label{eq:yukawa-fermion-induced2}
\end{align}
Comparing Eq.~(\ref{eq:yukawa-fermion-induced}) with Eq.~(\ref{eq:yukawa-fermion-induced2}), we obtain the self-energy,
\begin{align}
\label{eq:yukawa-selfenergy-p}
\varSigma^\R(p)=\cp^2\int \frac{d^4k}{(2\pi)^4} \frac{\Slash{k}\rho^0(k)(\nb(k^0)+\nf(k^0))}{(2k\cdot p+\delta m^2)}.
\end{align}
This equation coincides with Eq.~(\ref{eq:one-loop-pc}), which is the expression of the retarded fermion self-energy~\cite{Hidaka:2011rz} except for the absence of the damping rates of the hard particles in the denominator.
As mentioned in the previous subsection, the damping rates of order $\cp^4T\ln(1/\cp)$
is neglected when the external momentum is of order  $g^2T$;
one can include them by taken into account  the imaginary part of Eq.~(\ref{eq:yukawa-result}) if one is interested in the damping rate.

\begin{figure}[t]
\begin{center} 
\includegraphics[width=0.4\textwidth]{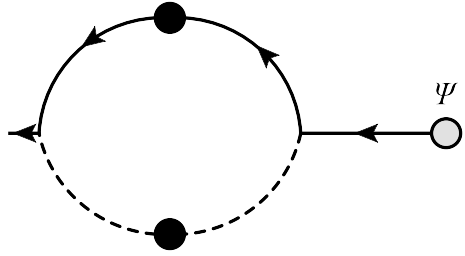}
\caption{The fermionic induced source $\etaind(X)$ in the leading order in the Yukawa model.
This diagram is obtained by connecting two ends in Fig.~\ref{fig:yukawa-K}.
By truncating $\varPsi$, we obtain the diagram in Fig.~\ref{fig:fermion-selfenergy}, which expresses $\varSigma^\R(p)$ in the resummed one-loop analysis~\cite{Hidaka:2011rz} written in Chapter~\ref{chap:ultrasoft}.}
\label{fig:yukawa-induced-source}
\end{center}
\end{figure}

The diagrammatic representation of the fermion retarded self-energy in the present approach is the same as that in the resummed perturbation theory~\cite{Hidaka:2011rz}, which is explained as follows:
The off-diagonal density matrix $\varLambda_\pm(\vk,X)$, which follows the generalized kinetic equation, is represented by Fig.~\ref{fig:yukawa-K}.
The fermionic induced source $\etaind(X)$, shown in Fig.~\ref{fig:yukawa-induced-source}, is diagrammatically obtained by connecting the ends of fermion and boson propagators in the right-hand side of Fig.~\ref{fig:yukawa-K}.
This diagram is the resummed one-loop diagram~\cite{Hidaka:2011rz} appeared in Fig.~\ref{fig:fermion-selfenergy} itself except for the fermion average field $\varPsi$.

\section{QED} 
\label{sec:kinetic-QED}

In this section, we deal with QED.
First we introduce the background field method, which is useful to construct the equations for the average fields and the Kadanoff-Baym equation in a gauge-covariant form.
Next, we derive the generalized kinetic equation in the linear response regime adopting the Coulomb gauge fixing, in which the transversality of the free photon propagator simplifies the analysis.
After the derivation, we show the equivalence between the generalized kinetic equation and the resummed perturbation theory~\cite{Hidaka:2011rz, Lebedev:1989ev}, and discuss the interpretation of the terms in the kinetic equation.
We also check that the Ward-Takahashi identity for the self-energy, consequence of the $U(1)$ gauge symmetry, is satisfied in our formalism.
Finally we discuss how to compute the higher-point-vertex function whose external momenta are all ultrasoft, and make an order estimate of it in the weak coupling regime.

\subsection{Background field gauge method} 
\label{ssc:kinetic-QED-background}

In the derivation of the average field equation and the Kadanoff-Baym equation in QED, 
it is convenient to formulate them in a covariant form under gauge transformations.
For this purpose, we employ the background field gauge method~\cite{\BackgroundFieldMethod, Blaizot:2001nr}.
In this method, the following generating functional is employed:
\begin{align}
\begin{split}
\tilde{Z}[j,\eta,\overline{\eta}; A,\varPsi,\overline{\varPsi}]=\int [{\cal D}{a}][{\cal D}\overline{\psi}][{\cal D}\psi]e^{ iS},
\label{eq:generatingFunctionalQED}
\end{split}
\end{align}
with
\begin{equation}
\begin{split}
S&= \int_C d^4x 
\left[  {\cal L}[A_\mu+a_\mu, \varPsi+\psi, \overline{\varPsi}+\overline{\psi}]+{\cal {L}}_{\mathrm {GF}}
-(j^\mu a_\mu+\overline{\psi}\eta+\overline{\eta}\psi)\right] ,
\end{split}
\end{equation}
where we dropped the ghost  term, which is not coupled with the other fields.
$A^\mu$ and $a^\mu$ are vector fields, $\varPsi$ ($\overline{\varPsi}$) and $\psi$ ($\overline{\psi}$) are (anti-) spinor fields, and $j^\mu$ is the external current, respectively.
The Lagrangian of QED has the form,
\begin{align}
{\cal L}[a, \psi, \overline{\psi}]=-\frac{1}{4}F^{\mu\nu}[a]F_{\mu\nu}[a] +i\overline{\psi}\Slash{D}[a]\psi,
\end{align}
where $F_{\mu\nu}[a]\equiv \partial_\mu a_\nu-\partial_\nu a_\mu $ is the field strength and $D_\mu[a]\equiv \partial_\mu +i\cp a_\mu$ is the covariant derivative.
In the background field method, the fields in the Lagrangian are decomposed to the classical field, identified as the average fields later, and fluctuations in Eq.~(\ref{eq:generatingFunctionalQED}). 
The external sources are chosen to be coupled to $a^\mu$, $\psi$, and $\overline{\psi}$, but not to $A^\mu$, $\varPsi$, and $\overline{\varPsi}$. 
We impose the following conditions:
\begin{align}
\label{eq:QED-condition}
\langle a_\mu\rangle=\langle\psi\rangle=\langle\overline{\psi}\rangle=0,
\end{align}
which implies that $A_\mu$ and $\varPsi$ ($\overline{\varPsi}$) can be interpreted as the average parts of the photon and (anti) electron field, respectively, and $-\ln \tilde{Z}$ coincides with the effective action~\cite{\BackgroundFieldMethod, Blaizot:2001nr}.

In the background gauge field method, the gauge-fixing term is chosen to be a functional of $a_\mu$ such as
\begin{align}
{\cal {L}}_{\mathrm {GF}}= -\lambda\frac{(\GaugeFixing[a])^2}{2},
\end{align}
where $\GaugeFixing[a]$ is the gauge-fixing function and $\lambda$ is the gauge-fixing parameter.
Although $a_\mu$ is fixed by the gauge fixing term,
the generating functional, Eq.~(\ref{eq:generatingFunctionalQED}),  is invariant under the background field gauge transformations defined by
\begin{align}
\label{eq:qed-gauge-trans}
\begin{split}
\varPsi(x)&\rightarrow h(x)\varPsi(x),~~ 
A_\mu(x)\rightarrow A_\mu(x)-\frac{i}{\cp}h(x)\partial_\mu h^\dagger(x), ~~\\
\psi(x)&\rightarrow h(x)\psi(x) ,~~ 
\overline{\psi}(x)\rightarrow  \overline{\psi}(x)h^\dagger(x),~~
a_\mu(x)\rightarrow a_\mu(x),\\
\eta(x)&\rightarrow h(x)\eta(x) ,~~
\overline{\eta}(x)\rightarrow \overline{\eta}(x)h^\dagger(x),~~
j^\mu(x)\rightarrow j^\mu(x),
\end{split}
\end{align}
where $h(x)\equiv \exp[{i\theta(x)}]$.

Since the fluctuations covariantly transform under Eq.~(\ref{eq:qed-gauge-trans}), the propagators also covariantly transform as
\begin{align}
D_{\mu\nu}(x,y)&\equiv  \langle\Tc a_\mu(x)a_\nu(y)\rangle_c\to D_{\mu\nu}(x,y),\\
S(x,y)&
\to h(x)S(x,y) h^\dagger(y),\\
K_\mu(x,y) &\equiv \langle\Tc\psi(x)a_\mu(y)\rangle_c \to  h(x)K_\mu(x,y).
\end{align}

Also, the Wigner transformed off-diagonal propagator in the leading order of $\cp$ is covariant, which can be confirmed by performing the gradient expansion~\cite{Blaizot:2001nr}:
\begin{align}
\label{eq:QED-K-transform}
K_\mu(k,X)\rightarrow h(X) K_\mu(k,X),
\end{align}
which implies that the Kadanoff-Baym equation covariantly transforms  with respect to the background gauge transformations as will be seen in the next subsection.

We note that, apart from the covariance with respect to the background field gauge transformation, the possible gauge-fixing dependence, which will be confirmed only in the Coulomb gauge and temporal gauge in this thesis, should be analyzed. 

\subsection{Derivation of the kinetic equation} 
\label{ssc:kinetic-QED-derivation}
We work in the Coulomb gauge-fixing condition because this gauge fixing makes the analysis simple owing to the transversality of the free photon propagator.
The gauge-fixing condition is $\GaugeFixing[a]=\partial_i a^i$ and $\lambda\rightarrow \infty$, which constrains the off-diagonal propagator as
\begin{align}
\label{eq:qed-coulomb-condition}
\partial^i_y K_i (x,y)=0.
\end{align}

The equations of motion for the average fields are given by
\begin{align}
\label{eq:qed-meanfield-fermion}
i\Slash{D}_x[A]\varPsi(x)&=\eta(x)+\etaind(x),\\ 
\label{eq:qed-meanfield-boson}
\partial^2A^\mu(x)-\partial^\mu\partial^\nu A_\nu(x)-\cp\overline{\varPsi}(x)\gamma^\mu\varPsi(x)
&=j^\mu(x)+\jind^\mu(x).
\end{align}
Here we have used Eq.~(\ref{eq:QED-condition}), and the induced fermionic source and the induced current are defined as
\begin{align}
\etaind(x)&\equiv \cp\langle\Slash{a}(x)\psi(x)\rangle_c =\cp\gamma^\mu K_\mu(x,x) ,\\
\jind^\mu(x)&\equiv \cp\langle\overline{\psi}(x)\gamma^\mu\psi(x)\rangle_c =\cp \text{Tr}(\gamma^\mu S^<(x,x)),
\end{align}
which transform 
\begin{align}
\etaind(x)\rightarrow h(x)\etaind(x),~~
  \jind^\mu(x)\rightarrow \jind^\mu(x)
\end{align}
under the background gauge transformations.
Therefore, Eqs.~(\ref{eq:qed-meanfield-fermion}) and (\ref{eq:qed-meanfield-boson}) transform covariantly with respect to the background gauge transformation.

The equations for the propagators are given by 
\begin{align}
\notag
&\Slash{D}_xK^\mu(x,y)+i\cp\gamma_\nu D^{\nu\mu}(x,y)\varPsi(x)\\
\label{eq:qed-fermion}
&\quad=-i\int^\infty_{-\infty} d^4z (\varSigma^\R(x,z)K^\mu (z,y)
+\VC^\R_{\nu}(x,z) D^{< \nu\mu}(z,y)) , \\ 
\notag
&(\partial^2 g^{\mu \nu}-\partial^\mu\partial^\nu)_yK_\nu(x,y)+\cp S^<(x,y)\gamma^\mu\varPsi(y)\\
\label{eq:qed-boson}
&\quad=\int^\infty_{-\infty} d^4z (\varPi^{\A\mu\nu}(z,y)K_\nu(x,z)-S^<(x,z)\VC^{\R\mu}(z,y)),
\end{align}
where  $\varPi_{\mu\nu}(x,y)$ and $\VC_\mu(x,y)$ are 
  the photon and the off-diagonal self-energies, respectively.  
Here we set $x^0\in C^+$ and $y^0\in C^-$.
The Wigner transformed equations read 
\begin{align}
\notag
&\left(-i\Slash{k}+\frac{\Slash{\partial}_X}{2}+i\cp\Slash{A}(X)\right)K^\mu(k,X)
+i\cp\gamma_\nu D^{<\nu\mu}(k,X)\varPsi(X) \\
\label{eq:qed-wigner-fermion}
&\quad=-i(\varSigma^\R(k,X)K^\mu (k,X)+\VC^\R_{\nu}(k,X) D^{< \nu\mu}(k,X)), \\
\notag
&(-k^2+ik\cdot\partial_X)K^{\mu}(k,X)
+\left[k^\mu k^0-\frac{i}{2}\left(\partial^{\mu}_{X}k^{0}+\partial^{0}_{X}k^{\mu}\right)\right]K_0(k,X) 
+\cp S^<(k,X)\gamma^\mu\varPsi(X) \notag\\
\label{eq:qed-coulomb-wigner-boson}
&\quad=\varPi^{\A\mu\nu}(k,X)K_\nu(k,X)-S^<(k,X)\VC^{\R \mu}(k,X), 
\end{align}
where Eq.~(\ref{eq:qed-coulomb-condition}) and the gradient expansion have been used.

Let us show that $K^0$ is negligible compared with the spatial components.
We get the following equation by multiplying Eq.~(\ref{eq:qed-wigner-fermion}) by 
\begin{align}
\left(-i\Slash{k}+\frac{\Slash{\partial}_X}{2}+i\cp \Slash{A}(X)+i\varSigma^\R(k,X)\right)
\end{align} 
from the left, subtracting Eq.~(\ref{eq:qed-coulomb-wigner-boson}) from the quantity obtained above, and setting $\mu=0$: 
\begin{align}
\begin{split}
(2ik\cdot\partial_X-2\cp k\cdot A(X)-\{\Slash{k},\varSigma^\R(k,X)\}
&+ (k^0)^2)K_0(k,X)
-\varPi^{\A 0\nu}(k,X)K_\nu(k,X)\\
&=\cp S^<(k)\V^0(k,X)\\ 
 (k^0)^2 K_0(k,X)
 -\varPi^{\A 0i}(k,X)K_i(k,X)
 &=\cp S^<(k)\V^0(k,X).
\end{split}
\end{align}
Here we have introduced 
\begin{align}
{\cp}\V^\mu(k,X)\equiv \cp\gamma^\mu\varPsi(X)+\VC^{\R\mu}(k,X).
\end{align}
Since $k^0\sim T$, we see that
\begin{align}
\label{eq:K0-estimate}
K^0\sim \cp^2 K^i.
\end{align}

Using Eq.~(\ref{eq:K0-estimate}), Eq.~(\ref{eq:qed-coulomb-wigner-boson}) becomes
\begin{align}
\label{eq:qed-coulomb-wigner-boson2}
&(-k^2+ik\cdot\partial_X)K^{\mu}(k,X)+k^\mu k^0K_0(k,X) 
+\cp S^<(k,X)\gamma^\mu\varPsi(X) \notag\\
&\quad=\varPi^{\A\mu\nu}(k,X)K_\nu(k,X)-S^<(k,X)\VC^{\R \mu}(k,X).
\end{align}
Multiplying Eq.~(\ref{eq:qed-wigner-fermion}) by 
\begin{align}
\left(-i\Slash{k}+\frac{\Slash{\partial}_X}{2}+i\cp\Slash{A}(X)+i\varSigma^\R(k,X)\right)
\end{align}
 and Eq.~(\ref{eq:qed-coulomb-wigner-boson2}) by $P^{T}_{\mu i}(k)$ defined below, and subtracting the latter from the former, we obtain
\begin{align}
\label{eq:qed-coulomb-K}
\begin{split}
&(-2ik\cdot\partial_X+2\cp k\cdot A(X)+\{\Slash{k},\varSigma^\R(k,X)\})K^i(k,X)
+P^{T }_{\alpha i}(k)\varPi^{\A \alpha\nu}(k,X) K_\nu(k,X)\\
&\quad=
-\cp(\Slash{k}D^{<  \nu i} (k,X)+P^{T  \nu i} (k)S^<(k,X))\V_\nu(k,X),
\end{split}
\end{align}
where we have used 
\begin{align}
P^{T}_{\mu i}(k)K^\mu(k,X)=-K_i(k,X)=K^i(k,X).
\end{align} 
One can show that the background fields and the coupling dependences in the diagonal propagators are weak, so that we can replace the electron and the photon propagator, which is given as follows, by that in the free limit at equilibrium:
\begin{align}
D^{0<}_{\mu\nu}(k)=&\rho^0(k)\nb(k^0)P^T_{\mu\nu}(k).
\end{align}
The diagonal self-energies at on-shell in the leading order are given by 
\begin{align}
\{\Slash{k}, \varSigma^{\R ({\mathrm {eq}})}(k) \}&=\me^2-2i\zetae k^0,\\
P^{T }_{\alpha i}(k)\varPi^{\A  ({\mathrm {eq}}) \alpha\nu}(k)&= -\mg^2 P^{T \nu}_i(k).
\end{align} 
In contrast to the case of Yukawa model, the damping rate of the electron $\zetae$ cannot be neglected because $\zetae k^0\sim \me^2$, while 
 the photon damping rate of order $\cp^4 T\ln(1/\cp)$ can be neglected~\cite{\HardPhotonDamping}.
We note that the longitudinal part of the photon self-energy does not contribute because the projection operator $P^{T i}_{\alpha}(k)$ is multiplied.

The off-diagonal self-energy in the leading order has the following expression, which is similar to that in the Yukawa model:
\begin{align}
\label{eq:qed-vertex-correction}
\VC^{\R\mu}(k,X)= -\cp^2\int\frac{d^4 k'}{(2\pi)^4}\gamma^\nu S^{0\R}(k+ k')\gamma^\mu K_\nu(k',X) .
\end{align}
The off-diagonal self-energy  in $\V^\mu(k,X)$ has to be retained in the case of QED in contrast to the case of the Yukawa model since there is no special suppression mechanism.

\begin{figure*}[t]
\begin{center}
\includegraphics[width=0.9\textwidth]{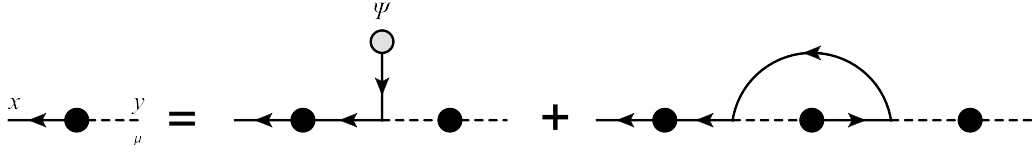}
\caption{The diagrammatic representation of self-consistent equation for $K^\mu(x,y)$ in the leading order.
For simplicity, $A^\mu$ is not drawn.
}
\label{fig:QED-K}
\end{center}
\end{figure*}

By substituting these expressions in Eq.~(\ref{eq:qed-coulomb-K}), we obtain
\begin{align}
\label{eq:qed-coulomb-result-K}
\begin{split}
&(-2ik\cdot D_X-2i\zetae k^0-\delta m^2)K^i(k,X)
=-\cp\Slash{k} P^{T \nu i}(k) \rho^0(k)(\nb(k^0) +\nf(k^0))\\
&\qquad\times\left(\gamma_\nu\varPsi(X)
+ \cp\int\frac{d^4 k'}{(2\pi)^4}\frac{k^\alpha \gamma_\nu+\gamma^\alpha k'_\nu}{k\cdot k'} K_\alpha(k',X)\right).
\end{split}
\end{align}
Here we have used $P^T_{\mu\alpha}(k) P^{T \alpha}_{\nu}(k)=-P^T_{\mu\nu}(k) $ and introduced $\delta m^2\equiv \mph^2-\me^2$.
In this gauge-fixing condition, it is apparent that only the transverse component of the thermal photon contributes to $K^i(k,X)$ because of the projection operator $P^{T \nu i}(k)$ appearing in the right-hand side of Eq.~(\ref{eq:qed-coulomb-result-K}). 
We note that this equation transforms covariantly with respect to the background gauge transformation from Eq.~(\ref{eq:QED-K-transform}).
The diagrammatic representation of this equation is shown in Fig.~\ref{fig:QED-K}.

From Eq.~(\ref{eq:qed-coulomb-result-K}), we write $K^i(k,X)$ in terms of the off-diagonal self-energy for later use:
\begin{align}
\label{eq:qed-coulomb-K-vertex}
&K^i(k,X)
={\cp}\frac{\Slash{k}P^{T i}_{\nu}(k)  \rho^0(k)(\nb(k^0) +\nf(k^0))}{2ik\cdot D_X +2i\zetae k^0+\delta m^2}\V^\nu(k,X).
\end{align}

\subsection{Kinetic interpretation}
\label{ssc:kinetic-QED-kinetic}

Next, we derive  the linearized kinetic equation.
Multiplying Eq.~(\ref{eq:qed-coulomb-result-K}) by $\gamma_i$ from the left and setting $A^\mu(X)=0$, we obtain
\begin{align}
\label{eq:QED-coulomb-result-Kslash}
\begin{split}
&(-2ik\cdot \partial_X-2i\zetae k^0-\delta m^2)\Slash{K}(k,X)\\
&\quad=-\cp 2\Slash{k} \rho^0(k)(\nb(k^0) +\nf(k^0))\varPsi(X)\\
&\qquad+\gamma_i\Slash{k}P^{T i}_{\nu}(k)  \rho^0(k)(\nb(k^0) +\nf(k^0))
\cp^2\int\frac{d^4 k'}{(2\pi)^4}\frac{k^\alpha\gamma^\nu+k'^\nu\gamma^\alpha}{k\cdot k'}K_\alpha(k',X).
\end{split}
\end{align}
We  decompose $\varLambda^\mu_\pm(\vk,X)$ into positive and negative energy terms as 
\begin{align}
K^\mu(k,X)\equiv2\pi\delta(k^2)[\theta(k^0)\varLambda^\mu_+(\vk,X)+\theta(-k^0)\varLambda^\mu_-(-\vk,X)],
\end{align}
 so that we arrive at the kinetic equation\footnote{We note that the equation which is similar to Eq.~(\ref{eq:QED-kineticeq}) was obtained in Ref.~\cite{Markov:2001ke}.
However, the authors of Ref.~\cite{Markov:2001ke} failed to write down all of the leading contributions:
their equation does not have the third term in the left-hand side and the second term in the right-hand side of Eq.~(\ref{eq:QED-kineticeq}).} from Eq.~(\ref{eq:QED-coulomb-result-Kslash}):
\begin{align}
\label{eq:QED-kineticeq}
\begin{split}
&\left(2iv\cdot\partial_X\pm\frac{\delta m^2}{|\vk|}+2i\zetae\right)\Slash{\varLambda}_{\pm}(\vk, X)
=2\cp \Slash{v}[\nb(|\vk|)+\nf(|\vk|)]\varPsi(X)\\
&~~~-\cp^2\gamma_{i}\Slash{v}[\nf(|\vk|)+\nb(|\vk|)]P^{\nu i }_{\mathrm T}(v)
\sum_{s=\pm}\int\frac{d^3\vk '}{(2\pi)^3}\frac{1}{2|\vk '|}\frac{s|\vk|v^\alpha\gamma_\nu\pm |\vk'|v_{ \nu}'\gamma^\alpha}{|\vk| |\vk '|v\cdot v'}\varLambda_{s \alpha}(\vk',X),
\end{split} 
\end{align}
where we have introduced $v^{\prime\mu}\equiv (1, \hat{\vk} ')$.
There are two terms that do not appear in the Yukawa model analyzed in Sec.~\ref{sec:kinetic-yukawa}.
One is the last term in the right-hand side.
The first term in the right-hand side is interpreted as the counterpart of the force term in the diagonal case~\cite{\HTLVlasov, Blaizot:2001nr}, so the last term in the right-hand side acts like ``{\it  the correction to the force term},'' at least in the linear response regime.  
Note that this term mixes the positive and negative energy modes, in contrast to the case of the Yukawa model.

The other is the third term in the left-hand side.
 This term has a similar form to the collision term in the relaxation time approximation of the diagonal case, i.e., pure imaginary constant ($2i\zetae$) times $\varLambda_\pm(\vk,X)$.
For this reason, we call this term ``{\it collision term}.''
We note that this term is negligible in the case of the Yukawa model as shown in the previous section.

In the diagonal case~\cite{\UltrasoftGluon, Blaizot:2001nr}, the collision term contains momentum integral for the diagonal density matrix.
In contrast, such term does not survive in the linearized equation in the off-diagonal case.
As a result, the collision term in the off-diagonal kinetic equation has a similar form to that in the relaxation time approximation.

We emphasize that the off-diagonal self-energy is not negligible in the off-diagonal kinetic equation, while negligible in the diagonal one because the terms containing that quantity are higher order in $\varPsi$. 
This fact makes the correction to the force term, which is absent in the diagonal case, appears in Eq.~(\ref{eq:QED-kineticeq}).

 As we discussed in Sec.~\ref{ssc:kinetic-yukawa-kinetic}, both of the usual Boltzmann equation and our generalized and linearized kinetic equation are composed of the non-interacting part, the interaction part between the hard particle and the average ultrasoft field, and the interaction part among the hard particles.
Which part is the counterpart of the mass difference term, the collision term, and the correction to the force term?
Because the mass difference and the collision term come from the self-energies at equilibrium, they correspond to the interaction part among the hard particle.
The correction to the force term is a part of the interaction part between the hard particle and the average ultrasoft field.


\begin{table}[t]
\begin{center}
\caption{The correspondence between the resummed perturbation theory and the generalized and linearized kinetic equation.}
\begin{tabular}{c c}
\hline
Diagrammatic method & kinetic equation  \\ \hline \hline
thermal mass difference & mass difference term \\
damping rate& collision term \\
ladder diagrams&  correction to force term \\
\hline
\end{tabular}
\label{tab:correspondence}
\end{center}
\end{table}

\subsection{Correspondence between kinetic theory and resummed perturbation theory}
Here, let us show the equivalence between  Eq.~(\ref{eq:qed-coulomb-result-K}) and the self-consistent equation in the resummed perturbation theory~\cite{Hidaka:2011rz, Lebedev:1989ev} in Chapter~\ref{chap:ultrasoft}.
To this end, we rewrite Eq.~(\ref{eq:qed-vertex-correction}) in terms of the off-diagonal self-energy using Eq.~(\ref{eq:qed-coulomb-K-vertex}):
\begin{align}
\begin{split}
\V^\mu(k,X)&=\gamma^\mu\varPsi(X)
 -\cp^2\int\frac{d^4 k'}{(2\pi)^4}\gamma^\nu S^{0\R}(k+ k')\gamma^\mu\\
 &\quad\times \frac{\Slash{k}'P^{T }_{\alpha\nu}(k')  \rho^0(k')[\nb(k'^{0}) +\nf(k'^{0})]}{2ik'\cdot \partial_X +2i\zetae k'^{0}+\delta m^2}
\V^\alpha(k',X) .
\end{split}
\end{align}
Here we set $A^\mu=0$.
By performing the Fourier transformation, Eq.~(\ref{eq:fourier-p}), we get
\begin{align}
\label{eq:QED-BS}
\begin{split}
\varGamma^\mu(k,p)&=\gamma^\mu
 -\cp^2\int\frac{d^4 k'}{(2\pi)^4}\gamma^\nu S^{0\R}(k+ k')\gamma^\mu\\
 &\!\!\!\quad\times \!\frac{\Slash{k}'\! P^{T }_{\alpha\nu}(k')  \rho^0(k')[\nb(k'^{0}) +\nf(k'^{0})]\varGamma^\alpha(k',p)}{2k'\cdot p +2i\zetae k'^{0}+\delta m^2} ,
\end{split}
\end{align}
where $\V^{\mu}(k,p)\equiv \varGamma^\mu(k,p)\varPsi(p)$.
We note that $\varGamma^\mu(k,p)$ is the vertex function~\cite{Hidaka:2011rz, Lebedev:1989ev} introduced in Chapter~\ref{chap:ultrasoft} whose momenta are hard and ultrasoft.
Equation~(\ref{eq:QED-BS}) is none other than Eq.~(\ref{eq:vertex}), which is the self-consistent equation in the resummed perturbation~\cite{Hidaka:2011rz, Lebedev:1989ev}.

The retarded fermion self-energy is also written in terms of the vertex function:
from Eq.~(\ref{eq:qed-coulomb-K-vertex}), we arrive at
\begin{align}
\begin{split}
\varSigma^\R(p)&= \cp\int \frac{d^4k}{(2\pi)^4} \frac{\delta\Slash{K}(k,p)}{\delta\varPsi(p)}\\
&=\cp^2\int \frac{d^4k}{(2\pi)^4}\frac{\gamma_i\Slash{k}P^{T i}_{\nu}(k)  \rho^0(k)[\nb(k^0) +\nf(k^0)]}{2k\cdot p +2i\zetae k^0 +\delta m^2}
 \varGamma^\nu(k,p).
\end{split}
\end{align}
This expression equals to Eq.~(\ref{eq:selfEnergy}), which is the expression of the fermion retarded self-energy in the resummed perturbation theory\footnote{Equation~(\ref{eq:QED-BS}) is analytically solved for the energy region $|p_0+i\zetae |\ll \cp^2T$ in the previous chapter~\cite{Hidaka:2011rz}.}~\cite{Hidaka:2011rz, Lebedev:1989ev}.
Thus we see that Eq.~(\ref{eq:qed-coulomb-result-K}), derived in the non-equilibrium state in a linear response regime from the Kadanoff-Baym equation, is equivalent to 
the self-consistent equation in the resummed perturbation theory, which is constructed in the thermal equilibrium state.

Here we discuss the correspondence between each prescription in the resummed perturbation theory~\cite{Hidaka:2011rz, Lebedev:1989ev} and each term in the kinetic equation.
As in the Yukawa model, the resummation of the thermal mass difference in the resummed perturbation theory corresponds to the mass difference term in the kinetic equation.
The damping rate corresponds to the collision term since both of them contain the damping rate of the hard electron, $\zetae$.
The ladder summation in the resummed perturbation theory~\cite{Hidaka:2011rz, Lebedev:1989ev} is caused by the correction to the force term in the kinetic equation.
Thus, the ladder summation corresponds to the correction to the force term.
These interpretations are summarized in the Table.~\ref{tab:correspondence}.

\subsection{$n$-point functions $(n\geq 3)$} 

The fermionic induced source $\etaind$ generates the $n$-point functions with $n\geq 3$, not only the fermion self-energy.
In this subsection, we derive the self-consistent equation determining the $n$-point function
whose external lines consist of two fermions ($\varPsi$) and $(n-2)$ bosons ($A^\mu$) with ultrasoft external momenta, and make an order estimate of the quantity.
For example in the case of $n=3$, we obtain the correction to the bare three-point function, $\cp\gamma_\mu \delta^{(4)}(p-q-r)$, from $\etaind$~\cite{\HTLVlasov, Blaizot:2001nr, Taylor:1990ia, Braaten:1991gm, Frenkel:1991ts}:
\begin{align}
\label{eq:QED-ultrasoft-vertex}
\begin{split}
\delta^{(4)}(p-q-r)\cp\delta\varGamma^\mu(p,-q,-r)
&\equiv\left.\frac{\delta^2 \etaind(p)}{\delta\varPsi(q) \delta A_\mu(r)}\right|_{A=0}\\
&= \cp \int\frac{d^4k'}{(2\pi)^4}\frac{\delta^2 }{\delta\varPsi(q) \delta A_\mu(r)}\Slash{\delta K}(k,p).
\end{split}
\end{align}
Here we have expanded $K^\mu(k,X)$ around $A^\mu=0$: 
\begin{align} 
K^\mu(k,X)=K^\mu(k,X)_{A=0}+\delta K^\mu(k,X) +O(A^2\varPsi),
\end{align} 
where $K^\mu(k,X)_{A=0}$ contains one $\varPsi$ while $\delta K^\mu(k,X)$ contains one $\varPsi$ and one $A^\mu$.

$\delta K^\mu(k,X)$ can be obtained by the following way.
Collecting terms that contain one $A^\mu$ in Eq.~(\ref{eq:qed-coulomb-result-K}), we obtain
\begin{align}
\begin{split}
&(-2ik\cdot \partial_X-2i\zetae k^0-\delta m^2)\delta K^i(k,X)
+2\cp k\cdot A(X) K^i(k,X)_{A=0}\\
&\quad=-\cp^2\Slash{k} P^{T \nu i}(k) \rho^0(k)[\nb(k^0) +\nf(k^0)]
 \int\frac{d^4 k'}{(2\pi)^4}\frac{k^\alpha \gamma_\nu+\gamma^\alpha k'_\nu}{k\cdot k'} \delta K_\alpha(k',X). \label{eq:deltaKi}
\end{split}
\end{align}
Since $K^\mu(k,X)_{A=0}$ is determined by setting $A^\mu=0$ in Eq.~(\ref{eq:qed-coulomb-result-K}), this equation is closed and $\delta K^\mu(k,X)$ can be determined.

Let us estimate  the order of $\delta K^\mu(k,X)$.
From Eq.~(\ref{eq:qed-coulomb-result-K}) and $K^\alpha_{A=0}\sim \cp^{-1}T^{-3}\varPsi $, we 
find 
\begin{align}
\begin{split}
\delta K^\mu&\sim \cp^{-1}T^{-1}A^\mu K^\alpha_{A=0}\\
&\sim \cp^{-2}T^{-4}\varPsi A^\alpha.
\end{split}
\end{align}
Therefore, the vertex correction is estimated as 
\begin{align}
\cp\delta\varGamma^\mu\sim \cp^{-1},
\end{align}
which is much larger than the bare vertex, $\cp\gamma^\mu\sim\cp$.
Similar order estimate for the $n$-point function with $n>3$ can be done 
with the same procedure; as a result, we find that the order of the $n$-point-vertex function is $\cp^{2-n}$ .

\subsection{Ward-Takahashi identity}

We show that the off-diagonal self-energy given in Eq.~(\ref{eq:qed-vertex-correction}) satisfies the Ward-Takahashi (WT) identity.
From Eq.~(\ref{eq:qed-vertex-correction}), we get
\begin{align}
\label{eq:qed-WT}
\begin{split}
k_\mu\VC^{\R \mu}(k,X)
&= \cp^2\int\frac{d^4 k'}{(2\pi)^4}\gamma^\nu \frac{\Slash{k}+\Slash{k}'}{(k+k')^2}(\Slash{k}+\Slash{k}'-\Slash{k}')K_\nu(k',X) \\
&= \cp^2\int\frac{d^4 k'}{(2\pi)^4}\Slash{K}(k',X)=\cp\etaind(X).
\end{split}
\end{align}
Here we have used $\Slash{k}'K_\nu(k',X)=0$, which can be confirmed by multiplying Eq.~(\ref{eq:qed-coulomb-result-K}) by $\Slash{k}$ from the left.
This equation generates the WT identity (Eq.~(\ref{eq:mod-WT})) derived in Chapter~\ref{chap:ultrasoft} by setting $A^\mu=0$ and differentiating with respect to $\varPsi$.
The WT identity implies that the vertex correction is not negligible because the identity relates the vertex correction to the fermion self-energy, which is much larger than the inverse of the free fermion propagator with an ultrasoft momentum.
In the Yukawa model, the WT identity associated with gauge symmetries is absent from the outset, so the smallness of the vertex correction is not in contradiction with any identity.

Equation~(\ref{eq:qed-WT}) can be derived from the conservation law of the induced current, 
\begin{align}
-i\cp(\overline{\eta}_{\text {ind}}\varPsi-\overline{\varPsi}\etaind)(x)
=\partial_\mu \jind^\mu(x).
\label{eq:conservationLaw}
\end{align}
By differentiating Eq.~(\ref{eq:conservationLaw}) with respect to $ \overline{\varPsi}(y)$, we obtain
\begin{align}
\label{eq:QED-WT-pre}
\begin{split}
&-\cp\left(\frac{\delta{\overline{\eta}_{\text {ind}}}(x)}{\delta\overline{\varPsi}(y)}\varPsi(x)-\delta^{C}(x^0-y^0)\delta^{(3)}(\vx-\vy)\etaind(x)\right)
=\partial^x_\mu \VC^\mu(y,x).
\end{split} 
\end{align}
Here $\delta^{C}(x-y)$ is the delta function defined along the contour $C$.
By multiplying this equation by $\int d^4s \exp({ik\cdot s})$, we get
\begin{align} 
\begin{split}
&-\cp\int d^4s e^{ik\cdot s}\left(\frac{\delta{\overline{\eta}_{\text {ind}}}(x)}{\delta\overline{\varPsi}(y)}\varPsi(x)
-\delta(x^0-y^0)\delta^{(3)}(\vx-\vy)\etaind(x)\right)\\
&\quad=\int d^4s e^{ik\cdot s}\partial^s_\mu \VC^\mu(y,x).
\end{split}
\end{align}
Here we have set $x^0$, $y^0\in C^+$  and neglected the sub-leading terms.
The first term in the left-hand side has the same order of magnitude as the hard fermion self-energy $\varSigma(k)$ times $\varPsi(X)$, so that term is negligible.
Thus the left-hand side becomes $\cp\etaind(X)$.
The right-hand side becomes
\begin{align}
\begin{split}
&\int^0_{-\infty} d^4s e^{ik\cdot s}\partial^s_\mu \VC^{>\mu}(y,x)
+\int^\infty_0 d^4s e^{ik\cdot s}\partial^s_\mu \VC^{<\mu}(y,x)\\
&\quad=\int^\infty_{-\infty} d^4s e^{ik\cdot s}\partial^s_\mu \VC^{>\mu}(y,x)
+\int^\infty_0 d^4s e^{ik\cdot s}\partial^s_\mu (\VC^{<\mu}(y,x)-\VC^{>\mu}(y,x))\\
&\quad=-ik_\mu \VC^{>\mu}(-k,X)-k_\mu \VC^{\A\mu}(-k,X).
\end{split}
\end{align}
We see that the first term in the last line is negligible because of the on-shell condition.
Thus we obtain
\begin{align}
\label{eq:QED-WT2}
\cp\etaind(X)
=- k_{\mu}\VC^{\R \mu}(-k,X),
\end{align} 
if we remember that $\VC^\R(k,X)\simeq \VC^\A(k, X)$.
This equation is nothing but  Eq.~(\ref{eq:qed-WT}).

\section{Brief summary}
\label{sec:kinetic-summary}

In this chapter, we derived the linearized off-diagonal Boltzmann equation for the ultrasoft fermion average field in the Kadanoff-Baym formalism, and showed that the resultant equation is equivalent to the self-consistent equation in the resummed perturbation theory at the leading order of the perturbative expansion, while it has been known that the Vlasov equation is equivalent to the basic equation of the HTL approximation. 
This equivalence shows that the resummation scheme of that perturbation theory contains the interaction effect among the particles beyond the mean field approximation.
We note that the off-diagonal Boltzmann equation contains not only the collision term but also the thermal mass terms and the off-diagonal self-energy term, in contrast to the usual Boltzmann equation.
We also derived the equation which determines the $n$-point functions with external lines for a pair of fermions and $(n-2)$ bosons with ultrasoft momenta, by considering the non-linear response regime using the gauge symmetry. 
We also derived the Ward-Takahashi identity from the conservation law of the electromagnetic current in the Kadanoff-Baym formalism.

\chapter{Summary and Outlook}
\label{chap:summary}
\thispagestyle{headings}

In this thesis, we developed  the resummed perturbation theory which takes into account the separation of the scale and enables us to successfully regularize the infrared singularity.
By using that method, we analyzed the spectral properties of the ultrasoft fermion, including its existence itself, in the Yukawa model and QED/QCD.
The procedure of that method consists of the resummation of the asymptotic thermal masses.
In QED/QCD, the summation of the ladder diagrams is also necessary.
As a result of the analysis, we established the existence of the novel fermionic mode in that energy region, and obtain the expressions of the pole position and the residue of that mode, which are summarized in Table.~\ref{tab:ultrasoft-expression}.
We also showed that the resultant fermion propagator and the vertex function satisfy the Ward-Takahashi identity in QED/QCD. 
We also derived the linearized and generalized Boltzmann equation for ultrasoft fermion excitations, which describe the time-evolution of the amplitude of the process of changing the fermion to the boson, from the Kadanoff-Baym equation in a Yukawa model and QED.
We showed that this equation is equivalent to the self-consistent equation in the resummed perturbation theory used in the analysis of the ultrasoft fermion spectrum at the leading order. 
It helps us to establish the foundation of the resummed perturbation theory.
We derived the Ward-Takahashi identity from the conservation law of the electromagnetic current in QED.
Furthermore, we derived the equation that determines the $n$-point function with external lines for a pair of fermions and $(n-2)$ bosons with ultrasoft momenta in QED, by using $U(1)$ gauge symmetry.

We analyzed the finite temperature and zero chemical potential ($\mu$) case in this thesis, so it is quite natural to ask how our result on the ultrasoft fermion mode is modified in other region.
Such analysis is already done by the author and his collaborator~\cite{Satow-Blaizot}, and we hope to report its results elsewhere.
We just describe the analysis we performed and its results in the following:
We constructed the resummed perturbation theory in high-$\mu$ and $T=0$ case, and analyzed the fermion spectrum with the ultrasoft momentum ($\lesssim \cp^2\mu$) in the Yukawa model.
As a result, we found that the ultrasoft fermion mode does not exist in this case.
This result is consistent with the absence of the charge symmetry in high-$\mu$ case, because the charge symmetry is essentially important for generating the ultrasoft fermion mode, as discussed in Sec.~\ref{ssc:intro-ultrasoft}.
We also investigate how large $\mu$ kills the ultrasoft fermion mode, and found that the maximum chemical potential in which the mode persists is of order $T$.

The resummed perturbation scheme used in this thesis can be applied to other quantities in the low energy region, in general multi-component system.
We briefly discuss some examples in the following:
\begin{itemize}
\item As we wrote in Chapter~\ref{chap:intro}, the computation of the transport coefficient also needs taking into account the interaction effect among the hard particles.
Usually, only the collision effect need to be resummed.
However, our analysis in this thesis suggests that the difference of the dispersion relations of the particles also needs to be taken into account, when two kinds of particle appear in the loop integral.
It is the case in the computation of the flavor diffusion constant~\cite{Arnold:2000dr}: the quarks with different flavor appear in the integral.
Since the current quark masses are different in each flavor, it is quite interesting to investigate the effect of the difference of the current quark mass by performing the analysis which takes into account the current quark mass.

\item The perturbative analysis of the neutrino spectrum~\cite{\Neutrino} around the critical temperature of the electroweak transition, which is expected to be realized at the early universe, needs the resummation used in this thesis since the bare masses of the weak bosons becomes negligible compared with the temperature.
It is an interesting task to perform such analysis, and investigate the effect of the fermion spectrum in the ultrasoft region, on the properties of the early universe.

\item Another example is the supersymmetric system in both of the relativistic and nonrelativistic system.
In such system, the SUSY is spontaneously broken by the finite temperature/density effect~\cite{Buchholz:1997mf, Yu:2007xb}, and as a result, the fermionic zero mode, goldstino, appears~\cite{\Goldstino, Yu:2007xb}.
As is described in Sec.~\ref{ssc:intro-ultrasoft}, the analysis of the goldstino, which uses the resummation scheme, was performed in relativistic system~\cite{\Goldstino}. 

Though the SUSY has not been observed as a fundamental symmetry in the real world, it was suggested to simulate the supersymmetric system using the cold atom system~\cite{\SUSYColdAtom}, which is nonrelativistic.
Such system also needs the resummation scheme~\cite{Shi:2009ak}, and it may be surprising since the power counting rule in the nonrelativistic system is different from that in the relativistic system.
We note that, however, there is an important difference between the relativistic system and the nonrelativistic one:
Since the energy-momentum tensor and the supercurrent are not in the same supermultiplet in the nonrelativistic system, the goldstino can not be interpreted as the phonino.
The dynamical analysis on the goldstino in the nonrelativistic system was performed~\cite{Shi:2009ak}, but without the Bose-Einstein condensation (BEC).
The analysis in the system with BEC is quite attractive task.

\end{itemize}

\chapter*{Acknowledgments}
\thispagestyle{headings}
The author is very grateful to his advisor, Teiji Kunihiro, for the collaboration to this work and the supervision.
He supported and encouraged the author's research.

The author thanks Yoshimasa Hidaka for the collaboration to this work, and his kind encouragement to the author.
The author acknowledges Jean-Paul Blaizot, who gave the author the hospitality during the author's staying in his group, and contributed to the author's research as a collaborator.
The author also thanks the members of Nuclear Theory Groups in Kyoto University and in Yukawa Institute for Theoretical Physics.

The author is supported by Japan Society for the Promotion of Science and a Grant-in-Aid for Scientific Research [(C) No.~24$\cdot$56384].

\appendix 
\newpage
\thispagestyle{headings}
\chapter{Kadanoff-Baym Equation in Temporal Gauge}  
\thispagestyle{headings}
\label{app:temporal}

In this thesis, the calculation was performed in the Coulomb gauge.
However, since the resummed perturbation theory was first proposed in the temporal gauge~\cite{Lebedev:1989ev}, we need to check that the results in this thesis also can be obtained in the temporal gauge.
To this end, we show that the equation determining $K^\mu$ in the temporal gauge is the same as that in the Coulomb gauge, Eq.~(\ref{eq:qed-coulomb-result-K}), in QED in this Appendix.
Showing that equation can be obtained also in the temporal gauge suffices to show that the expression of the properties of the ultrasoft fermion mode obtained in the Coulomb gauge do not change in the temporal gauge, because that equation is equivalent to the self-consistent equation in the resummed perturbation theory, which was used to analyze the ultrasoft fermion mode.

The gauge-fixing condition in the temporal gauge is $\GaugeFixing[a]=a^0$ and $\lambda\rightarrow \infty$. 
This condition is equivalent to the constraint $a^0=0$.
Because of this constraint, we have
\begin{align}
K^0(x,y)=D^{0\mu}(x,y)=0.
\end{align}
The equations governing $K^i$ are 
\begin{align}
\label{eq:qed-temporal-wigner-fermion}
&\left(-i\Slash{k}+\frac{\Slash{\partial}_X}{2}+i\cp\Slash{A}(X)\right)K^i(k,X)
+i\cp\gamma_j D^{< j i}(k,X)\varPsi(X)\\
\notag
&\quad=-i(\varSigma^\R(k,X)K^i (k,X)+\VC^\R_{j}(k,X) D^{< j i}(k,X)), \\
\label{eq:qed-temporal-wigner-boson}
&(-k^2+ik\cdot\partial_X)K^{i}(k,X)
+\left(k-\frac{i\partial_X}{2}\right)^i \left(k-\frac{i\partial_X}{2}\right)^jK_j(k,X)\\
\notag
&\qquad+\cp S^<(k,X)\gamma^i\varPsi(X)\\
\notag
&\quad=\varPi^{\A i j}(k,X)K_j(k,X)-S^<(k,X)\VC^{\R i}(k,X).
\end{align}
From these equations, we obtain
\begin{align}
\label{eq:qed-temporal-K}
\begin{split}
&(-2ik\cdot\partial_X+2\cp k\cdot A(X)+\{\Slash{k},\varSigma^\R(k,X)\})K^i(k,X)\\
&\qquad-\left(k-\frac{i\partial_X}{2}\right)^i \left(k-\frac{i\partial_X}{2}\right)^jK_j(k,X)
+\varPi^{\A ij}(k,X) K_j(k,X)\\
&\quad=
-(\Slash{k}D^{<  j i} (k,X)+\delta^{ij}S^<(k,X))\V_j(k,X).
\end{split}
\end{align}

Here let us evaluate $k^i K_i(k,X)$, which is the longitudinal component of $K^i(k,X)$.
By multiplying Eq.~(\ref{eq:qed-temporal-K}) by $k_i$, we get
\begin{align}
\begin{split}
&|\vk|^2 \left(k-\frac{i\partial_X}{2}\right)^jK_j(k,X)+k_i\varPi^{\A ij}(k,X) K_j(k,X)
=
- S^<(k,X)k_i\V_i(k,X).
\end{split}
\end{align}
Here we have neglected the terms that are of order $\cp^2 T^2 k^i K_i(k,X)$.
We see that $k^i K_i(k,X)\sim\cp^2 T K_i(k,X)$ and thus the longitudinal component of $K_i(k,X)$, $\hat{k}^i K_i(k,X)$, is negligible compared with the transverse component of $K_i(k,X)$.
We note that $K^0=0$, which is the result of the gauge-fixing condition, and $k^i K_i=0$ are valid also in the Coulomb gauge in the leading order.
Furthermore, also the free photon propagator at equilibrium is the same as that in the Coulomb gauge. 
Thus, we can obtain Eq.~(\ref{eq:qed-coulomb-result-K}) in the same way as in the Sec.~\ref{ssc:kinetic-QED-derivation}.

\chapter{Formulae}  
\thispagestyle{headings}
\label{app:formula}

In this appendix, we write some formulae used in the text.

\begin{align}
\label{eq:integral-formula-f1}
 \int^\infty_0 d|\vk| |\vk| \nf(|\vk|) &= \frac{\pi^2 T^2}{12},\\
 \label{eq:integral-formula-b1}
 \int^\infty_0 d|\vk| |\vk| \nb(|\vk|)&= \frac{\pi^2 T^2}{6},\\
\label{eq:integral-formula-f3}
 \int^\infty_0 d|\vk| |\vk|^3 \nf(|\vk|) &= \frac{7\pi^4 T^4}{120}, \\
 \label{eq:integral-formula-b3}
 \int^\infty_0 d|\vk| |\vk|^3 \nb(|\vk|)&= \frac{\pi^4 T^4}{15}.
\end{align}

\begin{align}
\label{eq:integral-formula-cos2}
\int dx \frac{x^2}{x-a}&=\frac{(x-a)^2}{2} +2a(x-a) +a^2\ln(x-a).
\end{align}


\end{document}